\documentstyle[psfig,twoside,a4]{article} 
\newcommand{\vbluw}{$V\! BLUW$}
\newcommand{\uvby}{{\it uvby}}
\newcommand{\UBV}{{\it UBV}}
\newcommand{\HIP}{HIP}
\newcommand{\FHIP}{HIP}
\newcommand{\HD}{HD}
\newcommand{\IC}{IC}
\newcommand{\FIC}{IC}
\newcommand{\NGC}{NGC}
\newcommand{\FNGC}{NGC}
\newcommand{\BD}{BD}
\newcommand{\CD}{CD}
\newcommand{\SAO}{SAO}
\newcommand{\HR}{HR}
\newcommand{\FHR}{HR}
\newcommand{\Sh}{S}

\newcommand{\Kh}{Kh}
\newcommand{\FKh}{Kh}
\def\spose#1{\hbox to 0pt{#1\hss}}
\def\la{\mathrel{\spose{\lower 3pt\hbox{$\sim$}}
    \raise 2.0pt\hbox{$<$}}}
\def\ga{\mathrel{\spose{\lower 3pt\hbox{$\sim$}}
    \raise 2.0pt\hbox{$>$}}}

\def\fm{\hbox{$.\!\!^{\rm m}$}}
\def\fdg{\hbox{$.\!\!^\circ$}}
\def\farcs{\kern 0.08ex\hbox{$.\!\!^{\prime\prime}$}\kern -0.08ex}
\begin{document}
%
%
\thispagestyle{empty}

\centerline{{\Large\bf A Hipparcos Census of the Nearby OB Associations}\footnote{Based on data from the Hipparcos astrometry satellite.}}
\vskip 1truecm
\centerline{P.T.\ de Zeeuw, R.\ Hoogerwerf, J.H.J.\ de Bruijne,}
\centerline{Sterrewacht Leiden, Postbus 9513, 2300 RA Leiden, The Netherlands}
\vskip 0.3truecm
\centerline{A.G.A.\ Brown,}
\centerline{Instituto de Astronom{\'\i}a U.N.A.M., Apartado Postal 877, Ensenada, 22800 Baja California, M\'exico}
\vskip 0.3truecm
\centerline{and}
\vskip 0.3truecm
\centerline{A.\ Blaauw}
\centerline{Kapteyn Instituut, Postbus 800, 9700 AV Groningen, The Netherlands and}
\centerline{Sterrewacht Leiden, Postbus 9513, 2300 RA Leiden, The Netherlands}
\vskip 1.0truecm
\centerline{Accepted for publication in the Astronomical Journal, January 1999 issue}

\vskip 1truecm

\section*{\centerline{\large\bf ABSTRACT}}
A comprehensive census of the stellar content of the OB associations
within 1 kpc from the Sun is presented, based on Hipparcos positions,
proper motions, and parallaxes. It is a key part of a long-term
project to study the formation, structure, and evolution of nearby
young stellar groups and related star-forming regions.

OB associations are unbound `moving groups', which can be detected
kinematically because of their small internal velocity dispersion. The
nearby associations have a large extent on the sky, which
traditionally has limited astrometric membership determination to
bright stars ($V \la 6^{\rm m}$), with spectral types earlier than
$\sim$B5. The Hipparcos measurements allow a major improvement in this
situation. Moving groups are identified in the Hipparcos Catalogue by
combining de Bruijne's refurbished convergent point method with the
`Spaghetti method' of Hoogerwerf \& Aguilar. Astrometric members are
listed for 12 young stellar groups, out to a distance of
$\sim$650~pc. These are the 3 subgroups Upper Scorpius, Upper
Centaurus Lupus and Lower Centaurus Crux of Sco~OB2, as well as
Vel~OB2, Tr~10, Col~121, Per~OB2, $\alpha$~Persei (Per~OB3), Cas--Tau,
Lac~OB1, Cep~OB2, and a new group in Cepheus, designated as
Cep~OB6. The selection procedure corrects the list of previously known
astrometric and photometric B- and A-type members in these groups, and
identifies many new members, including a significant number of F
stars, as well as evolved stars, e.g., the Wolf--Rayet stars
$\gamma^2$~Vel (WR11) in Vel~OB2 and EZ~CMa (WR6) in Col~121, and the
classical Cepheid $\delta$~Cep in Cep~OB6. Membership probabilities
are given for all selected stars.  Monte Carlo simulations are used to
estimate the expected number of interloper field stars. In the nearest
associations, notably in Sco~OB2, the later-type members include T
Tauri objects and other stars in the final pre-main sequence
phase. This provides a firm link between the classical high-mass
stellar content and ongoing low-mass star formation. Detailed studies
of these 12 groups, and their relation to the surrounding interstellar
medium, will be presented elsewhere.

Astrometric evidence for moving groups in the fields of R~CrA,
CMa~OB1, Mon~OB1, Ori~OB1, Cam~OB1, Cep~OB3, Cep~OB4, Cyg~OB4,
Cyg~OB7, and Sct~OB2, is inconclusive. OB associations do exist in
many of these regions, but they are either at distances beyond
$\sim$500~pc where the Hipparcos parallaxes are of limited use, or
they have unfavorable kinematics, so that the group proper motion does
not distinguish it from the field stars in the Galactic disk.

The mean distances of the well-established groups are systematically
smaller than the pre-Hipparcos photometric estimates. While part of
this may be caused by the improved membership lists, a recalibration
of the upper main sequence in the Hertzsprung--Russell diagram may be
called for. The mean motions display a systematic pattern, which is
discussed in relation to the Gould Belt.

Six of the 12 detected moving groups do not appear in the classical
list of nearby OB associations. This is sometimes caused by the
absence of O stars, but in other cases a previously known open cluster
turns out to be (part of) an extended OB association. The number of
unbound young stellar groups in the Solar neighbourhood may be
significantly larger than thought previously.

%
%
\twocolumn
\section*{\centerline{\normalsize 1. INTRODUCTION}}

Ever since spectral classifications for the bright stars became
available, and in particular with the publication of Cannon and
Pickering's monumental Henry Draper Catalog of stellar spectra in
1918--1924, it has been evident that O and B stars are not distributed
randomly on the sky --- and hence not uniformly among the stellar
population of the Galaxy --- but instead are concentrated in loose
groups. This inspired research on their individual properties, and on
their motions and space distribution. Studies of the stellar content,
the internal velocity distribution, and the common motion of the group
members with respect to the ambient stellar population naturally
relied on the capability of identifying the stars belonging to the
group, their `members'. Among extensive early investigations along
these lines in the wake of work on `moving groups' and compact stellar
clusters (summarized in Eddington 1914), we note Kapteyn's (1914,
1918) work on `the brighter Galactic helium stars' --- i.e., the B
stars --- and the work by Rasmuson (1921, 1927). Among the early
investigations of the space distribution we note in particular the
comprehensive work by Pannekoek (1929), carried out at the time of the
breakthrough of the modern concept of the Galaxy as an isolated,
rotating system. Pannekoek's summarizing table~15 lists 37 groups of
B stars, among which several, called by him `Lacerta', `Cep I', `Cep
II', `$\zeta$~Pers', `Lupus', `Scorpio', and others, may be considered
as `forerunners' of the objects of modern research. His diagram of the
groups of B stars, projected on the Galactic plane ({\it loc.\ cit.}\
p.\ 65) is a remarkable foreshadow of some of the figures presented in
this paper. In the early 1950s the work of W.W.\ Morgan and
collaborators provided the first identification of the Galactic spiral
structure in the distribution of stellar `aggregates' (Morgan,
Sharpless \& Osterbrock 1952a, b; Morgan, Whitford \& Code 1953).
Herschel (1847) and Gould (1874) had already noticed that the
brightest stars are not distributed symmetrically with respect to the
plane of the Milky Way, but seemed to form a belt that is inclined
$\sim$$18^\circ\!$ to it. This became known as the Gould Belt, and was
subsequently found to be associated with a significant amount of
interstellar material (Lindblad 1967; Sandqvist, Tomboulides \&
Lindblad 1988; P\"oppel 1997).

Ambartsumian (1947) introduced the term `association' for the groups
of OB stars; he pointed out that their stellar mass density is usually
less than 0.1 ${\rm M}_\odot$~pc$^{-3}$. Bok (1934) had already shown
that such low-density stellar groups are unstable against Galactic
tidal forces, so that OB associations must be young (Ambartsumian
1949), a conclusion supported later by the ages derived from
color-magnitude diagrams. This is in harmony with the fact that these
groups are usually located in or near star-forming regions, and hence
are prime sites for the study of star formation processes and of the
interaction of early-type stars with the interstellar medium (see
e.g., Blaauw 1964a, 1991 for reviews). Ruprecht (1966) compiled a list
of OB associations, with field boundaries, bright members, and
distance. He introduced a consistent nomenclature, approved by the
IAU, which we adopt here. Ruprecht's list is based on the massive
`Catalogue of Star Clusters and Associations', put together by him and
his group, and maintained for many years (Alter, Ruprecht \& Van\'ysek
1970; Ruprecht, Bal\'azs \& White~1981).

Detailed knowledge of the stellar content, structure, and kinematics
of OB associations allows us to address fundamental questions on the
formation of stars. What is the initial mass function? What are the
characteristics of the binary population? What is the star formation
rate and efficiency? Do all stars in a group form at the same time?
What causes the distinction between the formation of bound open
clusters and unbound associations? What is the connection between the
stellar content of associations and the energetics and dynamical
evolution of the surrounding interstellar medium? What are the
properties of the ensemble of OB associations, and how do these relate
to the structure and evolution of the Galaxy? Answers to these
questions are essential for the interpretation of observations of
extragalactic star-forming regions and starburst galaxies.

Although OB associations are unbound, their velocity dispersions are
only a few ${\rm km~s}^{-1}$ (e.g., Mathieu 1986; Tian et al.\ 1996),
and so they form coherent structures in velocity space. The common
space motion relative to the Sun is perceived as a convergence of the
proper motions of the members towards a single point on the sky (e.g.,
Blaauw 1946; Bertiau 1958). This can be used to establish membership
based on measurements of proper motions. Whereas many such astrometric
membership studies have been carried out for open clusters (e.g., van
Leeuwen 1985, 1994; van Altena et al.\ 1993; Robichon et al.\ 1997),
there are few such studies for nearby OB associations because these
generally cover tens to hundreds of square degrees on the sky.
Ground-based proper motion studies therefore almost invariably have
been confined to modest samples of bright stars $(V \la 6^{\rm m})$ in
fundamental catalogs, or to small areas covered by a single
photographic plate. Photometric studies extended membership to later
spectral types (e.g., Warren \& Hesser 1977a, b, 1978), but are less
reliable due to, e.g., undetected duplicity, or the distance spread
within an association. As a result, membership for many associations
has previously been determined unambiguously only for spectral types
earlier than B5 (e.g., Blaauw 1964a, 1991). Although this covers the
important upper main-sequence turnoff region, it is not known what the
lower mass limit is of the stars that belong to the association, so
that our knowledge of these young stellar groups has remained rather
limited.

The Hipparcos Catalogue (ESA 1997) contains accurate positions, proper
motions, and trigonometric parallaxes, which are all tied to the same
global reference system (ICRS: see ESA 1997, Vol.\ 1 \S 1.2.2). This
makes these measurements ideally suited for the identification of
astrometric members of the nearby OB associations, with greater
reliability, and to much fainter magnitudes than accessible
previously. Accordingly, the {\tt SPECTER} consortium was formed in
Leiden in 1982, which successfully proposed the observation by
Hipparcos of candidate members of nearby OB associations. An
extensive program of ground-based observations was carried out in
anticipation of the release of the Hipparcos data (for a summary, see
de Zeeuw, Brown \& Verschueren 1994; \S 2.3). Here we present the
results of our census of the nearby associations based on the
Hipparcos measurements.

We have developed a new procedure to identify moving groups in the
Hipparcos Catalogue in an objective and reliable way. It combines a
refurbished convergent point method (de Bruijne 1998) and the
so-called `Spaghetti method' of Hoogerwerf \& Aguilar (1998), and in
this way significantly reduces the number of misidentifications. We
have estimated the remaining number of interlopers by means of Monte
Carlo simulations. The Hipparcos measurements are most valuable for
the nearest associations.  The individual parallaxes have typical
errors of $\sim$1~mas, and hence are of little value beyond
$\sim$500~pc. However, the proper motions can be quite significant to
much larger distances. For this reason we apply our procedure to
fields centered on the known associations and suspected groups at
pre-Hipparcos estimated distances of less than 1~kpc from the Sun.

This paper is organized as follows. In \S 2 we summarize the original
selection of fields, and the resulting Hipparcos sample, and in \S 3
we describe our member selection procedure. Then we discuss the
results for each of the fields, starting with the
Scorpio--Centaurus--Lupus--Crux complex in \S 4. \S 5 is devoted to
the Vela region. In \S 6 we discuss Canis Major, Monoceros, and
Orion. Then follow Taurus, Perseus, Cassiopeia, and Camelopardalis (\S
7), and Lacerta, Cepheus, Cygnus, and Scutum (\S 8). In \S 9 we derive
mean distances and motions for the nearby associations. We summarize
our overall conclusions in \S 10, and also outline the next steps.
Appendices give details of the member selection, of pitfalls in the
determination of mean distances, and lists of association members. The
selection procedure for the Scorpio--Centaurus--Lupus--Crux complex
(\S 4) is described in detail. It provides an example of how we have
analysed the other fields, for which we restrict ourselves to a
concise description of the pre- and post-Hipparcos results.

Preliminary results of this census were reported in de Bruijne et al.\
(1997), Hoogerwerf et al.\ (1997) and de Zeeuw et al.\ (1997). These
are superseded by the results presented here. Detailed studies of the
physical properties of the stars in the nearby OB associations will be
presented elsewhere. 

\section*{\centerline{\normalsize 2. THE NEARBY OB ASSOCIATIONS}} 

\begin{figure}[t]
\centerline{
\psfig{file=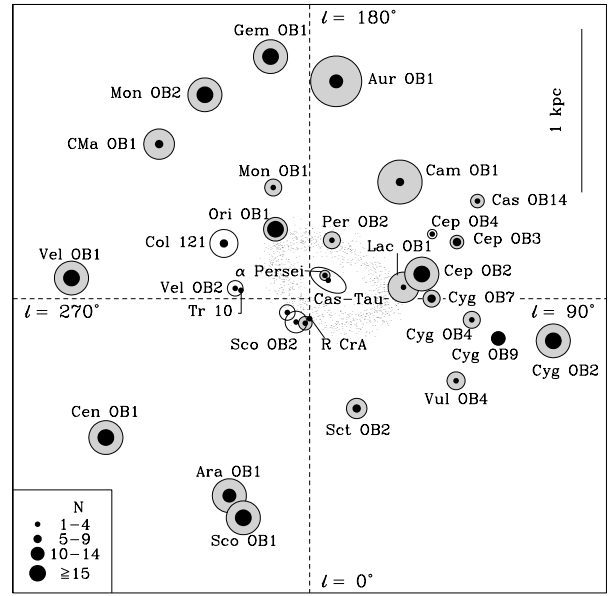,width=8truecm,silent=}}
\caption{\small Pre-Hipparcos locations of the OB 
associations within $\sim$1.5~kpc, projected onto the Galactic plane,
as listed by Ruprecht (1966). The Sun is at the center of the dashed
lines which give the principal directions in Galactic longitude,
$\ell$. The size of the circles represents the projected dimension of
the associations, enlarged by a factor 2 with respect to the distance
scale. The size of the central dots indicates the degree of current
or recent star formation activity, as given by the number $N$ of stars
more luminous than absolute magnitude $M_V$$\sim$$-5^{\rm m}$
(Humphreys 1978). Associations discussed in this paper which are
absent from Ruprecht's list are represented as open circles:
Scorpius~OB2 subgroups 3 and 4 (Blaauw 1946); Vela~OB2 (Brandt et
al.~1971); Trumpler~10 (Lyng\aa~1959, 1962); Collinder~121 (Feinstein
1967); Cassiopeia--Taurus (Blaauw~1956); Cepheus~OB4 (MacConnell
1968); R~Corona~Australis (Marraco \& Rydgren 1981). The distribution
of small dots indicates the Gould Belt (\S 9.2). Figure~29 presents
the post-Hipparcos map of the nearby OB associations.}
\end{figure}

\subsection*{\centerline{\normalsize \sl 2.1. The Solar neighbourhood}} 

Properties of the OB associations within 1.5~kpc from the Sun were
reviewed by Blaauw (1964a, 1991). Figure~1 shows the pre-Hipparcos
locations of these associations, as given in the official IAU list
(Ruprecht 1966). It also includes a few associations and subgroups
discussed here, but absent from Ruprecht's list. The large variety in
stellar content, projected dimension, clumpiness, age, and connection
to the interstellar medium introduces biases in the identification of
the associations. For example, associations containing several
luminous supergiants can be detected out to several kpc (e.g., Cep~OB1
in \S 8.2; Humphreys 1978; Garmany \& Stencel 1992), whereas
associations without such stars can easily escape attention.
Ruprecht's list is based on Ambartsumian's definition which requires
the presence of O stars and an open cluster as `nucleus' (e.g., Alter
et al.\ 1970). As a result the list is incomplete, e.g., two of the
subgroups of Sco~OB2 are not included (\S 4). Many of the groups in
Figure~1 are associated with concentrations of molecular material
(e.g., Dame et al.\ 1987). Most associations within 500~pc from the
Sun belong to the Gould Belt system (\S 9.2).

\subsection*{\centerline{\normalsize \sl 2.2. Hipparcos Proposal 141}}

When the call for proposals for the Hipparcos mission was released in
1982, the pre-launch specifications indicated absolute proper motions
and parallaxes with 1$\sigma$ accuracies of 1.5--2 mas~(yr$^{-1}$), a
limiting magnitude of $V$$\sim$$11^{\rm m}$, and a total mean stellar
density of about 3 stars per square degree. This clearly promised a
major step forward in the study of the nearby OB associations.
Accordingly, the {\tt SPECTER} consortium submitted a proposal to ESA
to include in the Hipparcos Input Catalogue candidate members of the
known OB associations (or subgroups) out to a distance of 1~kpc. We
took generous boundaries around the known locations. Table~1 lists
the associations, gives their pre-Hipparcos distance $D_{\rm clas}$,
the field boundaries in Galactic coordinates $(\ell, b)$, and the
proposed number $N_{\rm prop}$ of candidate members per association or
subgroup.
 
We selected all O and B stars from the Catalogue of Stellar
Identifications (CSI: Jung \& Bischoff 1971) within the assocation
boundaries, as well as later-type stars within certain magnitude
limits defined by the (pre-Hipparcos) distances of the associations,
so as to exclude foreground and background stars. This resulted in
13961 candidate member stars. We divided this sample initially into
three categories. Priority~1 contained stars of spectral type O and B
and stars that were considered as established or probable members of
the associations based on proper motion, radial velocity or
photometric work by previous authors. The remaining stars were divided
into two nearly equal groups, by assigning them priority 2 or 3
depending on whether their CSI number is odd or even. We took this
seemingly peculiar step because the number density of our proposed
objects was uncomfortably close to the limit of 3 per square
degree. We therefore suggested to the Input Consortium that in order
to satisfy the constraint on the number density of objects, while
preserving the statistical completeness of our sample, all priority 3
stars could be dropped, if necessary.

The {\tt SPECTER} proposal was approved as number 141, with the
comment that for groups beyond 600~pc the parallax measurements would
not be very significant, and that --- as we had expected --- the
number density of the fainter proposed stars was larger than was
possible to accept for the Hipparcos Input Catalogue. This resulted in
the inclusion of 9150 of our candidate members in the Hipparcos Input
Catalogue (Turon et al.\ 1992), amongst which were nearly all the
priority 1 stars. Table~1 lists the number $N_{\rm HIC}$ of the
originally proposed stars that were accepted per field, together with
the total number $N_{\rm HIP}$ of stars observed by Hipparcos in the
same field.

\begin{figure}[t]
\centerline{
\psfig{file=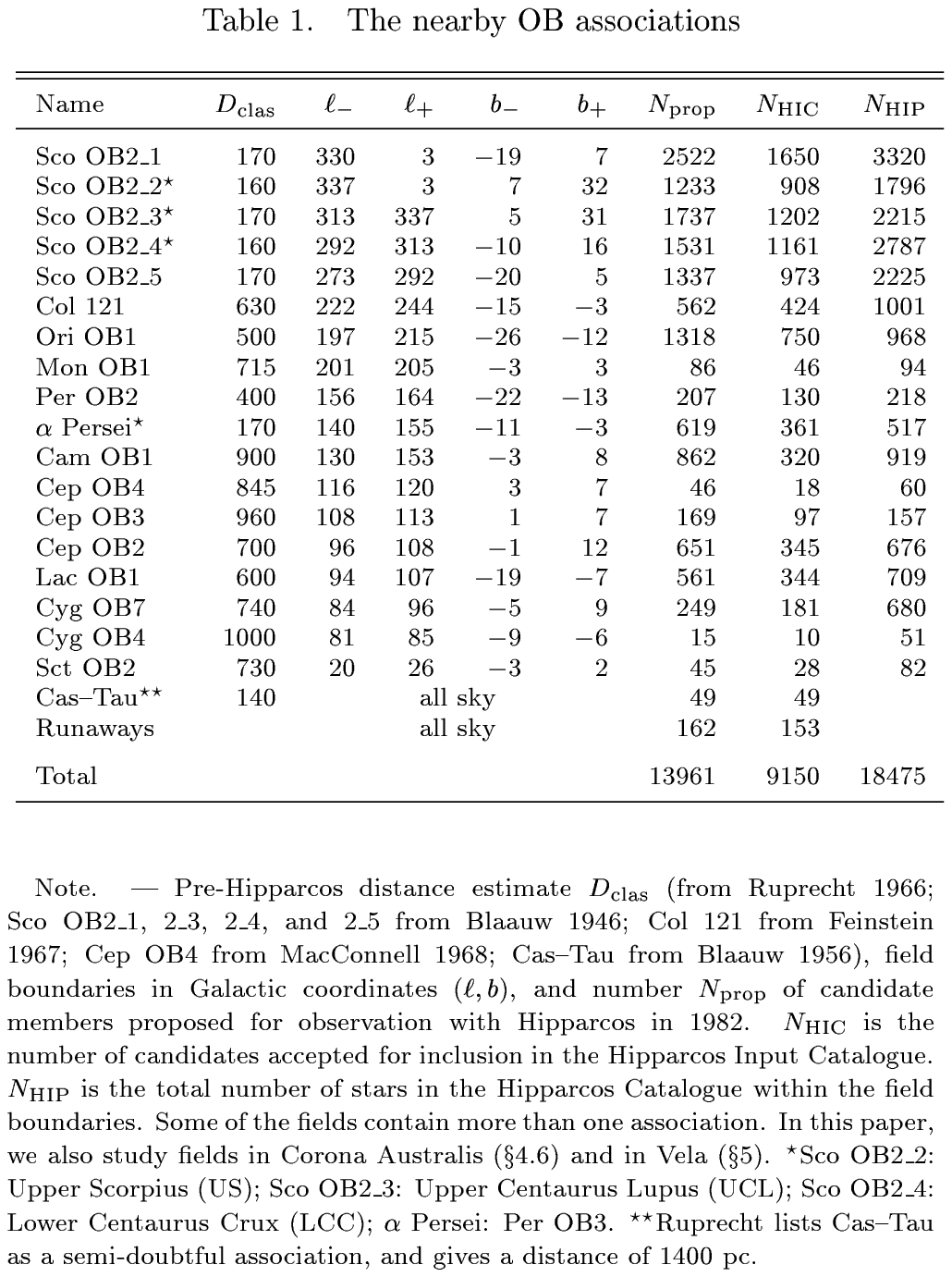,width=8truecm,silent=}}
\end{figure}

Even though we had distinguished priority 2 and 3 stars to allow
cutting of the sample by a factor 2 in a statistically unbiased way,
the final selection turned out to contain both, so that statistical
inferences, especially for stars fainter than the completeness limit,
should be made with care. For example, in some areas the inclusion of
many B-type stars resulted in a bias against stars of type A and
later, so as to stay within the limits on number density (Figure~2).

\begin{figure}[h]
\centerline{
\psfig{file=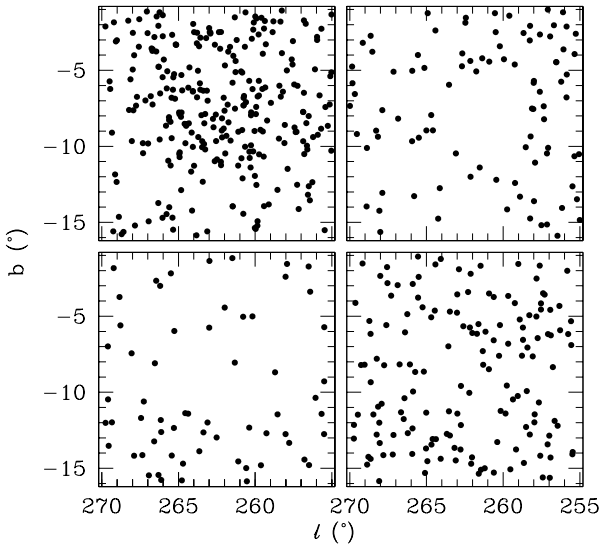,width=8truecm,silent=}}
\caption{\small An example of selection effects in the 
Hipparcos Catalogue. The four panels show all stars in the field
$255^\circ\!\! \leq \! \ell \! \leq \! 270^\circ\!$ and $-16^\circ\!\!
\leq \! b \! \leq \! -1^\circ\!$ in Vela (\S 5.1) with $V > 7^{\rm m}$
and spectral types O and B (top left; 277 stars), A (top right; 93
stars), F (bottom left; 62 stars), and G (bottom right; 153 stars).
The relative shortage of A, F, and G stars is caused by the large
numbers of B stars that were included, combined with the requirement
that the mean number density does not exceed about 3 per square
degree.}
\end{figure}

We reproposed our program in 1992, and took the opportunity to modify
the field boundaries slightly, and to include the groups R~CrA and
Vel~OB2. The preliminary results of our census, reported in de Bruijne
et al.\ (1997), Hoogerwerf et al.\ (1997), and de Zeeuw et al.\
(1997), are based on the resulting sample of 12842 stars, made
available to us in late 1996. During this investigation it became
clear that in some cases the associations seemed to `spill over' the
previously chosen boundaries in $\ell$, $b$, and $V$. Now that the
entire Hipparcos Catalogue is available, we can investigate the full set
of stars in the association fields, and also study the surrounding
areas, and we do so here.

The Hipparcos Catalogue also includes 153 of our original 162 proposed
candidate runaway OB stars. The study of these stars is of interest
for the question of their origin: supernova explosions in high-mass
binaries, or dynamical ejection, or both (Blaauw 1961, 1993; Poveda,
Ruiz \& Allen 1967). We are in the process of retracing the
three-dimensional paths of the runaways and the OB associations in the
Galactic potential in order to identify the parent associations, and
the age of the runaways. This is also of considerable interest for studies
of high-mass binary evolution (van Rensbergen, Vanbeveren \& de Loore
1996) and of high-mass X-ray binaries (Kaper et al.\ 1997). We will
report on this investigation elsewhere.

\subsection*{\centerline{\normalsize \sl 2.3. Photometry}}

The {\tt SPECTER} consortium carried out extensive photometric
observations of the southern associations between 1982 and 1989, with
the Walraven \vbluw\ photometer on the 91cm Dutch telescope at ESO (Lub
\& Pel 1977). As the total number of originally proposed candidate
stars in the southern fields approached 10$\,$000, we concentrated on the
priority 1 and 2 stars as defined in Proposal 141, and obtained
\vbluw\ measurements for a total of 5260 objects (de Geus, Lub \& van
der Grift 1990). A preliminary analysis for the priority 1 stars in
Sco~OB2 and in Ori~OB1 was reported by de Geus, de Zeeuw \& Lub (1989)
and Brown, de Geus \& de Zeeuw (1994), respectively.

The final set of candidates included in the Hipparcos Input Catalogue is
a mix of our original priority 2 and 3 objects (\S 2.2). By the time
this was known our Walraven program was concluded, so that we do not
have \vbluw\ photometry for the entire sample. The Walraven photometer
was retired before the release of the Hipparcos Input Catalogue. Our
member selection should not favor stars of either priority, so the
\vbluw\ photometry is incomplete also for the finally selected member
stars. Published \uvby$\beta$ and \UBV\ photometry is incomplete as
well (Hauck \& Mermilliod 1990; Mermilliod \& Mermilliod 1994). Here
we will therefore generally restrict ourselves to the $V$ and $B\!-\!V$
values listed in the Hipparcos Catalogue. We are taking steps to obtain
homogeneous intermediate band photometry for all the associations.

\section*{\centerline{\normalsize 3. SELECTION METHOD}}

The stars in an OB association have nearly identical space motions,
and can be recognized as a coherent structure in velocity space. We
identify the members of these moving groups by using Hipparcos
positions, proper motions {\it and$\,$} parallaxes. We do this by
combining two member selection methods. One is a modification of the
classical convergent point method, outlined in \S 3.1. The other
method is new, and we summarize it in \S 3.2. Then we describe the
selection procedure and membership definition (\S 3.3), and discuss
how we estimate the contamination by field stars (referred to as
interlopers, \S 3.4), the problems caused by marginally resolved
long-period binaries (\S 3.5), the derivation of mean distances (\S
3.6), and the prospects for resolving the internal structure and
motions (\S 3.7).

\subsection*{\centerline{\normalsize\sl 3.1. Convergent point method}}
 
The common space motion of stars in a moving group results in
converging proper motions on the sky. We employ a modern
implementation of a classical convergent point method (Brown 1950;
Jones 1971), which uses the Hipparcos positions and proper motions,
but not the parallaxes. The method is described and tested in full by
de Bruijne (1998). We first summarize the classical method, and then
describe the main modifications.

The classical convergent point method considers a set of stars $j$ at
positions $(\ell, b)_j$, with proper motions $(\mu_\ell \cos b,$
$\mu_b)_j$, and errors $(\sigma_{\mu_\ell \cos b},
\sigma_{\mu_b})_j$. The first step is to discard stars with 
insignificant proper motions, i.e., with
\begin{equation}
t \equiv {{\mu} \over {\sigma_{\mu}}}
  \equiv {{\sqrt{{\mu_{\ell}^{2} \cos^{2} b} 
                            + {\mu_{b}}^{2}}}\over
   {\sqrt{\sigma_{{\mu_{\ell} \cos b}}^{2} +
          \sigma_{{\mu_{b}           }}^{2}}}}
\leq t_{\rm min},
\label{deftpm}
\end{equation}
where $t_{\rm min}$ was typically chosen to be equal to 3--5 (Jones
1971). The next step is to search for the maximum likelihood
coordinates $(\ell, b)_{\rm cp}$ of the convergent point by minimizing
\begin{equation}
\chi^2 = \sum\limits_{j = 1}^{N} \, t_{\perp j}^2,
\label{basicsum}
\end{equation}
where $N$ is the number of stars in the sample, and $t_{\perp j}$ is
the value for star $j$ of the quantity $t_\perp$, defined as
\begin{equation}
t_\perp \equiv \mu_\perp/\sigma_\perp.
\label{deft}
\end{equation}
Here $\mu_\perp$ is the component of the proper motion perpendicular
to the direction towards the convergent point, and $\sigma_\perp$ is
its measurement error.

In case of a common space motion, the proper motion vectors will be
directed towards the convergent point, so that for all moving group
members the expectation value for $\mu_\perp$ equals 0. The sum
(\ref{basicsum}) is distributed as $\chi^2$ with $N \! - \! 2$ degrees of
freedom. If, after minimization with respect to $(\ell, b)_{\rm
cp}$, the value of $\chi^2$ is unacceptably high, the star with the
highest value of $\mu_{\perp} / \sigma_{\perp}$ is rejected, after
which minimization is repeated until a satisfactory value of $\chi^2$
is obtained. Subsequently, all non-rejected stars are identified as
members. This procedure allows for simultaneous convergent point
determination and member selection, and has been applied with success
to, e.g., the Hyades (van Bueren 1952; Perryman et al.\ 1998), and
Sco~OB2 (Jones 1971).

De Bruijne's (1998) modification of the convergent\break point method
consists of three steps. First, unlike previous astrometric catalogs,
the Hipparcos Catalogue gives the full covariance matrix for the
measured astrometric parameters. It is therefore possible to include
the full error propagation. Second, even for infinitely accurate
measurements, moving group members would not have proper motions
directed exactly towards the convergent point because of the intrinsic
velocity dispersion in the group. As a result, selecting only stars
with $t_\perp=0$ will not identify all members. For this reason
definition (\ref{deft}) of $t_\perp$ is changed to
\begin{equation}
t_\perp \equiv {{\mu_{\perp}}\over{\sqrt{
               {\sigma_{\perp}}^{2}+
               {\sigma_{\rm int}^\star}^{2}}}},
\label{deftnew}
\end{equation}
where $\sigma_{\rm int}^\star$ is an estimate of the intrinsic
one-dimensional velocity dispersion in the group, expressed in proper
motion units. Accordingly, the definition (\ref{deftpm}) of $t$ is changed
to
\begin{equation}
t \equiv {\mu\over{\sqrt{
               {\sigma_\mu}^{2}+
               {\sigma_{\rm int}^\star}^{2}}}}.
\label{defttt}
\end{equation}
Third, the current workstations are powerful enough to drop the
classical approach of evaluating the sum (\ref{basicsum}) on a grid. In
order to find the maximum likelihood convergent point, a global,
direct minimization routine can be applied. De Bruijne (1998) defines
a membership probability $p_{\rm cp}$ as $p_{\rm cp}=1-p$, where $p$
is the probability that the perpendicular component of the proper
motion has a value different from zero, given the covariance
matrix. He shows that
\begin{equation}
p_{\rm cp} = \exp (-t_\perp^2/2).
\label{defpcpm}
\end{equation}
The above procedure is biased towards inclusion of stars at larger
distances. It selects members based on the absolute value of
$t_{\perp}$, which is a function of distance: nearby stars generally
have large proper motions, and are more likely to be rejected than
those at larger distances, which generally have smaller proper
motions. Our combined member selection method (\S 3.3; Figure~4)
effectively deals with this bias.

\subsection*{\centerline{\normalsize\sl 3.2. Spaghetti method}}

In order to make optimal use of the Hipparcos data, we also apply a
kinematic member selection method developed by Hoogerwerf \& Aguilar
(1998, hereafter HA), which uses, besides positions and proper
motions, also parallaxes. We briefly summarize the main steps.

\begin{figure}[t]
\centerline{
\psfig{file=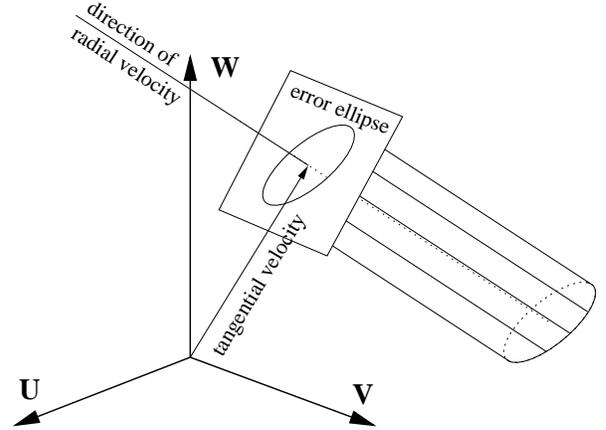,width=8truecm,silent=}}
\caption{\small The five astrometric parameters of a star
measured by Hipparcos define an elliptical cylinder in velocity space.
The offset from the origin is determined by the tangential velocity,
which follows from the proper motion and the parallax; the orientation
is set by the direction of the radial velocity (i.e., by the position
on the sky). The finite thickness of the cylinder is caused by the
measurement errors. All cylinders of stars in a moving group
intersect. $U$, $V$, and $W$ are the velocity components in standard
Galactic Cartesian coordinates: $U$ is directed towards the Galactic
center, $V$ in the direction of Galactic rotation, and $W$ towards the
north Galactic pole.}
\end{figure}

The Hipparcos measurements constrain a star to lie on a straight line
in velocity space: the proper motion and parallax determine the offset
from the origin (tangential velocity), while the sky position of the
star determines its direction (cf.\ Figure~3). The covariance matrix
of the astrometric parameters transforms this line into a probability
distribution in velocity space, which HA describe as a two-dimensional
Gaussian. Surfaces of equal probability are elliptic cylinders. HA
denote these as `spaghetti'. The cylinders of a set of stars with the
same space motion all intersect in one point. Thus, HA identify moving
groups by searching for maxima in the density of cylinders in velocity
space. This `spaghetti density' in velocity space is simply the sum
over all stars in the sample of the individual probability
distributions.

Measurement of the stellar radial velocities would reduce each
elliptic cylinder to an ellipsoidal probability distribution in
velocity space. In practice, high-quality radial velocities are
available only for a small subset of the stars in our sample, and we
therefore decided not to use them in the member selection.

In case of infinitely accurate astrometric measurements, moving group
members would not necessarily have cylinders coinciding exactly at the
space motion of the group because of the intrinsic velocity dispersion
in the group. This broadens the associated peak in velocity space.
Measurement errors broaden it further. HA therefore place a sphere at
the position of this peak, with a radius $\sigma_{\rm sp}$ given by
\begin{equation}
\sigma^2_{\rm sp} = \sigma^2_{\rm med} + \sigma^2_{\rm int},
\label{defsigtot}
\end{equation}
where $\sigma_{\rm med}$ is an estimate of the typical error in
tangential velocity for a star at the distance of the moving group,
taken as the median value of the semi-major axis length of the
$1\sigma$ cylinders of all stars in the range of distances where
moving group members can be expected (Appendix~A), and $\sigma_{\rm
int}$ is an estimate of the one-dimensional velocity dispersion in the
moving group. At a given distance $D$ in kpc, $\sigma_{\rm int}$ and
$\sigma_{\rm int}^\star$ (eq.\ \ref{deftnew}) are related through $\sigma_{\rm
int} = A\, \sigma_{\rm int}^\star D$, where $A = 4.74$~km~yr~s$^{-1}$ and
$\sigma_{\rm int}$ is in units of mas~yr$^{-1}$.

HA compute for each star the integral of its probability distribution
over three-dimensional velocity space restricted to the volume of the
sphere with radius $\sigma_{\rm sp}$ defined in eq.\ (7), and denote
this quantity by $S$. All stars with $S$ larger than a certain cutoff
$S_{\rm min}=0.1$ are accepted as members. This value is based on
extensive Monte Carlo simulations of associations, described in HA.
$S$ cannot be interpreted as a simple probability as it depends on
distance. Stars with small parallaxes, at the far side of a group,
will have smaller values of $S$ than those with large parallaxes at
the near side, because the error in tangential velocity, i.e., the
thickness of the cylinder, depends on parallax.

\subsection*{\centerline{\normalsize\sl 3.3. Combined method and search strategy}}

We combine the results of the two methods described above to define
our membership list: we consider as secure members those stars that
are selected by both methods. Figure~4 illustrates the power of this
combined approach. The left panel shows the secure members for Vel~OB2
(\S 5.1). Both the proper motions and parallaxes show a coherent
structure confirming Vel~OB2 as a physical group. The additional
stars selected by the Spaghetti method only (middle panel) have
parallaxes in the same range as those of the secure members. These
stars are rejected by the convergent point method because the proper
motions are inconsistent with the convergent point, even though the
tangential velocities used by the Spaghetti method are consistent with
that of the group within the errors. This occurs because the errors on
the tangential velocity components include the parallax error, which
dominates for the more distant groups like Vel~OB2. The fraction of
such rejected Spaghetti members drops to a few per cent in the nearest
associations. In turn, some of the stars selected by the convergent
point method are rejected by the Spaghetti method because the
corresponding tangential velocities are inconsistent with the cluster
motion, even though the proper motions are consistent with the
convergent point. These stars have deviating parallaxes (right panel);
the convergent point method is clearly biased towards selecting stars
with small parallaxes (\S 3.1).

\begin{figure*}[t]
\centerline{
\psfig{file=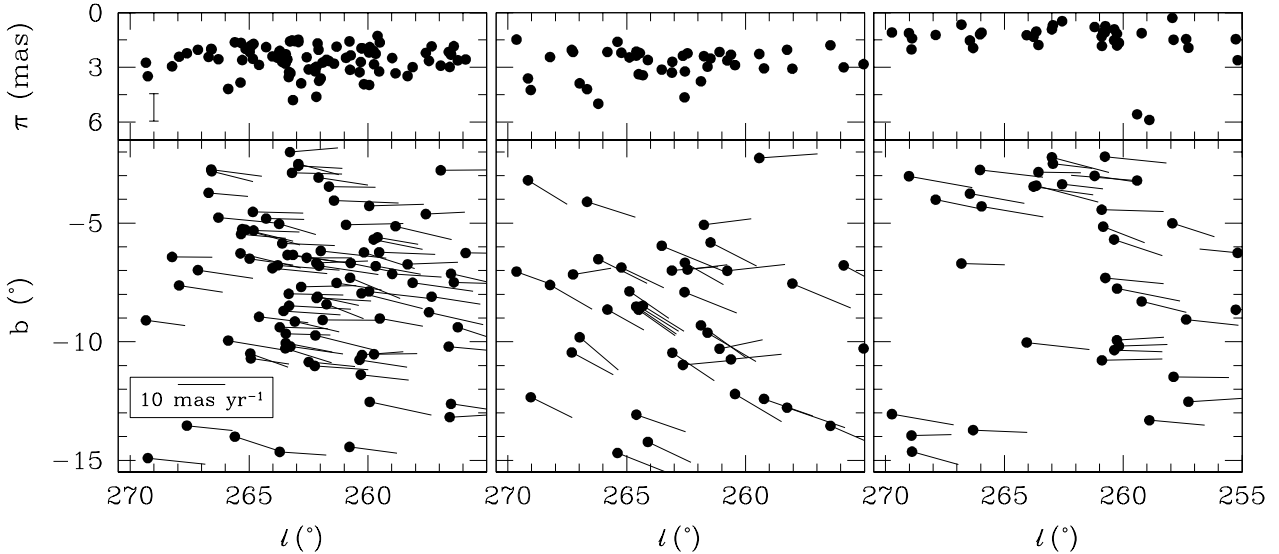,width=17truecm,silent=}}
\caption{\small Left: positions and proper motions
(bottom) and parallaxes (top) for the 93 secure members of Vela~OB2
(\S 5.1), selected by the convergent point method {\it and$\,$} the
Spaghetti method. Middle: same diagram for 41 additional stars
selected by the Spaghetti method only. The directions of the proper
motions differ considerably from those in the left panel. The group of
comoving stars around $(\ell,b)$$\sim$$(264\fdg5,-8\fdg5)$ consists of
members of \FNGC2547. Right: same diagram for 37 other stars selected
by the convergent point method only. The proper motions are parallel
to those in the left panel, but the parallaxes differ considerably,
with most of them being significantly smaller. Two stars move in the
opposite direction. The vertical bar in the left top panel corresponds
to the average $\pm$$1\sigma$ parallax range for the stars
shown.}
\end{figure*}

Many radial velocity and proper motion studies of moving groups have
been carried out for open and globular clusters (e.g., van Leeuwen
1994; Cudworth 1998). These systems are centrally concentrated and
gravitationally\break bound, have small internal velocity dispersions
$\sigma_{\rm int}$ (less than 1~km~s$^{-1}$ for most open clusters),
and a finite radius set by the Galactic tidal field. As a result,
kinematic membership is well-defined: in practice all stars that have
velocities within $3\sigma_{\rm int}$ of the mean motion are
considered members. It is sometimes possible to identify
`fellow-travelers', which at one time may have belonged to the
cluster, but have spilled over the tidal radius relatively recently
(e.g., Grillmair et al.\ 1995; Perryman et al.\ 1998).

By contrast, OB associations are low-density dispersed stellar groups
which have formed recently, but are unbound. The dispersion in
stellar velocities is at most a few km~s$^{-1}$ (Mathieu 1986; Tian et
al.\ 1996), but the velocity distribution may have substructure: there
may be subgroups with significantly smaller dispersions and/or
slightly different space velocities. The overall space motion can be
detected kinematically, but the definition of membership is less
well-defined, as it is not clear {\it a priori} what should be taken
as $\sigma_{\rm int}$ in eqs (\ref{deftnew}), (\ref{defttt}), and (\ref{defsigtot}).
A value that is too large will result in the inclusion of many
interloper field stars. A value that is too small may well exclude
many of the stars that were formed at the same time, and in the same
region. We take $\sigma_{\rm int}$ to be 3~km~s$^{-1}$. This value is
somewhat larger than the expected velocity dispersion in the OB
associations, and hence it reduces the number of genuine members that
is discarded by our selection procedure (but see \S 3.5).
Monte Carlo simulations show that a choice for $\sigma_{\rm int}$
consistent with the modeled intrinsic velocity dispersion will
typically miss less than $\sim$10~per cent of the members (de Bruijne
1998; Hoo\-ger\-werf \& Aguilar 1998).
If we used the convergent point or the Spaghetti method separately,
this high reliability would come at the price of an increased number
of interlopers. We rely on the combination of the two methods to
minimize the number of accidental interlopers (Figure~4).

It is difficult to define a membership probability for stars selected
by the Spaghetti method (\S 3.2). We therefore use $P = p_{\rm cp}$
defined in eq.\ (\ref{defpcpm}) as membership probability for the stars
selected by the combined method. In view of the above, this
probability should be interpreted with care.

The Hipparcos Catalogue is essentially complete to $V = 7\fm3$--$9\fm0$,
depending on Galactic latitude and spectral type. A few binaries and
stars in crowded fields brighter than this limit are missing. It
follows that many genuine association members do not appear in the
Catalogue, especially for later spectral types and/or associations at
large distances. Therefore, we decided on a search strategy in which
we identify the association among the early-type stars in the
Hipparcos Catalogue. This provides the common motion. We then find the
association members among the later spectral types by selecting those
stars that are consistent with the motion of the secure early-type
members. The full procedure is described step-by-step in Appendix~A. 

\subsection*{\centerline{\normalsize\sl 3.4. Contamination by field stars}}

Our kinematic selection method will accept some field stars as
association members. Here we describe our method for estimating the
expected number of these interlopers.

\subsubsection*{\centerline{\normalsize\sl 3.4.1. Principles}}

We consider a specific association, select all entries from the
Hipparcos Catalogue within the chosen field boundaries (Table~A1), and
optionally (see \S 3.4.2) delete the entries corresponding to the
secure members. Next, we delete all proper motions. Then we replace
these by synthetic values, consistent with the field star distribution
in the Solar neighbourhood. We draw these randomly from the kinematic
model of the Galactic disk derived from the Hipparcos Catalogue by
Dehnen \& Binney (1998). By construction, the magnitude, parallax,
position, and spectral-type distribution are reproduced consistent
with those in the Hipparcos Catalogue. We then apply our
selection procedure to 100 of these Monte Carlo realisations in the
same manner as for the corresponding association, to search for
comoving stars at the known space motion of the group. The resulting
median number of comoving stars is an estimate of the expected number
of interlopers. We summarize the results in Table~A2.

\subsubsection*{\centerline{\normalsize\sl 3.4.2. Removal of secure members?}}

A non-trivial aspect is whether or not to remove the secure members
from the field before performing the interloper analysis. Either
option leads to systematic errors. Two facts are important: (i) the
ratio of the number of association members and the number of field
stars in the Hipparcos Catalogue decreases strongly from spectral type
O to~M; and (ii) the ratio of the number of association members
observed by Hipparcos and the total number of association members is a
function of distance, and strongly decreases towards later spectral
types due to the completeness limit of the Catalogue.

Removing the secure members, including a number of interlopers, from a
sample leads to an underestimate of the expected number of interlopers
because the analysis takes into account too few stars. Not removing
the secure members leads to an overestimate of the expected number of
interlopers because the analysis takes into account too many stars,
namely the field stars {\it and$\,$} the genuine association members.
In both cases, the systematic under/overestimate is significant for
the earliest spectral types, and unimportant beyond spectral type
A--F; the precise transition depends on the distance of the
association. If one assumes that the stars in our membership lists are
primarily genuine association members, the option leading to the
smallest systematic errors is to remove the secure members before
performing the analysis. Table~A2 presents the results for both
options.

\subsubsection*{\centerline{\normalsize\sl 3.4.3. The $B\!-\!V$ versus age degeneracy}}

The velocity distribution in the Galactic disk is a function of
stellar age. Dehnen \& Binney (1998) used the $B\!-\!V$ color listed
in the Hipparcos Catalogue as a simple age indicator for main-sequence
stars. Our simulations accordingly use the $B\!-\!V$ color of each
star to assign it the appropriate proper motion. However, for giants
and supergiants, age and $B\!-\!V$ are not uniquely related.
Assignment of kinematics to stars based on $B\!-\!V$ color without
regard to luminosity class gives `old' kinematics to relatively young
giants and supergiants, leading to systematic effects. OB~associations
generally have small space motions relative to the Sun, and are thus
observed in velocity space around the reflex of the Solar motion plus
a small offset due to differential Galactic rotation. At similar
distances as the association, any unrelated population of evolved
young giants and supergiants will be clustered around the same
velocity as the association. These stars are then wrongly assigned
`old' kinematics, spreading them around in velocity space. This leads
to an underestimate of the number of interlopers.

In Table~A2, we list for each field and for each spectral type the
percentage of stars explicitly classified as main-sequence star
(luminosity class V). As expected, this percentage is roughly
constant for spectral types O through A--F and decreases going from F
to M due to the completeness limit of the Catalogue. The low percentages
found in the northern associations in Perseus, Cepheus and Lacerta are
due to the specific spectral classifications used by the Hipparcos
consortium, which in this region often lack an indication of
luminosity class (cf.\ ESA 1997, Vol.\ 1 \S 2.1, pp 134--135).
Although the percentages are lower limits to the actual values, they
show that the expected number of interlopers for spectral types F--G
and later should be interpreted with care (cf.\ ESA 1989, Vol.\ 2 \S
7.3, pp 90--94).

\subsubsection*{\centerline{\normalsize\sl 3.4.4. Solar neighbourhood}}

Dehnen \& Binney's (1998) determination of the local velocity
distribution is based on a kinematically unbiased sample of Hipparcos
main-sequence stars with relative parallax errors better than 10~per
cent. Using a median parallax accuracy of 0.97~mas, this translates
the term `local' to a volume centered on the Sun with a radius of
80--100~pc. However, even the nearest association, Lower Centaurus
Crux, lies outside this volume (\S 4.4). Furthermore, it is part of
the Gould Belt, which exhibits peculiar kinematics (\S 9.2). An improved
interloper analysis will require a more extended model of the velocity
distribution, which includes the Gould Belt.

\subsubsection*{\centerline{\normalsize\sl 3.4.5. Selection effects in the Hipparcos Catalogue}}

The Hipparcos Catalogue contains $\sim$52$\,$000 stars brighter than
the completeness limit ($V = 7\fm3$--$9\fm0$, depending on Galactic
latitude and spectral type). The remaining $\sim$66$\,$000 stars were
explicitly proposed to be observed by Hipparcos, and thus constitute a
special set. Two large-scale effects are the inclusion of high
proper-motion stars, and the inclusion of additional F- and G-type
stars in two strips from the north to the south Galactic pole (at
$\ell$$\sim$$0^\circ\!$ and $\ell$$\sim$$180^\circ\!$) for the purpose
of Galactic structure studies (ESA 1989, Vol.\ 2 \S 7.3, pp 94--100).
Dehnen \& Binney's (1998) model does not take these effects into
account, so it is to be expected that our synthetic data sets based on
their model differ kinematically from the Hipparcos measurements
beyond the completeness limits of the Catalogue. Tests indeed reveal
differences, particularly in the fourth Galactic quadrant in a strip
of width $60^\circ\!$ centered on the Galactic plane. In this region,
we estimate the resulting uncertainties in the expected number of
interlopers to be a factor of 2 at maximum.

\subsection*{\centerline{\normalsize\sl 3.5. Binaries}}

The Hipparcos observations were carried out over 3.3 years. The
measured motions of long-period binaries that are marginally resolved
can differ by many mas~yr$^{-1}$ from the mean motion of the system
(Lindegren 1997). Genuine association members of this kind can
therefore be missed by our selection method (even with our generous
choice of $\sigma_{\rm int}$, see \S 3.3), and some field binaries may
accidentally masquerade as members. This problem is most significant
for nearby binaries with massive components (Wielen et al.\ 1997).

Our kinematic member selection method is objective and consistent, and
is based only on the Hipparcos positions, parallaxes and proper
motions. At this stage we do not want to include physical properties
of the stars as further criteria for membership, so as not to bias
subsequent discussion of the stellar content of the associations. We
therefore do not discuss on a star-by-star basis whether its
assignment as a member should be trusted or not based on its spectral
type, colors, radial velocity, or multiplicity, also because this
additional information is generally not of homogeneous quality. We
make one exception to this approach, namely for the few cases where
our selection method rejects stars which were previously considered to
be solid astrometric members, based on proper motions derived from
measurements in fundamental catalogs, over time spans much longer than
three years. We do not include these stars in our member lists
(Table~C1), but identify them in the text. For many of these the
Hipparcos Catalogue turns out to give indications for `perturbed' proper
motions. The majority of these systems may therefore well be
association members.

\subsection*{\centerline{\normalsize\sl 3.6. Mean distances}}
 
Use of the Hipparcos trigonometric parallaxes of the secure members to
determine mean distances to the associations or their subgroups
requires some care, as the inverse of the parallax is a biased
distance indicator (Smith \& Eichhorn 1996; Brown et al.\ 1997b), and
the conversion of mean parallax to mean distance for a group of stars
depends on the distribution of stars within the group. We show in
Appendix~B that for all spherical groups the expectation value of the
mean of the measured parallaxes is equal to the true mean parallax,
and corresponds to the true distance of the group. For elongated
associations the bias in the mean parallax is small, typically less
than 1 per~cent. Hipparcos parallaxes measured in regions of high
stellar density (in the Catalogue) have to be interpreted with care
(Lindegren 1989; Robichon et al.\ 1997). This is not a problem for
the low-density associations (Appendix~B).

Interlopers generally have parallaxes similar to those of true members
(otherwise they would have been rejected; cf.\ Figure~4), so these
should not influence the mean distances significantly. However, the
observed distribution of parallaxes may not be representative of the
true underlying parallax distribution of an association. Magnitude
limits bias the selected members, and furthermore some stars in the
Hipparcos Catalogue have a negative measured parallax. Our member
selection method rejects these, which introduces a bias towards a
smaller mean distance. In order to estimate the magnitude of these
effects, we have carried out Monte Carlo simulations, set up as
follows. We considered associations with radius 25~pc, and then
created 200 members at random positions in a homogeneous sphere. We
chose masses, and corresponding magnitudes, using a power-law initial
mass function with exponent $-1.7$ (cf.\ Brown et al.\ 1994), and a
minimum and maximum mass of $1~{\rm M}_\odot$ and $120~{\rm M}_\odot$,
respectively. We took the parallax error distribution as a function of
magnitude from the Hipparcos Catalogue (ESA 1997, Vol.\ 1,
figure~3.2.39), using a standard deviation on the median parallax
error of 0.2~mas for $V \le 10^{\rm m}$ and 0.4~mas for $V>10^{\rm
m}$. We then retained all stars with $V < 7\fm3$, and discarded
fainter stars consistent with the magnitude distribution in the
Hipparcos Catalogue. We also discarded all stars with $\pi \le
0$~mas. We carried out this process as a function of distance, by
constructing 100 random realisations in intervals of 50~pc each. The
effect of the magnitude limit turns out to be negligible, but, as
expected, the bias caused by the exclusion of stars with negative
parallaxes increases with distance, and reaches nearly 30~per cent at
a distance of 1~kpc. This bias depends sensitively on the magnitude of
the parallax errors. For a mean parallax error of 0.97~mas with a
standard deviation of 0.2~mas independent of magnitude, the bias is
only 15~per~cent at 1~kpc.

We give the mean distances derived from the mean parallax of the
secure association members in \S\S 4--8. In all cases, these are
corrected for the above mentioned bias, and the error quoted is the
formal error on the mean. The results are summarized in Table~2.

\subsection*{\centerline{\normalsize\sl 3.7. Internal structure and motions}}

The OB associations studied here have distances larger than 100~pc,
and linear dimensions of order 50~pc or less. It is impossible to
resolve the depth of the associations with the median parallax error
of 0.97~mas. Estimates of the individual stellar distances (after
member selection) can be improved by deriving so-called secular
parallaxes from the measured proper motions (e.g., Jones 1971). A
sophisticated method to determine the improved parallaxes and space
motions was presented by Dravins et al.\ (1997). We will investigate
its application to the nearby associations in a future paper.

OB associations will expand due to their unbound nature (Ambartsumian
1949). The expansion can be detected if and only if the correct mean
streaming motion of the association with respect to the Sun is
subtracted from the observed proper motions {\it and$\,$} radial
velocities. What remains is the true expansion. Proper motions alone
cannot distinguish between a radial streaming motion and contraction
or expansion (e.g., Blaauw 1964b). Thus,\break knowledge of the radial
streaming velocity is essential for an unambiguous proof of
expansion. This requires radial velocities with errors smaller than
the expected expansion velocities, a few ${\rm km~s}^{-1}$. These are
generally not available, so that even with Hipparcos quality proper
motions the expected expansion of OB associations, and thus the
kinematic ages, cannot be determined (see also Brown, Dekker \& de
Zeeuw 1997). Our member selection method is not influenced by either
radial streaming or expansion/contraction (Blaauw 1952b; de Bruij\-ne
1998; Hoogerwerf \& Aguilar 1998). However, the resulting space
motions, and convergent points, not only reflect the mean streaming
motion of the association but may also include an extra component in
radial velocity.

\section*{\centerline{\normalsize 4. THE SCORPIO--CENTAURUS--LUPUS--CRUX} \break
\vskip -1.4truecm \centerline{\normalsize COMPLEX}}

In this section, and the ones that follow, we report the results of
applying our member selection method to the fields centered on the
nearby associations, starting with Sco~OB2 in the region of the
constellations Scorpius, Centaurus, Lupus, and Crux, and continuing
along the Galactic plane in the direction of decreasing $\ell$. We
have surveyed the literature, and in each case first review
pre-Hipparcos work on association membership. While in some cases this
has established a list of commonly accepted astrometric and/or
photometric members, for other associations there is considerable
confusion about membership, and sometimes even on the existence of an
underlying association. Our adopted fields contain all suspected
early-type members, and are small enough to minimize the number of
interlopers. Table~A1 summarizes the field boundaries, and specifies
the other parameters used in the member selection.

We review pre-Hipparcos work on Sco~OB2 in \S 4.1, and present our
analysis of the Hipparcos data for the three subgroups of this
association in detail in \S\S 4.2--4.4. We then briefly consider the
entire Sco~OB2 complex (\S 4.5), and conclude with an investigation of
an adjacent field, centered on the nearby star-forming region near
R~CrA (\S 4.6).

The presentation in this section also serves as an example of how we
have investigated the other fields of the census, for which we
restrict ourselves to a concise summary of the pre- and post-Hipparcos
results (\S\S 5--8). Unless stated other\-wise, we use spectral types,
and $V$ and $B\!-\!V$ photometry reported in the Hipparcos Catalogue.

\subsection*{\centerline{\normalsize\sl 4.1. Sco~OB2}}

In the lengthy paper `On the individual parallaxes of the brighter
Galactic Helium stars in the southern hemisphere, together with
considerations on the parallax of stars in general', Kapteyn (1914)
investigated the secular parallaxes of 319 OB stars brighter than
$V$$\sim$$6^{\rm m}$ contained in L.\ Boss' Preliminary General
Catalog in the region $216^\circ\!\! \la \! \ell^I \! \la \!
360^\circ\!$, $-30^\circ\!\! \la \! b^I \! \la \! 30^\circ\!$ (`old'
Galactic coordinates)\footnote{The approximate relation between
$(\ell^I, b^I)$ and $(\ell, b)$ is $\ell\approx\ell^I+32\fdg9$, and
$b\approx b^I$.}. He recognized the Sco~OB2 association as a moving
group of $\sim$150 stars spread over $70^\circ\! \times 40^\circ\!$ on
the sky (cf.\ Plummer 1913). Later kinematic studies confirmed Sco~OB2
as a moving group (e.g., Rasmuson 1921; Plaskett \& Pearce 1928, 1934;
Canavaggia \& Fribourg 1934; Kulikovsky 1940; but see Smart 1936,
1939; Petrie 1962). Blaauw (1946; 114 members), Bertiau (1958; 77
bright members and 37 additional faint B-type members), and Jones
(1971; 47 members), made detailed kinematic studies of Sco~OB2,
resulting in a total of 157 classical proper motion members with
spectral types earlier than $\sim$B9, of which 153 were observed by
Hipparcos. Several photometric studies suggested membership for
another 243 stars to spectral type $\sim$A8--F0, 215 of which are
included in the Hipparcos Catalogue (Hardie \& Crawford 1961; Garrison
1967; Guti\'errez--Moreno \& Moreno 1968; Glaspey 1971,
1972). Numerous authors have investigated radial and rotational
velocities, and the properties of the binary distribution (e.g.,
Blaauw \& van Albada 1963, 1964; van Albada \& Sher 1969; Rajamohan
1976; Levato et al.\ 1987; Verschueren, David \& Brown 1996; Brown \&
Verschueren 1997; Brandner \& K\"ohler 1998).

Blaauw (1960, 1964a) divided the association in three separate
concentrations or subgroups: Upper Scorpius\break (US), Upper Centaurus
Lupus (UCL), and Lower Centaurus Crux (LCC), and derived an expansion
age of $\sim$20~Myr for the entire complex (Blaauw 1964b). He
estimated the kinematic age of US to be $\sim$5~Myr (Blaauw 1978). At
$b$$\sim$$20^\circ\!$, US is well-separated from the early-type Galactic
field star distribution and is most concentrated, whereas UCL and LCC
are spread over a considerably larger area on the sky (see
Figure~9). As a result, UCL and LCC have received remarkably little
attention compared to US; they are not mentioned in Ruprecht's (1966)
list of nearby OB associations (\S 2.1).

De Geus (1992) summarized the characteristics of the interstellar
medium related to Sco~OB2. Whereas little interstellar matter is
associated with the subgroups UCL and LCC, filamentary material
connected to the Ophiuchus cloud complex ($345^\circ\!\! \la \! \ell \!
\la \! 10^\circ\!$, $0^\circ\!\! \la \! b \! \la \! 25^\circ\!$)
is observed towards US. This is also reflected in spatial variations
of extinction in this region. The densest part of this complex, the
$\rho$~Oph dark cloud, is a site of ongoing low-mass star formation
(e.g., Grasdalen, Strom \& Strom 1973; Greene \& Young 1992; Wilking
et al.\ 1997b), as well as interaction with the early-type stars of
US.

De Geus et al.\ (1989) analyzed Walraven as well as published
Str\"omgren photometry of the priority 1 stars in Sco~OB2. The
derived extinctions were compared with the IRAS 100$\mu$m emission,
and it was shown that the Ophiuchus dark clouds are on the near side
of US, at $\sim$125~pc. These authors derived ages of $\sim$5~Myr for
US, $\sim$13~Myr for UCL, and $\sim$10~Myr for LCC by isochrone
fitting in the Hertzsprung--Russell diagram. It follows that stars
with spectral types beyond the mid-F range may not yet have reached
the main sequence. Extensive H$\alpha$ and X-ray surveys of the
Sco~OB2 region indeed have revealed several dozens of pre-main
sequence objects (e.g., Walter et al.\ 1994; Feigelson \& Lawson 1997;
Preibisch et al.\ 1998; Sciortino et al.\ 1998).

Finally, we remark that $\zeta$~Oph (\HIP81377, O9.5V) has long been
recognized as a runaway star originating in Sco~OB2 (e.g., Blaauw
1952b, 1961, 1991; van Rensbergen et al.\ 1996). We will discuss it
elsewhere.

\begin{figure*}[t]
\centerline{
\psfig{file=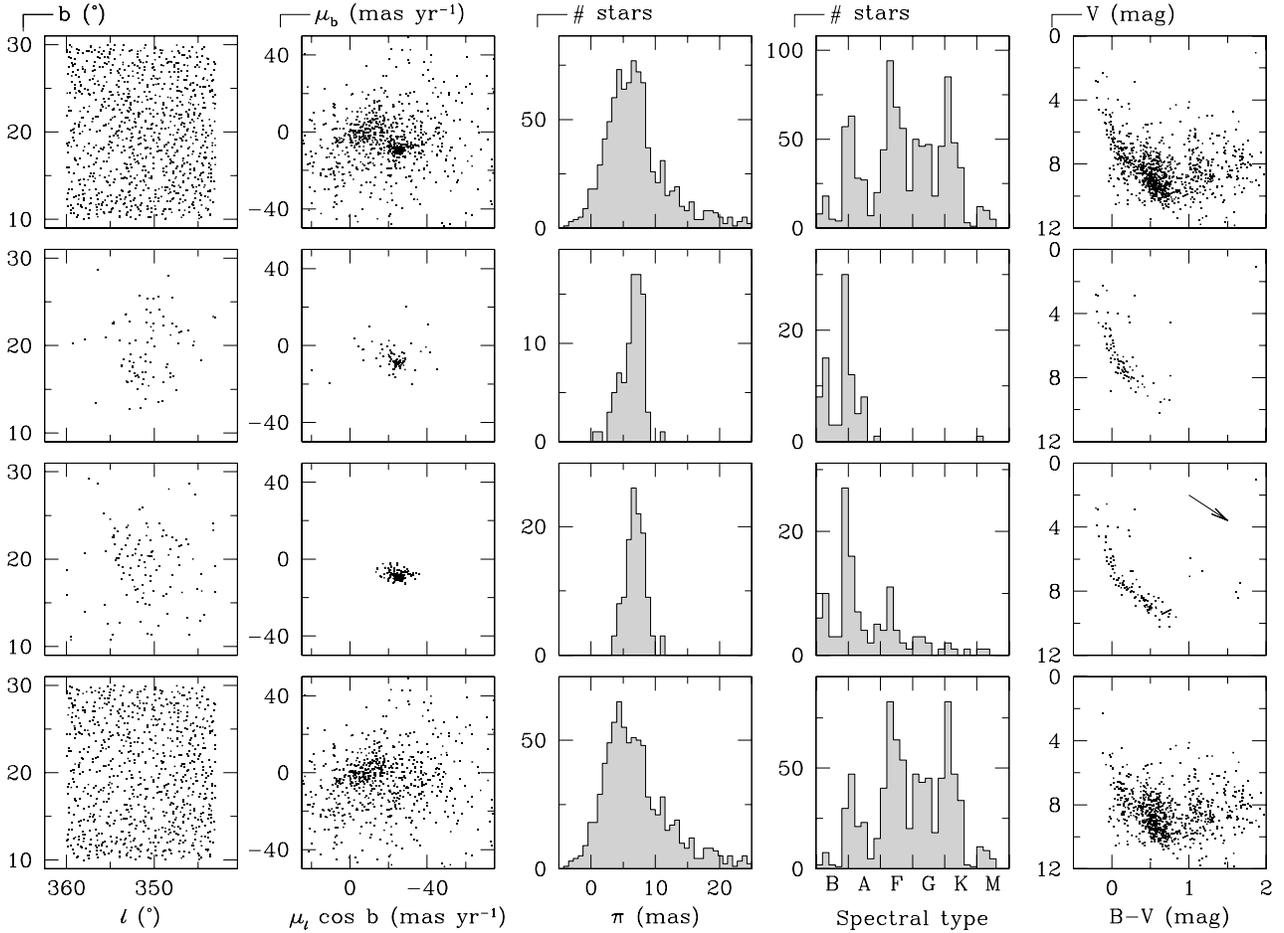,width=17truecm,silent=}}
\caption{\small Hipparcos measurements for Upper
Scorpius (from the top row down): (1) all 945 Hipparcos stars in the
region (cf.\ Table~A1); (2) the 86 pre-Hipparcos members; (3) the 120
Hipparcos members; (4) the remaining stars after member selection.
The columns show (from left to right): (1) positions in Galactic
coordinates; (2) Galactic vector point diagram; (3) trigonometric
parallax distribution; (4) spectral type distribution; (5)
color-magnitude diagram, not corrected for reddening. The arrow
indicates the direction of reddening for the standard value $R=3.2$ of
the ratio of total to selective extinction. The stars within the
apparent concentration around $(\mu_{\ell} \cos b, \mu_{b})$
$\sim$$(5,-5)~{\rm mas~yr}^{-1}$ in the first and fourth panels of the
second column do not form a moving group. }
\end{figure*}

\subsection*{\centerline{\normalsize\sl 4.2. Upper Scorpius}}

Figure~5 summarizes the results of our selection procedure for US.
The first row displays the Hipparcos measurements for all stars in our
US field (cf.\ Table~A1). The panels show no clear sign of a physical
group, except for the vector point diagram (panel two), which contains
a concentration around $(\mu_{\ell} \cos b, \mu_{b})$$\sim$$(-25,
-10)$~mas~yr$^{-1}$ superimposed upon the broader Galactic disk
distribution. The second row of Figure~5 shows only the 86 stars that
were proposed as members of US, based on pre-Hipparcos kinematic and
photometric studies (\S 4.1). They are mostly B- and A-type stars
concentrated towards the centre of the field, with the majority
contained in the same clump in the vector point diagram. Their
parallax distribution is narrower than the one in the first row, and
is peaked around 7~mas. The characteristics of the set of members
identified by our selection method are presented in the third row of
Figure~5. There is considerable overlap with the classical members,
but there are also significant differences. The vector point diagram
of these secure members is more concentrated than that of the
classical members, and the parallax distribution is narrower. This is
most likely due to a reduced contamination by field stars. The
observed spread and the elongated shape in the vector point diagram
are consistent with the combined effects of observational errors, our
choice of $\sigma_{\rm int}$, and projection on the sky. Finally, the
panels in the bottom row of Figure~5 show the not-selected stars, and
demonstrate that our procedure does not leave `holes' in the
distributions of positions, proper motions, and parallaxes. This
indicates that our method has separated US cleanly from the field
stars.

We find a total of 120 secure members of US: 49 B, 34 A, 22 F, 9 G, 4
K, and 2 M-type stars. Of these, 53 were previously classified as
member (39 kinematic, 14 photometric), while 67 are new. We confirm
the evolved supergiant Antares ($\alpha$~Sco, \HIP80763, M1Ib), as
well as the stars $\rho$~Oph (\HIP80473, B2V), $\chi$~Oph (\HIP80569,
B2Vne), and 48~Lib (\HIP78207, B8Ia/Iab), as members. The latter two
are classical proper motion members with peculiar positions in the
Hertzsprung--Russell diagram (e.g., de Geus et al.\ 1989). We select 2
A stars with peculiar spectra (\HIP78494, A2m...; \HIP79733, A1m...),
and reject the doubtful classical proper motion member $o$~Sco
(\HIP80079, A4II/III; see Blaauw, Morgan \& Bertiau 1955; de Geus et
al.\ 1989). We also do not select the classical proper motion member
$\delta$~Sco (\HIP78401). However, this star is a binary with a
separation of $0\farcs13$ (16~AU at $\pi = 8.12~{\rm mas}$). Hipparcos
detected a change in the position angle of the binary components of
$18^{\circ}~{\rm yr}^{-1}$, which indicates a period of
$\sim$20~yr. Therefore, the Hipparcos proper motion, observed during
the mission lifetime of $\sim$3.3~yr, does not necessarily reflect the
center-of-mass proper motion (\S 3.5). The typical differences between
the center-of-mass and observed proper motions are on the order of
2~${\rm mas}~{\rm yr}^{-1}$ (e.g., Wielen et al.\ 1997), enough to
explain why $\delta$~Sco is not selected.

\begin{figure}[t]
\centerline{
\psfig{file=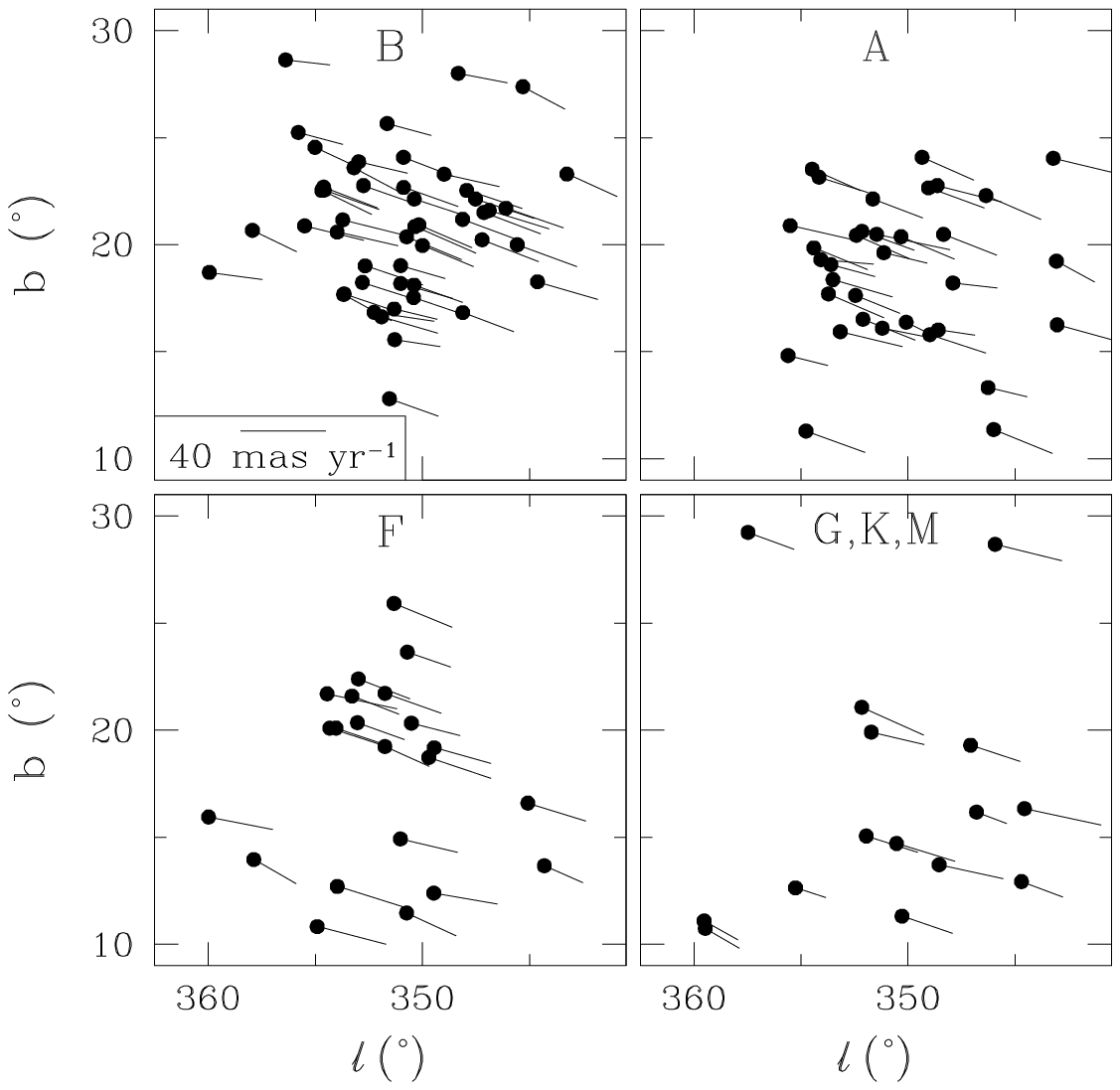,width=8truecm,silent=}}
\caption{\small Positions and proper motions for the
Hipparcos members of Upper Scorpius, as a function of spectral type:
49 B, 34 A, 22 F, 9 G, 4 K, and 2 M-type stars. The selected F, G, K,
and M stars co-move with the B and A stars. The distribution of F
stars is similar to that of the early-type stars, suggesting that the
stellar content of the association has now been established to
spectral type F. Some of the later-type stars could be
interlopers.}
\end{figure}

The mean distance of US is 145$\pm$2~pc, where the error is the
formal error on the mean, and the correction applied for the bias
caused by exclusion of stars with negative parallaxes is negligible
(\S 3.6). This value is consistent with the most recent photometric
determination of 160$\pm$40~pc by de Geus et al.\ (1989). The
accuracy of the individual parallaxes does not allow us to resolve the
internal structure. If the association extends about as far along the
line of sight as it does on the sky ($\sim$$14^\circ\!$), the parallax
spread is $\sim$1.6~mas, corresponding to $\sim$$0\fm5$. Taking into
account the individual measurement error of $\sim$1~mas, this is
consistent with the observed width of the parallax distribution.

Figure~6 shows the positions and proper motions for all secure
members, as a function of spectral type. There is a clear
concentration in the early-type stars. Some of the stars that lie
away from this main concentration could be interlopers. We have
estimated the expected number of interlopers by means of the Monte
Carlo simulations described in detail in \S 3.4. The result is given
in Table~A2, and shows that the observed clustering in the F-star
distribution is significant. The concentration of F-type members
coincides with a similar concentration of X-ray selected pre-main
sequence stars, around $(\ell,b)$$\sim$$(350^\circ\!,20^\circ\!)$ found
by Walter et al.\ (1994) and Preibisch et al.\ (1998). These generally
have $V$$\sim$$12^{\rm m}$--$17^{\rm m}$, and are too faint for
Hipparcos, so no reliable kinematic link is possible. The one
exception may be provided by the US member\break \HIP77960 (\SAO183927,
A4IV/V), which is listed as the possible optical identification of the
X-ray detected candidate pre-main sequence object Sco--014 (Sciortino
et al.\ 1998). The majority of the G- to M-type members probably do
not belong to US, although only 4 interlopers are predicted for these
spectral types. However, this number is an underestimate because of
the large fraction of giants and supergiants in the Hipparcos Catalogue
(\S 3.4.3).

The secure US members span a much larger range in spectral type than
the classical members, which are confined to stars earlier than
$\sim$A8--F0. The distribution of secure members in the
color-magnitude diagram extends to fainter magnitudes than for the
classical members (right panels in Figure~5). The main sequence is not
very narrow; we infer that this is caused by the distance spread along
the line of sight ($\sim$$0\fm5$), by unresolved binaries, and
probably by significant differential reddening, especially for the
bright early-type members (including $\rho$~Oph, $\sigma$~Sco,
$\pi$~Sco, and $\nu$~Sco, all of which are associated with
nebulosities; de Geus et al.\ 1989). Many fainter members are likely
missing, due to the completeness limit of the Hipparcos
Catalogue. Even so, the new late-type members allow a much improved
definition of the lower part of the main sequence (cf.\ \S 4.5).

\subsection*{\centerline{\normalsize\sl 4.3. Upper Centaurus Lupus}}

We started the analysis of UCL by applying our selection method to the
Hipparcos objects in the classical `field 3' defined by Blaauw
(1946). This resulted in the detection of a moving group, with members
spread over nearly the entire field. We subsequently modified it to
the field given in Table~A1, to be sure that we included the entire
subgroup (cf.\ Figure~9). The results are summarized in Figure~7,
which is similar to Figure~5, except that the improvement in the
member list for UCL provided by Hipparcos is more dramatic than for
the much better studied US subgroup. Our member selection method
identifies 221 members in this field, with a mean distance of
140$\pm$2~pc: 66 B, 68 A, 55 F, 25 G, 6 K, and 1 M-type star. Of
these, only 58 were previously classified as member (24 kinematic, 34
photometric), while 163 are new. We discarded 78 classical members.
The parallax distribution and vector point diagram of the new members
both show a significantly smaller spread than those of the classical
members. The new UCL members are not spread uniformly over the sky but
seem to be concentrated in `clumps'. Although some of this `clumping'
could be caused by selection effects in the Hipparcos Catalogue, it
seems likely that UCL has substructure (\S 4.5). The number of
expected interlopers (Table~A2) indicates that some of the G and K
stars may well be non-members.

The mean distance of UCL is consistent with the value 160$\pm$40~pc
found by de Geus et al.\ (1989). Just as for US, the width of the
parallax distribution is consistent with the measurement errors and
the angular extent of $\sim$$27^\circ\!$, which at this distance
corresponds to $\sim$3.6~mas ($1\fm1$) if UCL is roughly spherical.

\begin{figure*}[t]
\centerline{
\psfig{file=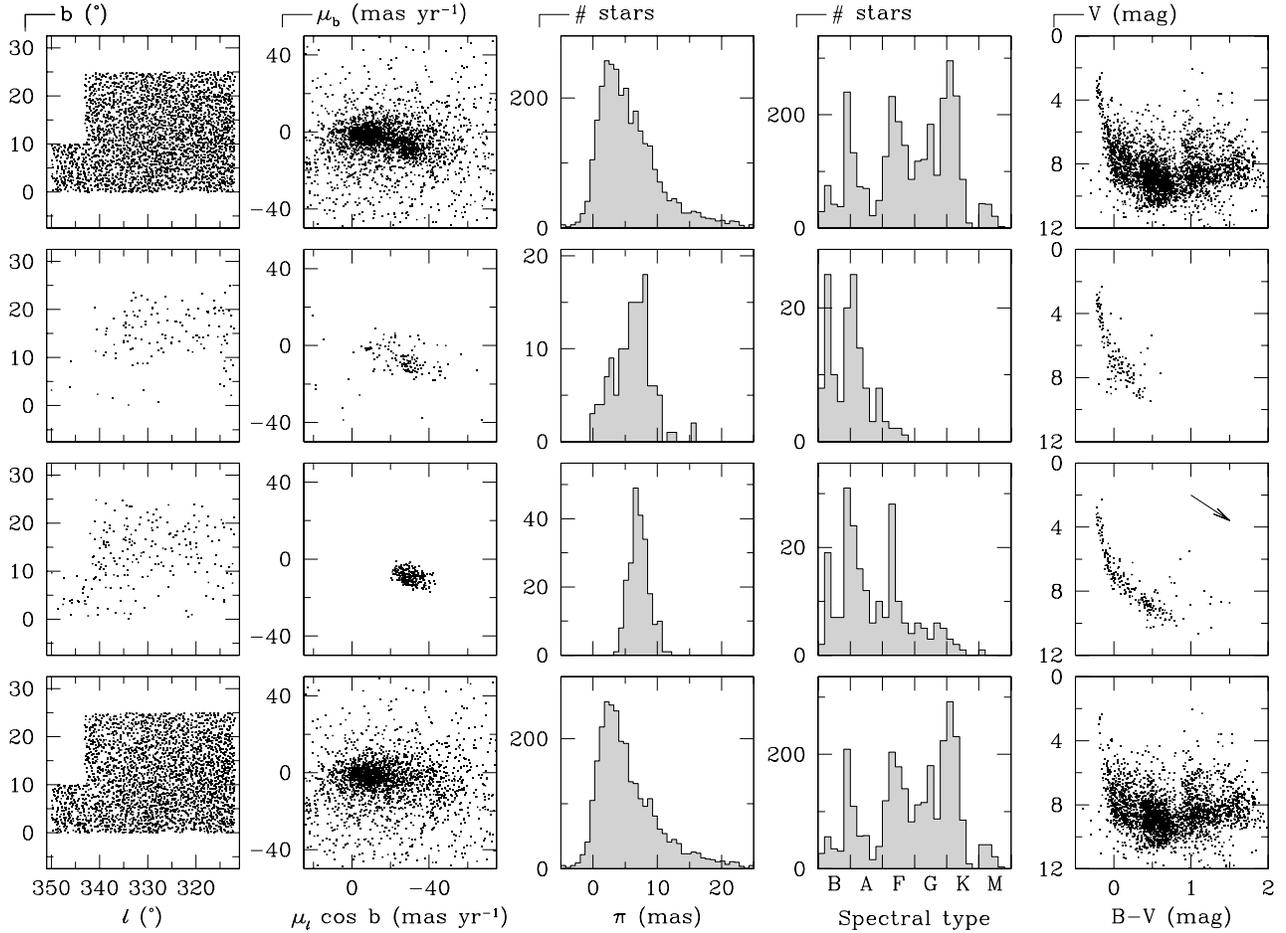,width=17truecm,silent=}}
\caption{\small As Figure~5, but for the 3132 stars in
the field of Upper Centaurus Lupus. The stars within the apparent
concentration around $(\mu_{\ell} \cos b,
\mu_{b})$$\sim$$(-20,-15)~{\rm mas~yr}^{-1}$ in the second panel of
the first row do not form a moving group.}
\end{figure*}

The panels in the fourth row of Figure~7 show smooth distributions of
Galactic disk field stars in position, proper motion, and parallax.
The color-magnitude diagram of the non-members contains five stars
with $V \la 4^{\rm m}$ which follow the main sequence of the secure
members remarkably well. With the exception of $\zeta$~Cen, these are
classical proper motion members. This suggests that $\mu^1$~Sco
(classified as `V'-type by Hipparcos\footnote{We use the
nomenclature for binaries that corresponds to the five parts of the
Double and Multiple Systems Annex of the Hipparcos Catalogue, where
`C' indicates component solutions, `G' indicates an acceleration
solution, `O' an orbital solution, `V' a variability-induced mover,
and `X' a stochastic solution.}), $\beta$~Lup, $\eta$~Cen (B1Vn+A),
and $\iota$~Lup, could be members of UCL, although their kinematic
properties as observed by Hipparcos are not consistent with the mean
motion of UCL. We suspect that perturbations of the stellar motion due
to marginally resolved multiplicity could be important (\S 3.5), but
note that the SIMBAD radial velocity for $\iota$~Lup differs
significantly from that of UCL (Table~2). 

The color-magnitude diagram of the secure members of UCL is cleaner
than that of US, which we attribute to the absence of significant
differential extinction (de Geus et al.\ 1989). The vertical spread of
the main sequence is consistent with the expected distance range of
$\sim$$1\fm1$, and the presence of unresolved binaries. The main
sequence extends nearly two magnitudes fainter than in the
pre-Hipparcos diagram, and indeed includes not only 8 stars with
peculiar and/or magnetic spectra, but also \HIP77157 (HT~Lup,
$V=10\fm31$, $B\!-\!V=1\fm264$, Ge) and \HIP82747 (AK~Sco, $V=9\fm21$,
$B\!-\!V=0\fm746$, F5V), which are binary T Tauri stars (e.g.,
Gregorio--Hetem et al.\ 1992), and \HIP74797, which is the A component
of the binary system CCDM J15172--3435, the B component of which is a
pre-main sequence T Tauri binary system (Brandner et al.\ 1996). We
also select another four pre-main sequence stars: \HIP75924 (G6V),
\HIP73777 (K(1)V+G), \HIP77524 (K0V:), and \HIP78684 (G8V) (Neuh\"auser \&
Brandner 1998).

The star-forming region associated with the Lupus\break clouds, at a
distance of 140$\pm$20~pc, contains several tens of low-mass pre-main
sequence stars with estimated ages of $\sim$3~Myr (e.g., Hughes,
Hartigan \& Clampitt 1993; Hughes et al.\ 1994; Chen et al.\ 1997;
Wichmann et al.\ 1997b). However, most of them have $V \ga 12^{\rm
m}$, and only 5 were observed by Hipparcos (HT~Lup, and \HIP78094,
78317, 79080, 79081). Wichmann et al.\ (1997a) derive a distance of
190$\pm$27~pc, based on Hipparcos parallaxes, and suggest that these
stars are members of UCL (cf.\ Murphy, Cohen \& May 1986). We select
only HT~Lup as secure UCL member.

\begin{figure*}[t]
\centerline{
\psfig{file=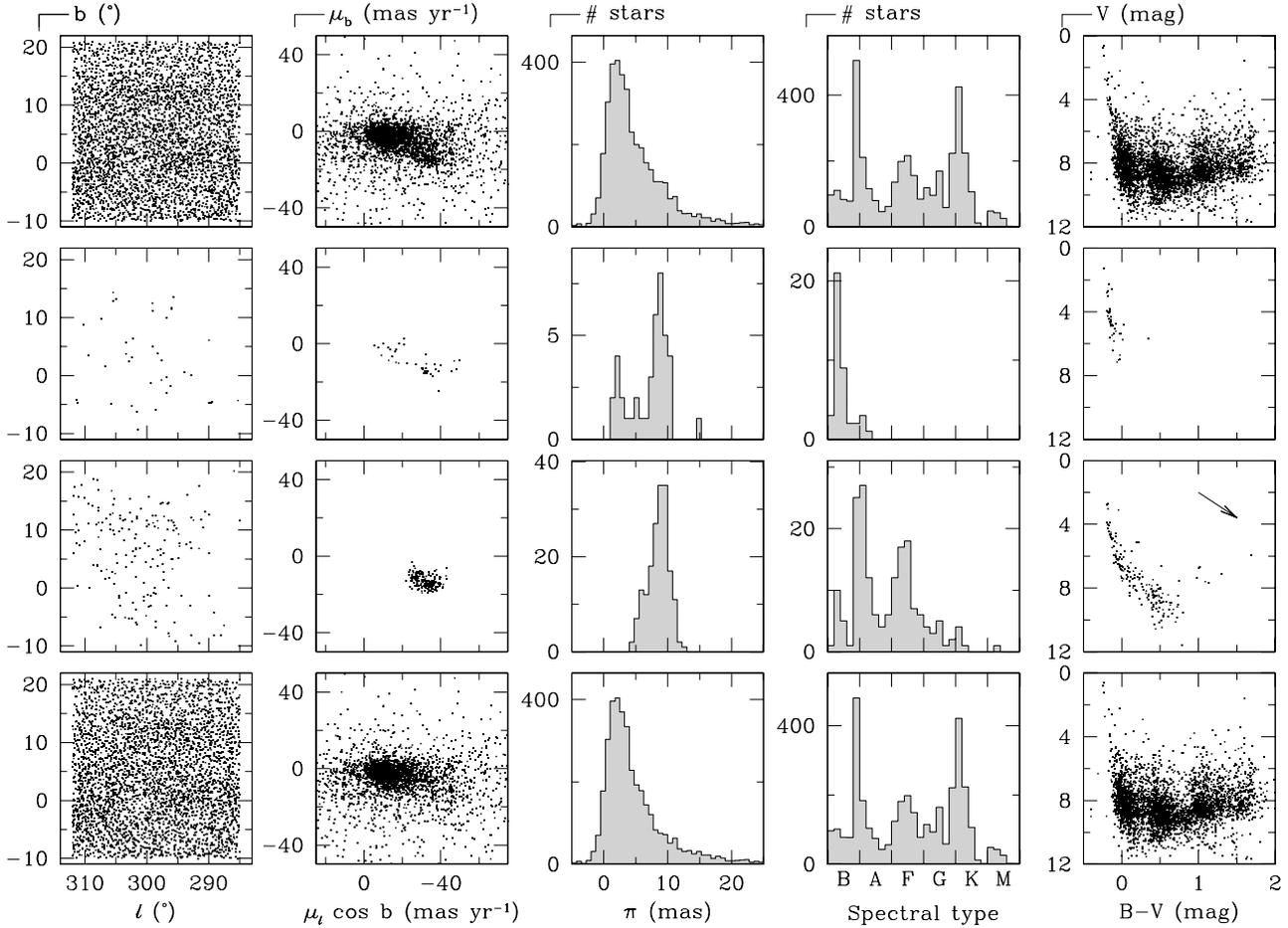,width=17truecm,silent=}}
\caption{\small As Figure~5, but for the 3897 stars in
the field of Lower Centaurus Crux. The stars in the concentration at
$(\ell, b)$$\sim$$(290^\circ\!, -5^\circ\!)$ in the first and fourth
panel of the first column are members of the open cluster \FIC2602
(cf.\ Hoogerwerf \& Aguilar 1998). }
\end{figure*}

\subsection*{\centerline{\normalsize\sl 4.4. Lower Centaurus Crux}}

We started our analysis of LCC in Blaauw's (1946) `field 4', and then
extended the field boundaries to those given in Table~A1 to be sure of
including the entire subgroup (cf.\ Figure~9). Our selection procedure
is summarized in Figure~8, which is organized in the same way as
Figures~5 and~7. LCC was never studied extensively, and only 41
classical members are known, all with spectral type earlier than B6.
Identification of later-type members in this rich field in the
Galactic disk was particularly difficult, and suggests that
interlopers are present even among the brightest members. The second
and third panels of row two in Figure~8 indeed show many deviating
proper motions and parallaxes. We confirm only 19 (16 kinematic, 3
photometric) of the classical members, but identify a total of 180
secure members: 42 B, 55 A, 61 F, 15 G, 6 K, and 1 M-type star, so
that Hipparcos has transformed this sparse subgroup into a moving
group that is very similar to UCL. The members are not spread
uniformly over the field, and show some evidence for substructure.

The panels in the bottom row of Figure~8 indicate a clean separation
of LCC from the field star distribution, but just as in the case of
UCL and US, the color-magnitude diagram of the non-members contains a
number of bright ($V \la 4^{\rm m}$) early-type stars which were
discarded by our selection method, and which deserve attention. The
classical proper motion member $\delta$~Cen (\HIP59196, B2IVne) is a
$\gamma$~Cas-type eruptive variable identified by Hipparcos as a
`variability-induced mover' (see footnote 2). Another is $\beta$~Crux
(\HIP62434, B0.5III), rejected as LCC member by Jones (1971). It was
observed by Hipparcos as `G'-type binary, and is probably an
astrometric binary with a period $\ga$10~yr. \HIP59449 (B3V) is
another `G'-type binary. We suspect that these are in fact members of
LCC. The stars $\beta$~Cen (\HIP68702, B1III, Agena) and \HIP60718
(B0.5IV) are Hipparcos `component binaries', but were never proposed
as members of LCC, and neither was \HIP51576 (B4Vne). We do not select
them either. \HIP66657 (B1III, `G'-type binary) and \HIP61199 (B5V,
$\gamma$~Mus) were already rejected by Jones (1971). Finally,
\HIP52419 (B0Vp, `O'-type binary) is a member of \IC2602 (Hoogerwerf
\& Aguilar 1998).

LCC is significantly closer than US and UCL: the Hipparcos mean
distance of 118$\pm$2~pc agrees well with earlier photometric
estimates (de Geus et al.\ 1989). The parallax distribution is
marginally broader than for UCL, but again does not allow us to
resolve the internal structure of the subgroup. The expected distance
spread corresponding to the $\sim$25$^\circ\!$ extent on the sky is
$\sim$3.7~mas (or $0\fm9$). Taking into account the individual
measurement errors, this is consistent with the observed width of the
parallax distribution.

Chereul, Cr\'ez\'e \& Bienaym\'e (1997) used a different method to
search for moving groups in the Hipparcos Catalogue, and applied it to
nearby A dwarfs with pre-Hippar\-cos distances less than 125~pc.
They identified a moving group of 33 stars at a mean distance of
105~pc in `Lupus--Centaurus', in a field which roughly coincides with
our LCC field. These are clearly LCC members, but the sample
definition biases them towards the near side of the subgroup, and
explains the somewhat smaller mean distance than we find.

\begin{figure*}[t]
\centerline{
\psfig{file=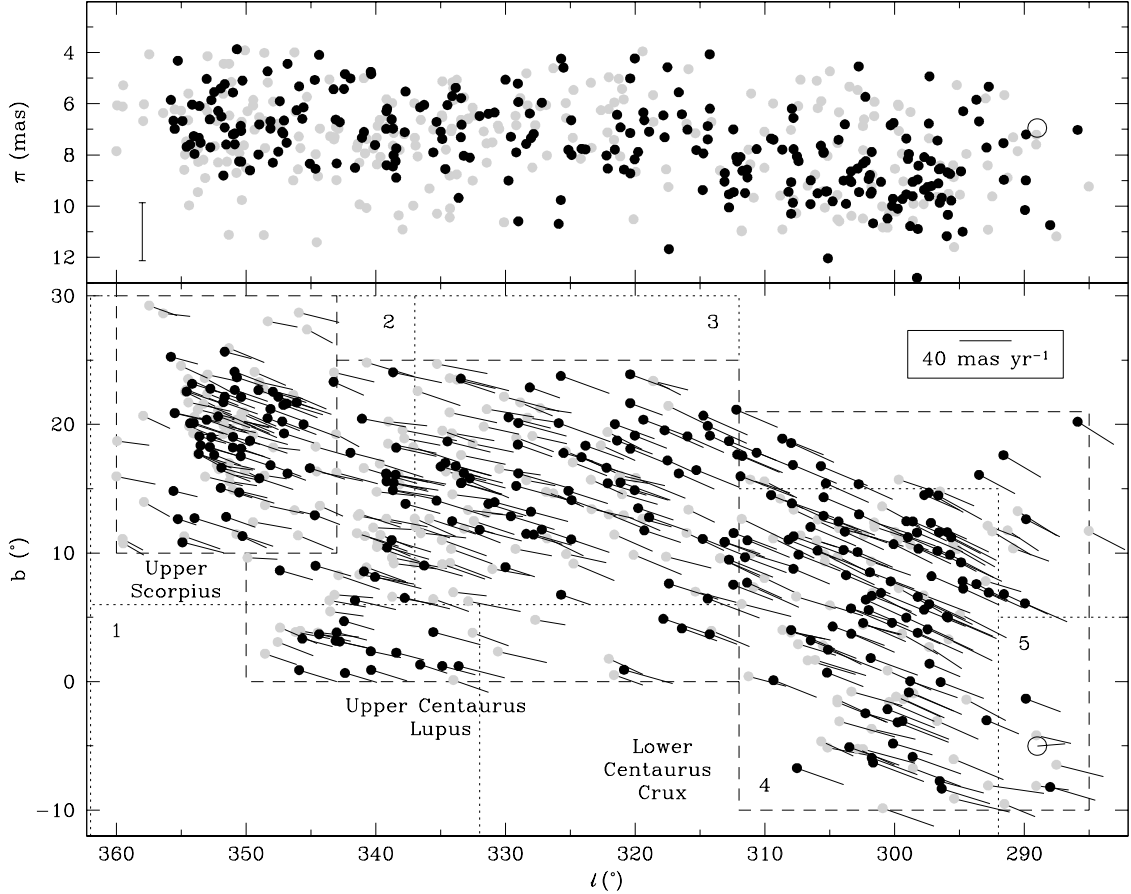,width=15truecm,silent=}}
\caption{\small Positions and proper motions (bottom), 
and parallaxes (top), for 521 members of Sco~OB2 selected from 7974
stars in the Hipparcos Catalogue in the area bounded by the dashed
lines. The vertical bar in the top panel corresponds to the average
$\pm$$1\sigma$ parallax range for the stars shown. The black dots
indicate stars with membership probability $P \geq 95$~per cent. Gray
dots indicate the remaining members. Many members near the association
boundary have a low membership probability. The dotted lines are the
schematic boundaries of the classical subgroups Upper Scorpius (2,
US), Upper Centaurus Lupus (3, UCL), Lower Centaurus Crux (4, LCC),
and the candidate subgroups (1 and 5) defined by Blaauw (1946). US is
identified as a subgroup based on the concentration of members on the
sky, and LCC can be distinguished from US and UCL based on its
significantly smaller distance. The large open circle represents the
open cluster \FIC2602.}
\end{figure*}

The color-magnitude diagram of our secure LCC members, not corrected
for reddening, is shown in the right-hand panel of the third row of
Figure~8. The vertical spread of the main sequence is consistent with
the combined effects of some differential reddening, the distance
range, and unresolved binaries. The main sequence extends a full
four magnitudes further towards fainter objects than the one for the
classical `members' shown in the panel above it, and indeed extends
into the regime where we would expect some pre-main sequence
objects. Unfortunately, none of the pre-main sequence members of LCC
found by Feigelson \& Lawson (1997) appear in the Hipparcos
Catalogue. However, we do select \HIP56420 ($V=11\fm56$, $B\!-\!V=0\fm789$,
$\pi=7.20\pm2.92$~mas, Gp), which might be a T Tauri star.

The LCC field also contains a concentration of stars in the direction
$(\ell,b)$$\sim$$(290^\circ\!,-5^\circ\!)$ (bottom left panel in
Figure~8). This is the open cluster \IC2602 (age 8--30~Myr; Whiteoak
1961; Braes 1962; Stauffer et al.\ 1997). Blaauw (1964a) suggested
that \IC2602 may belong to LCC. Hoogerwerf \& Aguilar (1998) show
that the Spaghetti method (\S 3.2) identifies \IC2602 flawlessly as a
separate group, containing 23 Hipparcos members at $\sim$144~pc (cf.\
Mermilliod et al.\ 1997), with a motion that differs significantly
from that of LCC. We do not discuss it further here.

\subsection*{\centerline{\normalsize\sl 4.5. The Sco~OB2 complex}}

Figure~9 illustrates the motions of all the Sco~OB2 members we have
identified, and also gives the subgroup boundaries defined by Blaauw
(1946, p.\ 43). UCL and LCC clearly extend beyond the classical
boundaries, and contain stars from Blaauw's `fields 1 and 5' (Sco~OB2$\_$1
and Sco~OB2$\_$5 in Table~1). We do not find astrometric evidence for
additional subgroups of Sco~OB2 in these fields.

The individual membership probabilities $P$ of the selected stars (\S
3.3) are listed in Table~C1. The cumulative distributions are
displayed in Figure~C1, and range from $\sim$50 to 100~per cent. The
black dots in Figure~9 indicate all members with $P \geq 95$~per
cent. These cover nearly the same area on the sky as the full set of
members, but the outlying members of US, and some of UCL and LCC, all
have low values of $P$. The stars in the concentrations near $(\ell,
b)$$\sim$$ (342^\circ\!, 3^\circ\!)$, $(338^\circ\!, 15^\circ\!)$ and
$(334^\circ\!, 16^\circ\!)$ have high probabilities, which suggests
that UCL has substructure.

Application of our member selection method to the entire field shown
in Figure~9 results in the identification of one coherent structure:
Sco~OB2. US stands out in the distribution of Sco~OB2 members on the
sky, and the parallax distribution clearly distinguishes UCL and
LCC. The differences between the Hertzsprung--Russell diagrams of the
groups (de Geus et al.\ 1989) also indicate that a division of Sco~OB2
into three separate subgroups is warranted. Our field boundary
separating UCL from US has, somewhat arbitrarily, been chosen in such
a way that US comprises the stellar concentration centered on
$(\ell,b)$$\sim$$(352^\circ\!,20^\circ\!)$ with radius $\sim$$5^\circ\!$ (\S
4.2), separated by a sparsely populated zone from the main body of UCL
(Figure~9). Analysis of the secular parallaxes (e.g., Dravins et al.\
1997) will allow an improved description of the internal structure of
Sco~OB2.

Accurate intermediate-band Walraven \vbluw\ photometry is available
for 1870 of the 7974 Hipparcos stars in the Sco~OB2 field (de Geus et
al.\ 1990). These include 261 of the 521 Hipparcos members. \UBV\
photometry is available for 327, and \uvby$\beta$ for 235, of the
Hipparcos members (Hauck \& Mermilliod 1990; Mermilliod \& Mermilliod
1994; cf.\ Slawson, Hill \& Landstreet 1992). We postpone a discussion
of the physical properties of the stars in Sco~OB2 based on this and
other material to a future paper, but refer to de Zeeuw et al.\ (1997;
figure~2) for an illustration of the improved quality of the
color-color diagrams due to the refined and extended astrometric
membership lists. Completion of the photometry will allow to correct
for differential reddening, to improve the provisional determination
of the initial mass function by Brown (1998), to determine whether an
up-turn of the main sequence occurs in the F-star regime, as expected
for groups of this age, and to investigate if small-scale structure is
associated with age differences.

\subsection*{\centerline{\normalsize\sl 4.6. Corona Australis}}

\begin{figure}[t]
\centerline{
\psfig{file=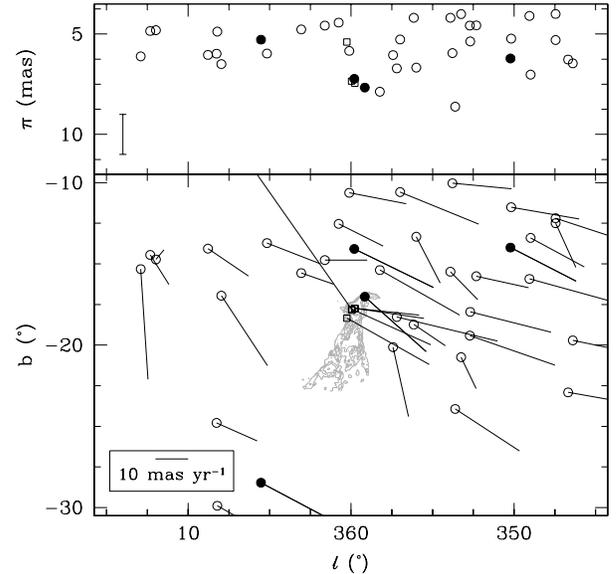,width=8truecm,silent=}}
\caption{\small Positions and proper motions (bottom)
and parallaxes (top) for all OB stars in the field $345^\circ\!\! <
\! \ell \! < \! 15^\circ\!$, $-30^\circ\!\! < \! b \! < \! -10^\circ\!$
with parallaxes $4 \! <\! \pi \!<\! 12$~mas. Filled circles denote
O--B3 stars. Open squares denote the 5 classical CrA members observed
by Hipparcos, which fall inside the shaded region representing the
IRAS 100$\mu$m flux measurements. The parallaxes for R~CrA and
\FHR7170 fall outside the plotted range. The deviating proper motion
of the star R~CrA is insignificant.}
\end{figure}

{\it Pre-Hipparcos}: Blaauw's (1946) `field 1' contains the Coro\-na
Australis cloud complex located at $(\ell,b)$$\sim$$(360^\circ\!,
-18^\circ\!)$, at 150$\pm$50~pc (Gaposchkin \& Greenstein 1936). It is
also called Corona Austrina (Botley 1980), and contains sites of
active intermediate- and low-mass star formation extending a few
degrees on the sky, which have been studied extensively (e.g., Rossano
1978; Loren 1979; Goss et al.\ 1980; Brown 1987; Cappa de Nicolau \&
P\"oppel 1991; Harju et al.\ 1993; Andreazza \& Vilas--Boas 1996; cf.\
Figure~10). The densest molecular core contains the emission-line
star R~CrA (A5IIe~var). The stellar content of the CrA complex was
studied through H$\alpha$ emission-line surveys, infrared surveys, and
X-ray observations (Knacke et al.\ 1973; Glass \& Penston 1975; Vrba,
Strom \& Strom 1976; Marraco \& Rydgren 1981; Wilking et al.\ 1985,
1992, 1997a; Koyama et al.\ 1996; Walter et al.\ 1997). These studies
revealed a mixed, predominantly pre-main sequence, stellar content:
several embedded and/or Herbig stars (e.g., R~CrA, T~CrA, \HD176386,
TY CrA) with the associated reflection nebulae \NGC6729 and 
\NGC6726/6727, as well as a loose association of T Tauri stars (e.g., Herbig 
\& Rao 1972), Herbig--Haro objects (Strom, Strom \&
Grasdalen 1974; Strom, Grasdalen \& Strom 1974; Wilking et al.\
1997a), and the embedded infrared Coronet cluster associated with
R~CrA (Taylor \& Storey 1984). The pair \HR7169/\HR7170 and
\SAO210888 may not have formed in the CrA cloud complex (Marraco 1978;
Marraco \& Rydgren 1981; Wilking et al.\ 1992). Str\"omgren photometry
for \HR7169/7170 and \HD176386 indicates a mean distance of approximately 
130~pc.

\begin{figure*}[t]
\centerline{
\psfig{file=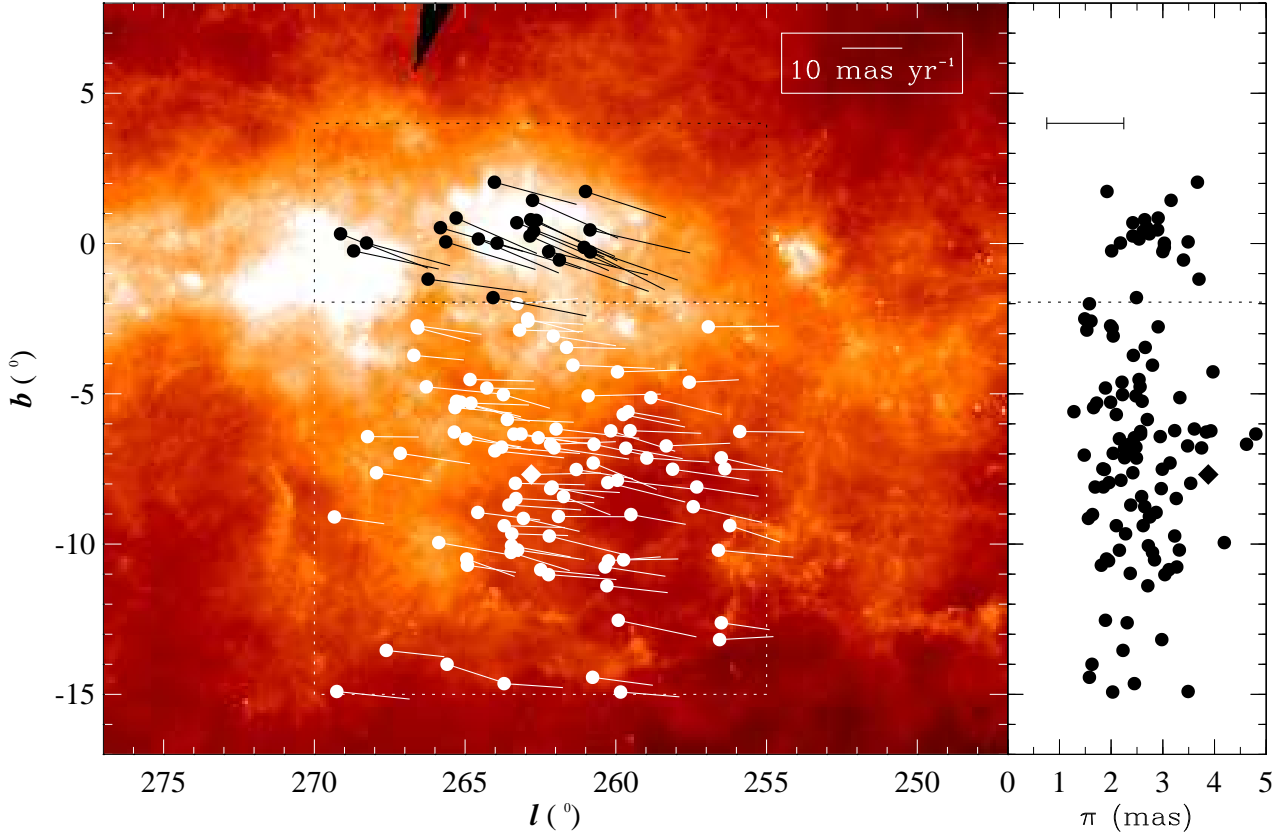,width=17truecm,silent=}}
\caption{\small Left: positions and proper motions for the 
Hipparcos members of Vela~OB2 (white) and Trumpler~10 (black). The
diamond denotes the Wolf--Rayet star $\gamma^2$~Velorum (WR11). The
dotted lines indicate the field boundaries (cf.\ Table~A1). The
gray-scale represents the IRAS 100$\mu$m skyflux. The IRAS Vela shell
is the ring-like structure centered on $(\ell, b)$$\sim$$(263^\circ\!,
-7^\circ\!)$ with a radius of $\sim$6$^\circ\!$ surrounding Vel~OB2.
The intense emission in the area $260^\circ\!\! \la \! \ell \! \la \!
273^\circ\!$, $-2^\circ\!\! \la \! b \! \la \! 2^\circ\!$ corresponds
to the Vela molecular ridge. Right: parallax distribution for Vel~OB2
and Tr~10}
\end{figure*}

\smallskip
{\it New results}: We have investigated whether there is astrometric
evidence for a moving group associated with the CrA star-forming
region. We have performed our membership analysis on Hipparcos data in
the field $345^\circ\!\! < \! \ell \! < \! 15^\circ\!$, $-30^\circ\!\!
< \! b \! < \! 0^\circ\!$. The few bright B0--B3 stars in the field
were already noted by Blaauw (1946, 1964a), who suggested they might
belong to a putative nearly dispersed `subgroup 1' of the Sco~OB2
association, at an estimated distance of $\sim$170~pc. We find no
evidence for the existence of a moving group. Changing the field size
does not alter this conclusion. Figure~10 shows the distribution of
all OB stars in the field $345^\circ\!\! < \! \ell \! < \! 15^\circ\!$,
$-30^\circ\!\! < \! b \! < \! -10^\circ\!$ with measured parallaxes $4 \!<
\! \pi\! <\! 12$~mas. Nearly all of these are B8--B9 stars with
parallaxes $\pi \la 6~{\rm mas}$; their proper motions simply reflect
the Solar motion.

The five stars observed by Hipparcos associated with the CrA clouds by
themselves also provide insufficient evidence for a moving group:
R~CrA (\HIP93449) is faint ($V = 11\fm57$): the measured parallax
($\pi = 121.75 \pm 68.24~{\rm mas}$) and proper motions are not
significant, and the `stochastic binary' and `suspected non-single'
flags are set. The pair \HR7169/7170 (\HIP93368/\-93371) has an
uncertain parallax. The remaining two, \SAO210888 (\HIP93689)
and\break \HD176386 (\HIP93425), have parallaxes consistent with the
classical distance estimate of $\sim$130~pc, and similar proper
motions. If these belong to a moving group associated with the CrA
complex, then the other members must be fainter than the magnitude
limit of the Hipparcos Catalogue.

\section*{\centerline{\normalsize 5. VELA}}

Blaauw's (1964a) map of the early-type stars along the Galactic plane
(his figure~3) contains a concentration of O--B3 stars brighter than
$V_{\rm pg}=7^{\rm m}$ in a field in Vela centered on old Galactic
coordinates $(\ell^I,b^I)$$\sim$$(230^\circ\!,-10^\circ\!)$, which had
already been noted by Kapteyn (1914). This field of low interstellar
extinction out to $\sim$1~kpc contains a number of nearby stellar
groups, sometimes referred to as the Vela Sheet (cf.\ Eggen 1980).
Here we discuss Vel~OB2 and Tr~10.

\subsection*{\centerline{\normalsize\sl 5.1. Vela~OB2}}

{\it Pre-Hipparcos}: In his 1914 paper, Kapteyn not only identified
Sco~OB2, but also discussed a group of bright stars in Vela (his \S 6,
p.\ 63) and, based on proper motions, listed 15 probable members (his
table~XIV) for this so-called Vela Group. Lacking radial velocities
for 14 of these stars, Kapteyn concluded that `the reality of the
group seems ... probable, though not beyond a doubt'. The
widely-spread radial velocities for 12 of the 15 stars used by Blaauw
(1946, his \S 22, p.\ 101) did not allow confirmation of Kapteyn's
Vela Group. It is not listed as an OB association by Ruprecht (1966).
Brandt et al.\ (1971) noted the presence of 17 bright early-type stars
within a few degrees of the Wolf--Rayet WC8$+$O9I binary system
$\gamma^2$~Velorum (\HIP39953, WR11). These authors took the similar
distance moduli for 10 of these stars (including $\gamma^2$~Vel) as
evidence for an OB association at $\sim$450~pc: Vel~OB2. Straka (1971,
1973) investigated SAO proper motions and Bright Star Catalog radial
velocities for these 10 stars. Only 5 of them, including the multiple
system $\gamma^1$--$\gamma^2$~Vel, turned out to share a common space
motion. Abt et al.\ (1976) discussed astrometric and photometric
observations of 7 stars around $\gamma^2$~Vel, and concluded that
$\gamma^1$~Vel and \CD$-$46~3848 are probably associated with
$\gamma^2$~Vel, and that $\gamma^2$~Vel, as well as
\HD68157 and \HD68324 (\HIP39970), are members of Vel~OB2. Photometric
evidence for the existence of an association at $\sim$450~pc was
provided by Upton (1971), Straka (1973), and Eggen
(1986)\footnote{Eggen (1982) also mentions Vel~OB2, but uses this
designation for another group of stars.}.

\smallskip
{\it New results}: Our selection confirms only 4 of the 10 Brandt et
al.\ stars as members. One of the 6 remaining stars, $\gamma^1$~Vel
(\HR3206), was not observed by Hipparcos. However, our member
selection procedure identifies 89 new members for Vel~OB2, which
brings the total number to 93: $\gamma^2$~Vel, 81 B, 5 early-A, and 3 G-
and 3 K-type giants. Figure~11 shows the distribution of positions and
parallaxes of the selected stars. The abrupt cutoff of the membership
list beyond spectral type B9 is artificial: inclusion of many B-type
stars in the Hipparcos Catalogue, and the restriction on the stellar
density of 3 stars per square degree, cause an absence of stars in
Vela of later spectral types, starting already with spectral type A
(cf.\ Figure~2 and \S 3.4). Vel~OB2 has a space motion which does not
clearly separate the association from the field star population.
Therefore, the expected number of interlopers is high: $\sim$30 B
stars and $\sim$5 A stars.

Figure~12 (left panel) shows the color-magnitude
dia\-gram\footnote{Three widely different $B\!-\!V$ values for the A0V
star \HIP38816 ($V=9\fm26$) are available: (i) the Hipparcos Catalogue
lists a ground-based $B\!-\!V = 1\fm020 \pm 0\fm020$; (ii) the SIMBAD
database gives $B\!-\!V =-0\fm9$; (iii) Tycho photometry
($B_{T}=9\fm399 \pm 0\fm024$, $V_{T} = 9\fm330 \pm 0\fm032$, $B_{T} -
V_{T} = 0\fm069$) can be converted into a Johnson $B\!-\!V$ color
using eq.~1.3.24 in ESA (1997; Vol.\ 1 \S 1.3, Appendix~4, p.\ 60),
resulting in $B\!-\!V = 0\fm062$. As neither POSS plates nor IRAS
maps reveal associated interstellar medium features which could affect
the $B\!-\!V$ color, we adopt $B\!-\!V = 0\fm062$, which places the
star on the main sequence.} for our Hipparcos members. No reddening
correction has been applied; we suspect this contributes significantly
to the horizontal width of the main sequence. The earliest spectral
type on the main sequence is B1, suggesting an age of $\la$10~Myr.

We find Collinder~173 (Collinder 1931; Catalog of Open Clusters
[1997]: $[\ell, b] = [261\fdg32, -8\fdg06]$; diameter
$\sim$6$^\circ\!$) to be identical to Vel~OB2
($[\ell,b]$$\sim$$[263^\circ\!,-7^\circ\!]$; diameter
$\sim$$10^\circ\!$). Collinder estimated the number of Col~173 members
to be 70, and quoted a distance of only 55~pc(!). Eggen (1980, 1983,
1986) found photometric evidence for 29 members of Col~173, including
our secure Vel~OB2 member $\gamma^1$--$\gamma^2$~Vel, and derived a
photometric distance of 350--525~pc (cf.\ $\sim$500~pc [Buscombe
1963]; 380$\pm$144~pc [Lyng\aa\ \& Wramdemark 1984]). He derived a
motion for 8 `cluster' members similar to that of our Vel~OB2
group. Only 7 of the 29 Eggen members are listed in the Hipparcos
Catalogue: we find that 3 of them are secure members of Vel~OB2.

\begin{figure}[t]
\centerline{
\psfig{file=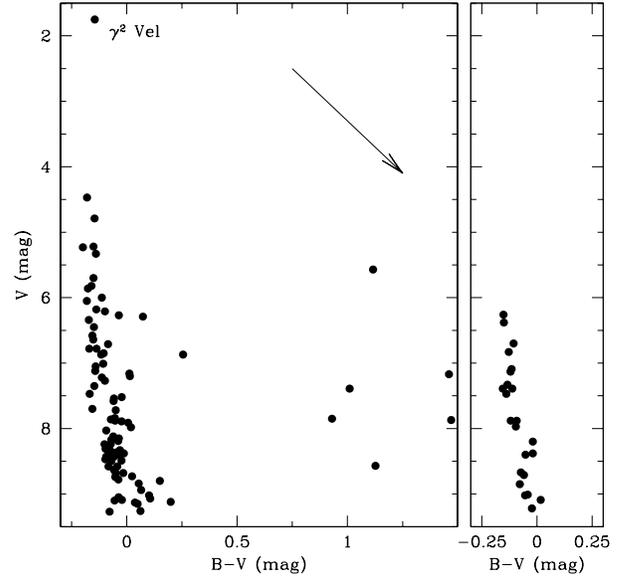,width=8truecm,silent=}}
\caption{\small  Left: color-magnitude diagram, not
corrected for reddening, for the Vela~OB2 members. The stars with
$B\!-\!V > 0\fm5$ are 3 G- and 3 K-giants. Right: color-magnitude
diagram for the Trumpler~10 members.}
\end{figure}

The $\gamma^2$~Vel system contains WR11, the nearest known Wolf--Rayet
star, and its distance is important for the absolute magnitude
calibration of WR stars. Classical distance estimates range from
$\sim$170--550~pc (Brandt et al.\ 1971; Sahu 1992; van der Hucht et
al.\ 1997); the most commonly assumed distance is 450~pc. Even though
its parallax of 3.88$\pm$0.53~{\rm mas} places $\gamma^2$~Vel at the
near edge of Vel~OB2, the direction and magnitude of the proper motion
leave little doubt about its membership. Schaerer, Schmutz \& Grenon
(1997) combined the Hipparcos parallax with single star evolutionary
models, and derived a weak upper limit of $\sim$6--10~Myr for the age
of the O9I companion of WR11.

We obtain a mean distance of 410$\pm$12~pc for Vel~OB2. The new
members are concentrated on the sky around $(\ell, b)$$\sim$$(263^\circ\!,
-7^\circ\!)$ within a radius of $\sim$5$^\circ\!$. Sahu (1992) reported
the detection of the so-called IRAS Vela shell in the IRAS Sky Survey
Atlas maps (cf.\ Sahu \& Blaauw 1994). This is an expanding shell,
centered on Vel~OB2, with a projected radius of $\sim$6$^\circ\!$
(Figure~11). Assuming that (i) the center of the IRAS Vela shell has
a distance of 450~pc, (ii) Vel~OB2 is a `standard association' with a
`normal' initial mass function, and (iii) Vel~OB2 has an age of
20~Myr, Sahu showed that the observed kinetic energy of the IRAS Vela
shell is of the same order of magnitude as the total amount of energy
injected into the interstellar medium through the combined effects of
stellar winds and supernovae. Our astrometric identification of
Vel~OB2 as a rich OB association, and the improved distance
determination, will allow a considerable refinement of Sahu's analysis
once photometry and radial velocities are available for all our
Hipparcos members.

\subsection*{\centerline{\normalsize\sl 5.2. Trumpler~10}}

{\it Pre-Hipparcos}: Based on relative proper motion data for 29
stars, Lyn\-g\aa\ (1959, 1962) identified 19 pro\-ba\-ble members of
the sparse open cluster Trumpler~10 (Tr~10; Catalog of Open Clusters
[1997]: $[\ell, b] = [262\fdg81, 0\fdg64]$; diameter $14^{\prime}$).
Photometry indicated a maximum age of $\sim$30~Myr and a distance of
$\sim$420~pc. Levato \& Malaroda (1975) derived a distance of
440$\pm$50~pc for a subset of these stars. Based on 15 members, Eggen
(1980) found a distance of 468$\pm$65~pc; this value is biased because
all stars out to 350~pc were considered foreground objects. Lyng\aa\
\& Wramdemark (1984) found 363$\pm$68~pc for 11 photometric
members. Stock (1984) carried out an extensive proper motion study of
979 stars near Tr~10 down to $V=12^{\rm m}$, and concluded that the
existence of a single cluster in this field was doubtful.

\smallskip
{\it New results}: During our investigation of the Vela region, our
member selection procedure not only identified Vel~OB2 in the
Hipparcos Catalogue, but it also `rediscovered' Tr~10. This resulted
in a moving group of 23 stars: 22 B-type stars (earliest spectral type
B3V) and 1 A0V star. We confirm 4 of the 7 classical members
(Lyng{\aa}; Levato \& Malaroda; Eggen) contained in the Catalogue.
The mean distance is 366$\pm$23~pc. The members are spread over
$\sim$8$^\circ\!$ in the sky ($\sim$50~pc at this distance;
Figure~11). Figure~12 shows the color-magnitude diagram for our 23
Hipparcos members. Even though the measurements are not corrected for
reddening, the main sequence is remarkably tight. Tr~10 is seen in
projection in front of the Vela molecular ridge (e.g., May, Murphy \&
Thaddeus 1988; Murphy \& May 1991). This group is clearly older than
Vel~OB2, and our provisional age estimate based on the earliest
spectral type is $\sim$15~Myr.

The many selected stars outside the classical diameter of $14^\prime$
cannot all be interlopers: we expect only 3--5 B-type interlopers and
1 A star (Table~A2). We conclude that Tr~10 is in fact not a tight
open cluster, but instead an intermediate age OB association. 

\section*{\centerline{\normalsize 6. CANIS MAJOR, MONOCEROS AND ORION}}

The region $245^\circ\!\! > \! \ell \! > \! 195^\circ\!$ contains two
nearby OB associations (Ruprecht 1966), Mon~OB1 and Ori~OB1, and the
interesting group Col~121, which we discuss first. Ori~OB1 lies in a
direction near the Solar antapex, and its space motion does not stand
out in the proper motion distribution, while its distance of
$\sim$400~pc makes it difficult to determine membership from the
Hipparcos parallaxes alone. We also discuss the well-studied
association CMa~OB1, which lies in the same region but at a distance
of $\sim$1~kpc.

\subsection*{\centerline{\normalsize\sl 6.1. Collinder~121}}

{\it Pre-Hipparcos}: Collinder (1931) studied the structural\break
properties and spatial distribution of Galactic open clusters, and
discovered a cluster of 20 stars at $\sim$590~pc in an area of
$1^\circ\! \times 1^\circ\!$ on the sky: Col~121. Schmidt--Kaler
(1961) noted a large number of evolved early-type stars in a field of
$10^\circ\! \times 10^\circ\!$ centered on the bright supergiant
$o$~CMa (\HIP33152, K3Iab) located in the central part of
Col~121. Using photometry, radial velocities, and proper motions, he
assigned membership of Col~121 to 10 of these stars, and derived a
mean distance of 630~pc. Feinstein (1967) found 12 photometric members
within $30^\prime$ of the cluster center: 11 main-sequence stars and
one evolved star, $o$~CMa. He extended this list with 14 bright,
mostly evolved, and 12 faint stars inside a circle of $10^\circ\!$
radius centered on the cluster. These include some of the
Schmidt--Kaler stars, in particular the supergiants $\delta$~CMa
(\HIP34444, F8Ia) and $\sigma$~CMa (\HIP33856, K4III), and the
Wolf--Rayet star WR6 (EZ~CMa, \HIP33165, WN5). Feinstein also put the
cluster at 630~pc. In another photometric study, Eggen (1981)
suggested a division of the stars in this direction into two groups:
an open cluster-like group of 13 stars at 1.17~kpc, identified as
Col~121, and a group of 18 stars at 740~pc resembling an OB
association.

\begin{figure*}[t]
\centerline{
\psfig{file=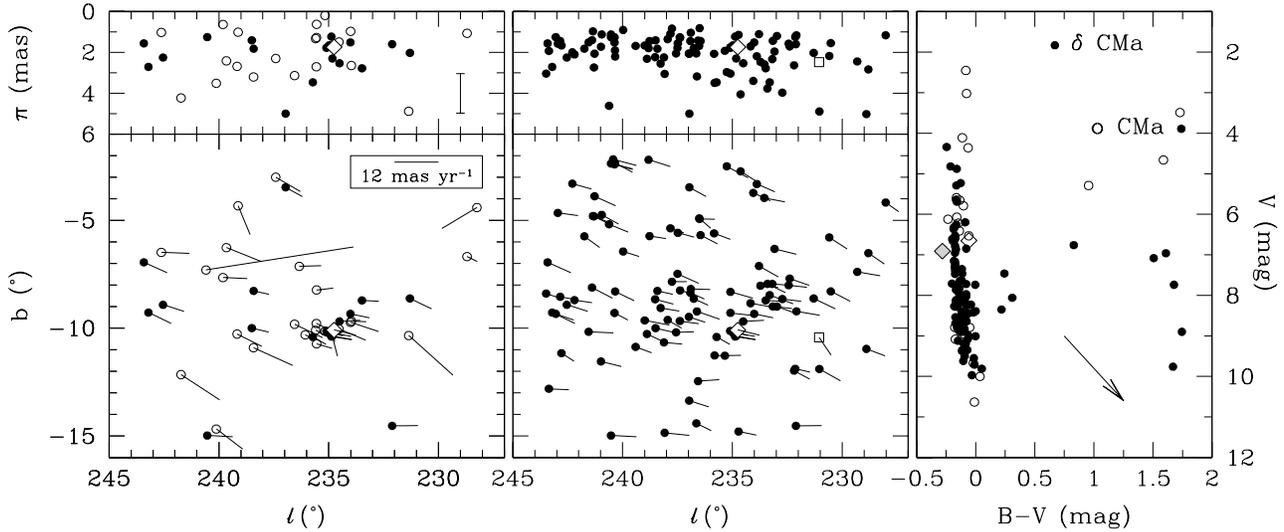,width=17truecm,silent=}}
\caption{\small Left: positions and proper motions (bottom) 
and parallaxes (top) of the pre-Hipparcos members of Collinder~121.
Filled circles are confirmed members. Open circles indicate stars
rejected by our selection procedure. Three of the latter fall outside
the plotted parallax range. Two of these have discrepant proper
motions as well. The confirmed member EZ~CMa (WR6) is indicated by an
open diamond. Middle: same diagram for all stars selected as member of
Col~121, illustrating the dramatic change from pre- to post-Hipparcos.
The open square represents the open cluster \FNGC2287: its position,
proper motion and parallax is an average of the measurements for 9
members observed by Hipparcos in an area of $0\fdg5$ diameter. We
expect $\sim$30 interlopers in the Col~121 membership list (see
Table~A2). These may be mostly the stars above $b$$\sim$$-6^\circ\!$ and
below $b$$\sim$$-12^\circ\!$ which lie outside the main
concentrations. Right: color-magnitude diagram, not corrected for
reddening, for the Col~121 members (filled circles), and rejected
classical members (open circles). The unusual position of EZ~CMa (open
diamond) may be caused by systematic effects in the $Hp$ to $V$ and
$B_T\!-\!V_T$ to $B\!-\!V$ transformation (cf.\ ESA~1997, Vol.\ 1 \S 1.3,
Appendix~4 and \S 2.1, pp~107--108). Feinstein (1967) obtained $V =
6\fm91$, $B\!-\!V = -0\fm28$ (gray diamond).}
\end{figure*}

\smallskip
{\it New results}: Figure~13 shows the proper motions for the 43 stars
of the combined list of members proposed by Collinder, Schmidt--Kaler,
Feinstein and Eggen which are listed in the Hipparcos Catalogue (5, 10,
34 and 13 stars, respectively). Evidence for a moving group can be
seen in the proper motions, but our selection procedure rejects 25 of
the 43 stars.

We select 103 stars in our Col~121 field (Table~A1) with a mean
distance of 592$\pm$28~pc; 1 WR, 1 O, 85 B, 8 A, 1 F, 1 G, 3 K, and 3
M-type. We reject $\sigma$~CMa, but include $\delta$ and $o$~CMa as
member. Only 6 of the 14 bright members and 5 of the 12 faint members
of Feinstein are selected by our procedure. One bright Feinstein
member (\HIP34489) has a negative parallax, while \HIP34045 (B8II) and
\HIP35205 (M2III) have parallaxes larger than 7~mas and discordant
proper motions. The classical member \HIP35904 (B5Ia; $\eta$~CMa) is
not selected because of its small, but significant, parallax: $\pi =
1.02\pm0.44$~mas. \HIP33347 (B3Ib/II), \HIP33856 (K4III), and
\HIP33977 (B3Ia) have incompatible or insignificant proper motions.
We select 4 of the 7 stars in Eggen's distant group and 2 of the 6
stars in his nearby group, suggesting that his proposed division in
two groups is unphysical. We do not confirm the possible connection
between Col~121 and \NGC2287 proposed by Eggen (1981): the directions
of the mean proper motions of these two groups differ by more than 45
degrees (see Figure~13).

Figure~13 (right panel) presents the Hipparcos color-magnitude
diagram, not corrected for reddening, for the secure members. It shows
that Col~121 contains a number of evolved stars. The abrupt cutoff of
the main sequence near $V=10^{\rm m}$ is caused by the completeness
limit of the Hipparcos Catalogue. Some of the late-type stars may
well be interlopers. The presence of an O star and early-type B stars
indicates this is a young group, of age $\sim$5~Myr.

Howarth \& Schmutz (1995) put a lower limit of 1800~pc on the distance
of EZ~CMa. This would imply that it is at least 320~pc away from the
Galactic plane. If it were a runaway star (as proposed by Skinner,
Itoh \& Nagase 1998) born in the Galactic plane, its observed proper
motion in $b$, $\mu_b = -1.46\pm0.6$~mas~yr$^{-1}$, puts a lower limit
on its age of $\sim$30~Myr, which is unlikely for a Wolf--Rayet star
(e.g., Maeder \& Meynet 1994). Although a distance of 1.8~kpc is
marginally consistent with $\pi=1.74$$\pm$0.76~mas, the
proper motion and parallax of EZ~CMa agree perfectly with those of
Col~121, leaving little doubt about its membership, and hence about
its distance. It follows that EZ~CMa is $\sim$10 times less luminous
than estimated by Howarth \& Schmutz.

Col~121 has completely changed its appearance compared to the
classical membership lists. Figure~13 clearly suggests two subgroups,
$(\ell,b)$$\sim$$(233^\circ\!,-9^\circ\!)$ and
$(238^\circ\!,-9^\circ\!)$. A possible third subgroup lies at
$(\ell,b)$$\sim$$(243^\circ\!,-9^\circ\!)$. We expect $\sim$30
interlopers among the B- and A-type stars in this large field
(Table~A2). It is likely that most of these lie outside the main
concentrations, where the membership probabilities are high
(Table~C1). The linear dimensions of this complex are $100~{\rm pc}
\times 30$~pc, similar to that of the Sco~OB2 association. The
color-magnitude diagram also resembles that of Sco~OB2, after taking
into account the four times larger distance, in the sense that the
earliest spectral type and amount of extinction are similar. A full
analysis of the stellar content of this OB association is clearly
warranted.

\subsection*{\centerline{\normalsize\sl 6.2. Canis~Major~OB1}}

{\it Pre-Hipparcos}: The high density of early-type stars in the Canis
Major region ($222\fdg0 \! < \!\ell\! <\! 226\fdg0$, $-3\fdg4 \!<
\!b\! <\! 0\fdg7$) led Ambartsumian (1949) to propose the existence of 
the OB association CMa~OB1. Markarian (1952) identified 11 probable
members, and derived an angular diameter of $\sim$4$^\circ\!$ and a
very uncertain distance of 960~pc\footnote{Schmidt (1958) copied
Markarian's distances but lists 950~pc.}. Ruprecht (1966) assumed that
the young open cluster \NGC2353 (age $\sim$12.6~Myr [Hoag et al.\
1961]; age $\sim$76~Myr [Fitzgerald, Harris \& Reed 1990])
constitutes a stellar concentration in CMa~OB1 (cf.\ Ambartsumian
1949, 1954), and therefore listed its distance as the open cluster
distance of 1315~pc (Becker 1963).

\begin{figure}[t]
\centerline{
\psfig{file=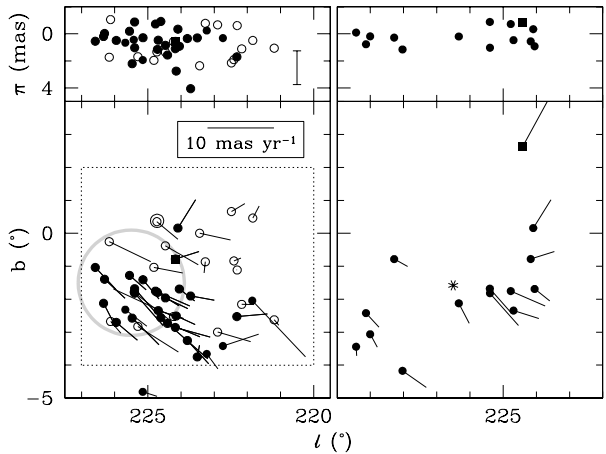,width=8truecm,silent=}}
\caption{\small Left: positions and proper motions (bottom)
and parallaxes (top) for 30 classical members of Canis Major~OB1--R1
(filled symbols). The 16 open symbols show all Hipparcos O--B5 stars
with $\pi \leq 5~{\rm mas}$ in the field $220^\circ\!\! \leq \! \ell \!
\leq \! 227^\circ\!$, $-4^\circ\!\! \leq \! b \! \leq \! 2^\circ\!$
(dotted lines), which are not classical members. The large open circle
at $(\ell, b) = (224\fdg72, 0\fdg38)$ denotes the open cluster
\FNGC2353. \FHIP33735 ($\ell = 221\fdg85$; $\pi = 6.16 \pm 1.29~{\rm
mas}$) falls outside the upper panel. \FHIP34536 (filled square) is
Herbst \& Assousa's (1977) runaway star, and the gray circle denotes
their supernova shell. Right: Comer\'on et al.'s (1998) 14 CMa~OB1--R1
members. The new candidate runaway star \FHIP35707 is the filled
square; its proper motion is 17~mas~yr~$^{-1}$. The asterisk denotes
the projected expansion center. The field has been shifted $2\fdg5$ in
Galactic longitude with respect to the left panel.}
\end{figure}

Based on \UBV\ measurements, Clari\'a (1974a, b) found 44 members of
CMa~OB1, derived a distance of\break 1150$\pm$140~pc and an age
$\sim$3~Myr, and confirmed its physical relation to the association of
reflection nebulae CMa~R1 (Racine 1968; Herbst, Racine
\& Warner 1978) and to the open cluster \NGC2353 (cf.\ Eggen 1978).
The CMa~OB1 members are the main exciting stars of the HII regions
\Sh292, \Sh296, and \Sh297, which are physically
related to the R association. Star formation in this CMa~OB1--R1
complex was initiated recently: very young objects are present, among
which the classical Herbig emission-line stars Z~CMa (\HIP34042) and
\HD53367 (\HIP34116). An upper limit of $3~{\rm
Myr}$ for the age of the complex was derived from the main-sequence
lifetime of the massive O-star progenitor of the carbon star W~CMa
(\HIP34413, CII...; Herbst, Racine \& Richer 1977).

Herbst \& Assousa (1977; cf.\ Reynolds \& Ogden 1978; Machnik et
al.\ 1980) found that CMa~R1 lies on the outer edge of an H$\alpha$
emission ring, centered on $(\ell, b) = (225\fdg5,$ $ -1\fdg5)$ with a
radius of $\sim$$1\fdg5$. The form of the ring, its position
with respect to CMa~R1, the small age of CMa~R1, the detection of the
ring in radio wavelengths, an expanding HI shell at the
same location, and the runaway star \HD54662 (\HIP34536, O6),
possibly related to a supernova explosion (Blaauw 1961), led these
authors to suggest that this ring is a supernova shell with an age of
$\sim$0.5~Myr.

\smallskip
{\it New results}: Based on Hipparcos proper motions (alone) for 14 O
and early B-type stars, corrected for Solar motion and Galactic
rotation, and {\it assuming$\,$} a distance of 1150~pc, Comer\'on,
Torra \& G\'omez (1998) claim the detection of an expanding structure,
with center $(\ell, b) = (226\fdg5, -1\fdg6)$, and an expansion age of
$1.5~{\rm Myr}$. According to these authors, the expansion of
$15~{\rm km}~{\rm s}^{-1}$, and the spatial arrangement of the stars,
confirm that a supernova was responsible for the formation of
CMa~OB1--R1. Furthermore, they propose a new candidate for the runaway
star related to the supernova: \HIP35707 (O9V), which has a residual
tangential velocity of $90~{\rm km}~{\rm s}^{-1}$ (at the assumed
distance of 1150~pc, cf.\ $\pi = -0.81$$\pm$$0.84$~mas) directed away
from the expansion center, and a corresponding runaway age of
$\sim$1~Myr.

Our member selection procedure is unable to detect a moving group in
the CMa~OB1--R1 field because of the small (often barely significant)
proper motions and parallaxes of the candidate members. Figure~14
(left panel) shows the 30 classical members of CMa~OB1--R1
contained in the Hipparcos Catalogue, 24 of which have spectral type
O--B5 (Racine 1968, and Herbst et al.\ 1978 [10]; Clari\'a
1974b [19]; Welin 1979 [1; Ape emission-line shell star \HIP33436]).
The 16 open symbols show all remaining Hipparcos O--B5 stars with
parallaxes $\pi \leq 5~{\rm mas}$ in the field $220^\circ\!\! \leq \! \ell
\! \leq \! 227^\circ\!$, $-4^\circ\!\! \leq \! b \! \leq \! 2^\circ\!$, a 
region somewhat larger than Ruprecht's and Clari\'a's fields. The
parallax distribution shows that the majority of stars lies beyond
$\sim$500~pc. The right panel is slightly shifted in $\ell$ with
respect to the left panel, and shows the 14 CMa~OB1--R1 members
suggested by Comer\'on et al.\ (1998).

\subsection*{\centerline{\normalsize\sl 6.3. Monoceros~OB1}}

{\it Pre-Hipparcos}: The Mon~OB1 association mainly consists of the
open cluster \NGC2264 and the R association Mon~R1, as well as a rich
molecular cloud complex (Herbst 1980) with numerous emission-line
objects (e.g., Herbig \& Rao 1972; Cohen \& Kuhi 1979; Ogura 1984) and
one detected outflow object (Margulis et al.\ 1990). \NGC2264 is an
open cluster at $\sim$950~pc with $\sim$150 members brighter than
$V=15^{\rm m}$, and age $\sim$3 Myr (P\'{e}rez 1991). The up-turn
around spectral type A0 in the color-magnitude diagram suggests that
low-mass stars are still contracting toward the zero-age main sequence
(e.g., Walker 1956; Vasilevskis, Sanders \& Balz 1965). The R
association Mon~R1 is located only $2^\circ\!$ from \NGC2264, and is
part of the same cloud complex (Racine 1968; Herbst et al.\
1982). Mon~R1 shows a similar up-turn of the main sequence for the
A-type stars. Turner (1976) published a new list of 14 photometric
Mon~OB1 members distributed over an area of $4^\circ\! \times 8^\circ\!$,
including the concentrations of \NGC2264 and Mon~R1. The area contains
two large HII regions, one centered on \NGC2264, the other on \NGC2244
(the Rosetta nebula), and a supernova remnant, the Monoceros
Ring. \NGC2244 is part of the Mon~OB2 association at 1.4~kpc (Ruprecht
1966). Several expanding shells have been observed in the Mon~OB1
region. Blitz (1978) reported an HI shell, expansion velocity of
$\sim$$15~{\rm km~s}^{-1}$, adjacent to the molecular clouds. The
shell is similar to the HI shell observed in CMa~OB1 (\S 6.2). Kutner
et al.\ (1979) reported a ring structure in the Mon~R1 molecular
clouds which appears to be kinematically distinct from the Mon~OB1
molecular cloud complex. They also found evidence for an expanding HI
shell associated with the molecular ring, and estimated an expansion
age of 1--3~Myr.

\begin{figure}[t]
\centerline{
\psfig{file=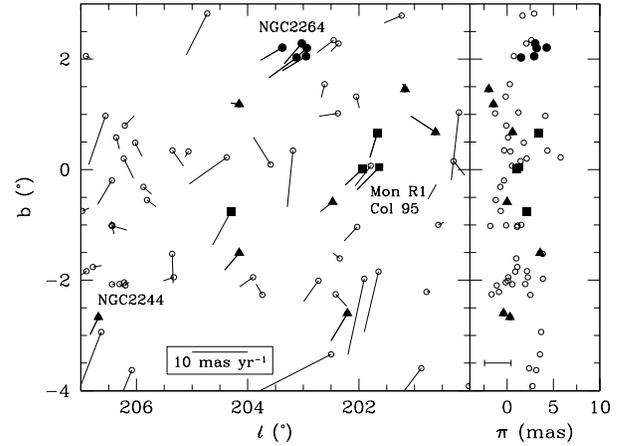,width=8truecm,silent=}}
\caption{\small Positions and proper motions (left) and
parallaxes (right) of the classical Monoceros~OB1 members in the
Hipparcos Catalogue (filled symbols), and all stars earlier than A0
with parallaxes $\pi < 10$~mas (open circles). One star lies outside
the plotted parallax range ($\pi = -5.93\pm6.94$~mas). The classical
\FNGC2264 members (Walker 1956) are denoted by filled circles. Their
parallaxes, $\pi$$\sim$$3$~mas, do not fit the distance estimate of
$\sim$1~kpc (e.g., Walker 1956; P\'erez 1991). The filled squares are
the classical Mon~R1 members (Racine 1968, Herbst et al.\ 1982). The
proper motion and parallax of \FHIP31065 ($[\ell,b]$
$\sim$$[202^\circ\!,0^\circ\!]$, open circle) indicate that it may
belong to Mon~R1 (also known as Col~95). The filled triangles are the
additional Mon~OB1 members proposed by Turner (1976). \FNGC2244, the
open cluster associated with the Rosetta nebula, is also
indicated. }
\end{figure}

\smallskip
{\it New results}: The Hipparcos data do not allow a rigorous
kinematic member selection. The parallel proper motions of the 5
\NGC2264 members contained in the Catalogue confirm \NGC2264 as a
moving group (Figure~15). However, four of these have parallaxes
larger than 2.5~mas, corresponding to a distance $\la$400~pc, which
is much smaller than previous estimates. One of these stars is S~Mon
(\HIP31978, O7). Due to the small angular size of the cluster the
Hipparcos measurements are correlated and must be treated with care.
The 4 Mon~R1 members observed by Hipparcos have parallaxes of
$\sim$$1.3 \pm 1.0$~mas, consistent with the classical distance of
1~kpc. Of the 10 stars in the Catalogue suggested by Turner (1976) one
is a member of Mon~R1 and two belong to \NGC2264. The proper motions
of the remaining 7 stars do not show a clear sign of a moving group,
and their parallaxes vary between $-1.97$ and $3.57$~mas (Figure~15).

\subsection*{\centerline{\normalsize\sl 6.4. Orion~OB1}}

{\it Pre-Hipparcos}: Of all OB associations, Ori~OB1 undoubtedly has
received most attention (e.g., Warren \& Hesser 1977a, b, 1978; Goudis
1982; Brown, Walter \& Blaauw 1998). It is related to the Orion
molecular cloud complex, which is a site of active star formation
(e.g., McCaughrean \& Burkert 1998). Blaauw (1964a) divided Ori~OB1
into four subgroups: 1a, northwest of the Belt stars; 1b, the Belt
region, including the Belt stars; 1c, the Sword region; and 1d, the
Orion Nebula cluster region, including the Trapezium stars.

The most recent photometric census of Ori~OB1 was carried out by Brown
et al.\ (1994), who determined membership, distances (1a:
380$\pm$90~pc; 1b: 360$\pm$70~pc; 1c: 400$\pm$90~pc; 1d: undetermined
due to nebulosity and small number of stars), ages (1a:
11.4$\pm$1.9~Myr; 1b: 1.7$\pm$1.1~Myr; 1c: 4.6$\pm$2~Myr; 1d:
$<$1~Myr), and found the initial mass function for subgroups 1a, 1b,
and 1c to be a single power law: $\xi(\log M) {\rm d}\log M \propto
M^{-1.7\pm0.2} {\rm d}\log M$. These authors showed that the total
energy output of the association over its lifetime can explain the
observed Orion--Eridanus HI shell surrounding Ori~OB1 (cf.\ Burrows et
al.\ 1993; Brown, Hartmann \& Burton 1995). In a proper motion study
of the Orion Nebula cluster, Tian et al.\ (1996) derived an upper
limit on the velocity dispersion of $\sim$$2~{\rm km~s}^{-1}$
(assuming a distance of 470~pc), and concluded that the cluster is
unbound.

Early evidence for pre-main sequence objects in Orion was discussed by
Haro (1953) and Walker (1969). H$\alpha$ and X-ray surveys (e.g.,
Nakano, Wiramihardja \& Kogure 1995; Alcal\'a et al.\ 1996; Walter,
Wolk \& Sherry 1998) have uncovered hundreds of emission-line and
X-ray sources in Ori~OB1, of which many are likely to be pre-main
sequence stars (Alcal\'a, Chavarr\'{\i}a--K.\ \& Terranegra 1998;
Brown et al.\ 1998). The population of low-mass pre-main sequence
stars clustered around $\sigma$~Ori (\HIP26549; O9.5V...) in subgroup
1b has a similar age as the high-mass members of the subgroup (Walter
et al.\ 1998). At $V>12^{\rm m}$, these stars are too faint to be
included in the Hipparcos Catalogue.

\begin{figure}[t]
\centerline{
\psfig{file=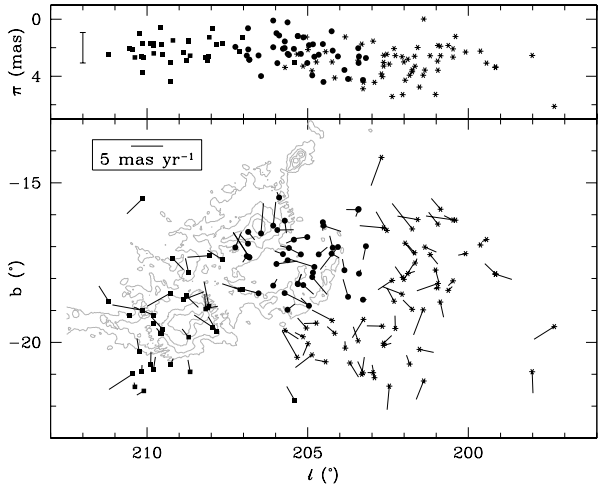,width=8truecm,silent=}}
\caption{\small Positions and proper motions (bottom) and
parallaxes (top) in the subgroups of Orion~OB1, for the stars selected
according to the proper motion criterion given in eq.~(8). The proper
motions are small because Ori~OB1 lies near the direction of the Solar
Antapex. The parallaxes in subgroup 1a (asterisks) are generally
larger than those in 1b (filled circles) and 1c (filled squares). The
contours present the 100$\mu$m IRAS flux map.}
\end{figure}

\smallskip
{\it New results:} Unfortunately, the relative motion of the Ori~OB1
association is mostly directed radially away from the Sun. This makes
it very hard to detect the association with our selection procedure
using the Hipparcos measurements, as is illustrated in
Figure~16. However, Ori~OB1 is such a well-studied association that
Brown et al.\ (1998) decided to analyse the Hipparcos data in a
provisional way in order to constrain its distance. They show that a
rough selection of Ori~OB1 members can be made by requiring:
\begin{equation}
(\mu_\alpha\cos\delta-0.44)^2+(\mu_\delta+0.65)^2\le 25,
\label{orioneq}
\end{equation}
where the units are mas~yr$^{-1}$. The resulting set of stars overlaps
the Brown et al.\ (1994) list of photometric members by 96~per cent.
Figure~16 shows the stars located close to the Orion molecular clouds,
selected according to eq.\ (\ref{orioneq}). The parallaxes of the members
of subgroup 1a are significantly larger than those of 1b and 1c
(Figure~16). Taking the parallax distributions at face value, the mean
distances to the subgroups (cf.\ \S 3.6) are: 336$\pm$16~pc for 1a,
473$\pm$33~pc for 1b and 506$\pm$37~pc for 1c, where the quoted errors
are the formal errors on the mean distances. The actual uncertainty is
larger due to the simplified member selection. Deriving more accurate
distance estimates requires better knowledge of membership. The only
firm conclusion we can draw at this point is that 1a is located
significantly closer to the Sun than 1b and 1c. The implications of
the smaller distance of 1a are discussed by Brown et al.\ (1998).

\section*{\centerline{\normalsize 7. PERSEUS, TAURUS, CASSIOPEIA AND} \break
\vskip -1.4truecm \centerline{\normalsize CAMELOPARDALIS}}

The Perseus region contains the well-studied association Per~OB2, as
well as the $\alpha$~Persei cluster, listed as Per~OB3 by Ruprecht
(1966). It may be related to the old Cas--Tau association, which we
discuss here also. Cam~OB1 lies at a larger distance, but in the same
general direction.

\begin{figure*}[t]
\centerline{
\psfig{file=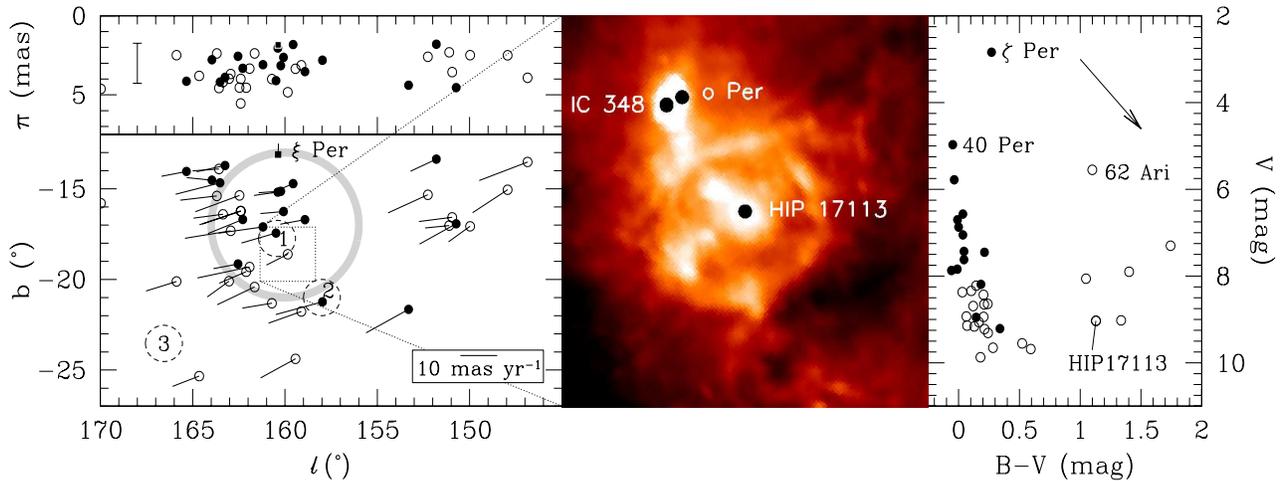,width=17truecm,silent=}}
\caption{\small Left: positions and proper motions
(bottom) and parallaxes (top) for the 41 Perseus~OB2 members. Filled
circles indicate OB-type members (17 stars), while open circles refer
to later-type members (24 stars). The filled square denotes the
runaway $\xi$~Per. The related HII region \FNGC1499 (`California
nebula') lies $\sim$1$^\circ\!$ above $\xi$~Per (e.g., Bohnenstengel
\& Wendker 1976). The gray ring denotes the expanding HI shell
detected by Sancisi et al.\ (1974). The small dashed circles indicate
the positions of (1) \FIC348, (2) \FNGC1333, and (3) the Pleiades. The
$3^\circ\! \times 3^\circ\!$ dotted box is centered on \FHIP17113
($\ell = 159\fdg85, b = -18\fdg61$), and is enlarged in the middle
panel. Middle: IRAS 60$\mu$m map of the cores \FIC348 (with
overlapping dots at the positions of the stars \FHIP17465 and 17468)
and B3/L1468--1470 in the Per~OB2 molecular cloud. The position of the
bright star $o$~Per is shown for reference. \FHIP17113 seems to be
related to a ring in the interstellar medium with a radius of
$\sim$$0\fdg6$. Right: color-magnitude diagram, not corrected for
reddening, for 40 of the 41 Per~OB2 members (without \FHIP17561).
Symbols as in the left panel. \FHIP20324 has $B\!-\!V = 2\fm020$. The
main sequence is broadened by differential reddening, stellar
multiplicity, and a distance spread of $\sim$$0\fm4$.}
\end{figure*}

\subsection*{\centerline{\normalsize\sl 7.1. Perseus~OB2}}

{\it Pre-Hipparcos}: Blaauw (1944; cf.\ Blaauw 1952a) noticed the
presence of 15 O--B3 stars concentrated in a region of $6^\circ\!$
diameter in Perseus: Per~OB2, sometimes called the $\zeta$~Persei
association or moving cluster after its brightest member. Other
special classical members are: $\xi$~Per (doubtful member because of
deviating radial velocity; $v_{\rm rad} = 67.1~{\rm km~s}^{-1}$),
X~Per (high-mass X-ray binary), $o$~Per, 40~Per, \HD23060 (unknown
radial velocity), AG~Per (double-lined eclipsing binary), \HD21483
(star in nebulosity with deviating radial velocity), and \BD+31~643
(heavily reddened B star with $\beta$~Pic-like circumstellar disk
[Kalas \& Jewitt 1997; Lissauer 1997]). Blaauw (1944) derived a
photometric distance of 267~pc, a mean radial velocity of $19.4~{\rm
km}~{\rm s}^{-1}$, and a mean space motion of $21.7~{\rm km}~{\rm
s}^{-1}$.

In a subsequent study, Blaauw (1952a; cf.\ Delhaye \& Blaauw 1953)
deduced a kinematic expansion age of\break 1.3~Myr. The star $o$~Per was
omitted from the analysis as it did not obey the expansion. Delhaye
\& Blaauw (1953) suggested that the proper motions of $o$~Per, as well
as of X~Per and \HD23478, deviate from the cluster mean because these
stars are actually multiple systems in which an invisible, massive
companion is still contracting towards the main sequence. Lesh (1969)
repeated Blaauw's expansion analysis using new proper motions for 9
bright Per~OB2 members, and obtained an expansion age of $1.3 \pm
0.1~{\rm Myr}$. The right ascension proper motions of the stars
$o$~Per and 40~Per deviated strongly from the expansion pattern and
were not taken into account.

Later studies of the stellar content of Per~OB2 were mainly
photometric (e.g., Morgan et al.\ 1953; Harris 1956; Hardie, Seyfert
\& Gren\-chik 1957; Seyfert, Hardie \& Gren\-chik 1960; Borgman \&
Blaauw 1964; Rydgren 1971; Guetter 1977; {\v C}ernis 1993). The
derived ages range from 2 to 15~Myr (e.g., Klochkova \& Kopylov 1985;
de Zeeuw \& Brand 1985; Gim\'enez \& Clausen 1994).

Blaauw (1952a) discussed the embedded, nebulous open cluster \IC348
$\sim$8$^{\prime}$ south of $o$~Per as a possible concentration in
Per~OB2 (Gingrich 1922; Hubble 1922a, b; Greenstein 1948; Harris,
Morgan \& Roman 1954, 1955; Strom, Strom \& Carrasco 1974; {\v C}ernis
1993). Proper motions revealed 8 members, the brightest of which is
\BD+31~643 (B5V). The proper motions used by Fredrick (1954, 1956)
showed that `... $o$~Per is not a member of the nucleus and perhaps is
not even a member of the association'. Later discoveries of H$\alpha$
emission-line T Tauri stars (Herbig 1954), reflection nebulae (Racine
1968), infrared sources (Ladd, Lada \& Myers 1993), X-ray selected T
Tauri stars (Preibisch, Zinnecker \& Herbig 1996), and photometrically
detected pre-main sequence stars have led to the recognition that star
formation in \IC348 has been going on continuously for 3--7~Myr (e.g.,
Strom et al.\ 1974; Lada \& Lada 1995; Trullols \& Jordi 1997).

The Per~OB2 region is rich in patches of bright and dark nebulosities.
The interstellar material in Perseus is distributed in various layers
at various distances, and the obscuring material has a patchy,
inhomogeneous distribution (e.g., Heeschen 1951; Lynds 1969; Rydgren
1971; {\v C}ernis 1990, 1993; Krelowski, Megier \& Strobel 1996).
Some of the material at 140 to 180~pc might simply be an extension of
the Taurus dark clouds to the northwest ({\v C}ernis 1990, 1993;
Cernicharo, Bachiller \& Duvert 1985). The elongated chain of dark,
molecular condensations call\-ed the Per~OB2 molecular cloud (e.g.,
Sargent 1979) hosts several isolated, opaque molecular cores, of which
only \IC348 and \NGC1333 (e.g., Lada et al.\ 1974; Lada, Alves \& Lada
1996), separated by $3\fdg5$, contain young B stars. Several hundreds of
pre-main sequence objects are known in and around \IC348 and \NGC1333
(e.g., Herbig 1954; Herbig \& Rao 1972; Preibisch et al.\ 1996;
Trullols \& Jordi 1997; Preibisch 1997). They generally have
$V$$\sim$$15^{\rm m}$--$20^{\rm m}$, and were hence not observed by
Hipparcos.

\begin{figure*}[t]
\centerline{
\psfig{file=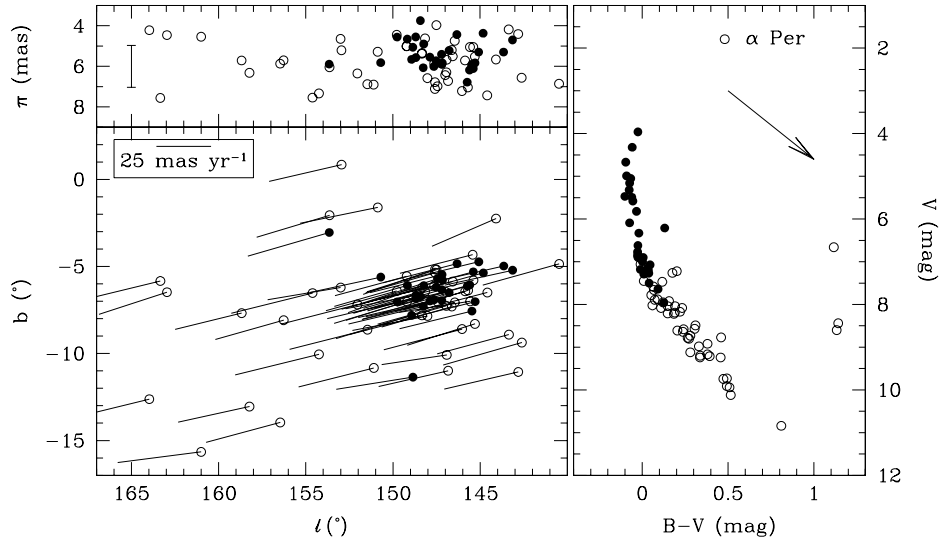,width=12.5truecm,silent=}}
\caption{\small Left: positions and proper motions
(bottom) and parallaxes (top) of the $\alpha$~Persei (Perseus~OB3)
members. Filled circles indicate B-type stars, and open circles the
later spectral types. Right: color-magnitude diagram, not corrected
for reddening, for the $\alpha$~Persei members. Symbols as in the left
panels. The main sequence extends from B3Ve ($V=3\fm96$) to G3V
($V=10\fm84$).}
\end{figure*}

A physical relation between Per~OB2 and the Per~OB2 molecular cloud
has never been proven beyond doubt, due to the uncertain distance
estimates to the cloud, which range from 150 to 500~pc (Rydgren 1971;
Strom et al.\ 1974). However, Sancisi (1970; cf.\ Sancisi
et al.\ 1974) found complex gas and dust clouds, which were
interpreted as an expanding supernova remnant shell of interstellar
matter physically connected with Per~OB2 and Gould's Belt. Sancisi
(1974; cf.\ Loren 1976; Hartquist \& Morfill 1983) suggested that the
stars in Per~OB2 formed 1--4~Myr ago in the densest parts of this
shell (e.g., \"Opik 1953). Indirect `evidence' for this supernova
explosion is provided by the existence of the runaway O star $\xi$~Per
(\HIP18614; Blaauw 1961, 1964a).

\smallskip
{\it New results}: We find 41 Hipparcos members of Per~OB2: 17 B, 16
A, 2 F, 2 G, 3 K, and 1 M-type. All 17 classical members (Blaauw
1944, 1952a) were observed by Hipparcos. We confirm only 8 of them,
among which $\zeta$~Per (\HIP18246, B1Ib), 40~Per (\HIP17313, B0.5V),
and AG~Per (\HIP19201, B5V:p; but see Gim\'enez \& Clausen 1994).
Three of the 9 rejected classical proper motion members have
unreliable or insignificant astrometric parameters: \HIP17631
(B3IV...), X~Per (\HIP18350, O9.5pe; also known as `non-conventional
member' [e.g., Borgman \& Blaauw 1964; Wackerling 1972; Sargent
1979]), and \BD+31~643\footnote{The parallax and proper motion of
\HIP17468 have been fixed to those of \HIP17465 as indicated by H$60$
= F.} (\HIP17465, B5V; brightest member of \IC348, of which
Gingrich~12 [\HIP17468, A2] is another member). Another three were
already suspected to be peculiar or even non members: $o$~Per
(\HIP17448, B1III; see e.g., Delhaye \& Blaauw 1953; Fredrick 1954,
1956; Sargent 1979; Snow et al.\ 1994), \HD21483 (\HIP16203, B3III;
cf.\ Schreur 1970; Sargent 1979), and the runaway $\xi$~Per
(\HIP18614, O7.5Iab:).

The early-type members give a mean distance of\break 318$\pm$27~pc, somewhat
smaller than most previous values. The interloper analysis predicts
one or two spurious B-type members and $\sim$10 A-type
members. Figure~17 (left panel) shows the positions and parallaxes of
the 41 members. The late-type members clumped around $(\ell,b)$$\sim$$
(151^\circ\!,-16^\circ\!)$,\break and separated by $\sim$5$^\circ\!$ from
the main body of the cluster, could be interlopers. The Per~OB2
molecular cloud roughly connects the dashed circles (1; \IC348) and
(2; \NGC1333).

Figure~17 (right panel) shows the color-magnitude diagram, not
corrected for reddening. The brightest member, $\zeta$~Per (B1Ib), is
an evolved star, leaving 40~Per (B0.5V) as the brightest main-sequence
member. The large and variable extinction in the Perseus region (e.g.,
Lynds 1969; {\v C}ernis 1990, 1993), combined with a large fraction of
spectroscopic binaries (e.g., Gim\'enez \& Clausen 1994) is probably
responsible for the broad main sequence. The 7 stars with $B -
V > 1^{\rm m}$ include all members later than spectral type F (the
brightest is 62~Ari [\HIP15696, G5III], most likely an interloper) and
\HIP17113. The Hipparcos Catalogue lists \HIP17113 as an F2 star, with $V
= 9\fm03$ and $B - V = 1\fm13$ based on Tycho photometry. 
The SIMBAD database gives a spectral type B5, with $V=10\fm6$ and
$B\!-\!V=-0\fm1$ based on aperture photometry. The IRAS 60$\mu$m map
(Figure~17) shows that the star appears to be located at the center of
a ring (or shell) of radius $\sim$$0\fdg6$ ($3.3~{\rm pc}$ at an
assumed distance of $318~{\rm pc}$). The star seems to be surrounded
by interstellar material related to the molecular globule
B3/L1468--1470 (Barnard 1927; Lynds 1962). We suspect this 
has influenced the photometric measurements reported in SIMBAD.

\subsection*{\centerline{\normalsize\sl 7.2. Cassiopeia--Taurus and $\alpha$~Persei (Per~OB3)}}

{\it Pre-Hipparcos}: The Cassiopeia--Taurus group was identified and
studied by Blaauw (1956). He proposed 49 members of spectral type B5
and earlier, in an area of $140^\circ\! \times 100^\circ\!$, which
showed remarkably parallel motions. Blaauw concluded that this group
formed an old, nearly dissolved, association at $\sim$140~pc, with an
expansion age of $\sim$50~Myr. Based on the similar velocity and
distance of the Cas--Tau group and the cluster $\alpha$~Persei, Blaauw
suggested that the two belonged to the same physical group. Rasmuson
(1921) already reported an extended stream of stars around
$\alpha$~Persei that shared its motion. Using new radial velocities,
Petrie (1958) claimed that Cas--Tau is not a physical group. Crawford
(1963) suggested that either the group did not exist, or that there
was a large contamination by non members, based on a large scatter in
the intrinsic color versus H$\beta$ relation.

The $\alpha$~Persei moving group was discovered independently by B.\
Boss (1910), Eddington (1910) and Kapteyn (1911) after publication of
the Preliminary General Catalog by L.\ Boss. They noted a group of
$\sim$16 bright early-type stars with large proper motions,
well-separated from the early-type field population. Later studies
extended and refined the list of members to $V \la 12^{\rm m}$ and
spectral types earlier than G5, using photometric, proper motion and
radial velocity measurements (e.g., Roman \& Morgan
1950; Harris 1956; Heckmann, Dieckvoss \& Kox 1956; Heckmann \&
L\"ubeck 1958; Artyukhina 1972; Fresneau 1980). CCD photometry
identified $\sim$300 candidate members to $V$$\sim$$19^{\rm m}$ (e.g.,
Stauffer et al.\ 1985; Stauffer, Hartmann \& Jones 1989;
Prosser 1992). Meynet, Mermilliod \& Maeder (1993) estimated an age of
$\sim$50~Myr for this group.

\begin{figure*}[t]
\centerline{
\psfig{file=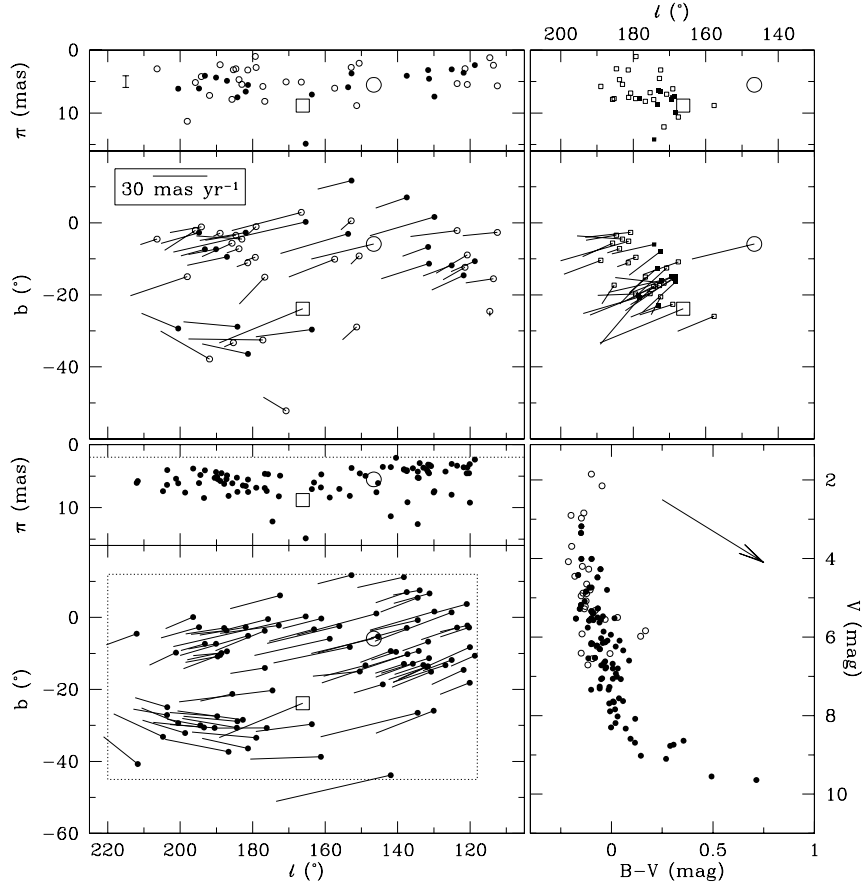,width=11.5truecm,silent=}}
\caption{\small Upper left panels: positions, proper
motions, and parallaxes for 47 classical members of the
Cassiopeia--Taurus association proposed by Blaauw (1956). The two
classical members $\alpha$~Ara (\FHIP85792) and $\eta$~UMa
(\FHIP67301) lie outside these panels (see \S 7.2). Filled circles are
confirmed Cas--Tau members. Open circles are stars rejected by our
selection procedure. The large open circle and open square represent
the $\alpha$~Persei and Pleiades cluster, respectively. Lower left
panels: positions, proper motions, and parallaxes for 83 Cas--Tau
members in the field $118^\circ\!\! \le \! \ell \! \le \! 220^\circ\!$
and $-45^\circ\!\! \le \! b \! \le \! 12^\circ\!$ with $\pi >
2$~mas. Field and parallax limits are indicated as dotted lines.
Lower right panel: color-magnitude diagram, not corrected for
reddening, for the 83 Cas--Tau members (filled circles), and rejected
classical members (open circles). Upper right panels: positions,
proper motions, and parallaxes for 12 T~Tauri stars (filled squares;
Frink et al.\ 1997), and 21 B-type members (open squares; Walter \&
Boyd 1991), associated with the Tau--Aur star-forming
region.}
\end{figure*}

\smallskip
{\it New results}: Only 52 of the 170 stars in the classical list of
bright $\alpha$~Persei members (Heckmann \& L\"ubeck 1958) are
contained in the Hipparcos Catalogue. Our selection method rejects 13 of
these, but adds another 40, resulting in a total of 79 members: 30 B,
33 A, 12 F, 2 G, and 2 K-type, nearly all of them with large
membership probability (Table~C1). We confirm $\alpha$~Per (\HIP15863,
F5Ib) itself as member. Two of the 13 rejected stars fall outside the
proper motion window for the secure members as defined in Table~A1,
and 7 have parallaxes smaller than 4~mas, and discordant proper
motions. Three of the remaining stars are labeled as binary in the
Hipparcos Catalogue (two `G'-type and one component `C' solution). Most
of the members are located in a region of $3^\circ\! \times 3^\circ\!$,
consistent with the $1\fdg25$ and $2\fdg5$ radii for the `nucleus' and
`corona', respectively, found by Artyukhina (1972). However, we find a
`halo' of $10^\circ\!$, consisting mostly of A-type stars
(Figure~18). We expect $\sim$5 interloper A stars (Table~A2) which may
partly explain this extended component of the moving group (but see
below). We select 10 of the 16 stars proposed by Eddington
(1910). This high success rate is mainly due to the small distance
(177$\pm$4~pc), and large tangential streaming velocity
($\sim$$30~{\rm km~s}^{-1}$) with respect to the Sun.

The $\alpha$~Persei main sequence in the color-magnitude diagram
shown in the right-hand panel of Figure~18 is remarkably narrow. It is
the result of nearly negligible differential reddening, and a small
distance spread. The turnoff at the bright end is consistent with an
age of $\sim$50~Myr. The absence of late-type members is only
apparent, and is due to the completeness limits of the Hipparcos
Catalogue.

Unlike previous fundamental catalogs, the Hipparcos Catalogue is free of
regional systematic errors, which makes it possible to establish
reliable kinematic membership of groups covering a large area on the
sky, such as Cas--Tau. We first applied our selection procedure to all
OB stars in the field $118^\circ\!\! \le \! \ell \! \le \! 220^\circ\!$ and
$-45^\circ\!\! \le \! b \! \le \! 12^\circ\!$. We excluded the secure
$\alpha$~Persei members in order to prevent confusion between the
Cas--Tau group and $\alpha$~Persei. To avoid a large contamination by
distant objects in this very large field we confined the search to all
stars with parallaxes $\pi > 2$~mas. We did not consider later-type
stars for the same reason. The refurbished convergent point method (\S
3.1) recognizes Cas--Tau as a moving group and finds a convergent
point $(\ell_{\rm cp},b_{\rm cp}) = (243\fdg6,-13\fdg1)$.
The Spaghetti method (\S 3.2) finds multiple peaks in velocity
space. The highest peak corresponds to the field B stars at a velocity
of $(U, V, W) = (-10.19,-6.36,-7.08)~{\rm km~s}^{-1}$. However, there is
a significant secondary peak at a velocity $(U, V, W) = (-13.24, -19.69,
-6.38)~{\rm km}~{\rm s}^{-1}$, which corresponds to Cas--Tau. The
resulting convergent point and streaming velocity with respect to the
Sun are $(\ell_{\rm cp}, b_{\rm cp}) = (236\fdg1, -15\fdg1)$
and $S = 24.6~{\rm km}~{\rm s}^{-1}$, respectively. The
convergent points found by both methods are remarkably similar to each
other and to the results of Blaauw (1956), $(\ell_{\rm cp}, b_{\rm
cp}) = (233\fdg9, -12\fdg0)$ and $S = 23.9~{\rm km}~{\rm s}^{-1}$,
Petrie (1958), $(\ell_{\rm cp}, b_{\rm cp}) = (225\fdg1, -10\fdg2)$
and $S = 20.1~{\rm km}~{\rm s}^{-1}$, and Eggen (1961), $(U, V, W) =
(-15.1, -19.6, -5.6)~{\rm km}~{\rm s}^{-1}$.

We select 83 stars which are consistent with the derived convergent
point and space velocity (Figure~19). The expected number of
interlopers in the large field is $\sim$50 (Table~A2). While this
number is substantial, it is significantly smaller than 83, which
strengthens the conclusion that Cas--Tau is a physical group. Our
procedure selects 16 of the 49 classical Cas--Tau members proposed by
Blaauw. One other classical member (\HIP19343) is a secure member of
$\alpha$~Persei. Six classical members brighter than $V = 4^{\rm m}$
are rejected by our selection scheme. Two of these are foreground
objects: $\eta$~UMa (\HIP67301; B3V; $[\ell,b] =
[100\fdg69,65\fdg32]$; $\pi = 32.39$~mas; $V = 1\fm85$) and
$\alpha$~Ara (\HIP85792; B2Vne; $[\ell,b] = [340\fdg75,-8\fdg82]$;
$\pi = 13.46$~mas; $V = 2\fm84$). Another three have indications of
binarity (\HIP4427, \HIP18532, \HIP26451). The distances of the
Cas--Tau stars vary systematically with $\ell$, ranging between 125
and 300~pc. This is not unexpected, given the large physical size and
unknown orientation of this group.

The color-magnitude diagram of the 83 B-type Cas--Tau members is
narrower than the similar diagram for the classical members.
The expected distance spread is $\sim$$1\fm9$. Three of the six
highly reddened stars ($B\!-\!V > 0\fm25$) are located within
5$^\circ\!$ of each other, and are probably obscured by the same
`cloud complex' ($[\ell,b]$$\sim$$[133^\circ\!,6^\circ\!]$).

The motion of the $\alpha$~Persei cluster is consistent with that of
Cas--Tau, which thus confirms the physical relation between the two
groups. It also suggests that the halo of $\alpha$~Persei members
mentioned above is the inner region of the Cas--Tau group. These
results provide kinematic support for the hypothesis that the
$\alpha$~Persei cluster and the Cas--Tau association have a common
origin, with $\alpha$~Persei surviving as a bound structure. This is
in harmony with the age estimates of $\sim$50~Myr for both the cluster
and the association.

The Taurus--Auriga clouds lie in this same area, near $(\ell, b)$$\sim$$
(170^\circ\!, -20^\circ\!)$, at a distance of $\sim$140~pc (e.g., Kenyon,
Dobrzycka \& Hartmann 1994). It is natural to ask whether this
well-known site of low-mass star formation is also kinematically
related to the Cas--Tau association. Walter \& Boyd (1991) searched
for B stars with similar kinematics as the low-mass stars in the
Tau--Aur clouds. They identified 29 such stars; 21 are listed in the
Hipparcos Catalogue, and 7 of these are in the original Cas--Tau list of
Blaauw (1956). We reject all of the latter, and select only 2 of the
remaining 14 as Cas--Tau members. Figure~19 shows that the proper
motions of the Walter \& Boyd stars have a large dispersion in
magnitude and direction. Frink et al.\ (1997) used proper motions
with accuracies of $\sim$5~mas~yr$^{-1}$ to identify a moving group of
pre-main sequence objects in Tau--Aur. Twelve of their objects are
listed in the Hipparcos Catalogue. The directions of the Hipparcos
proper motions of these stars are reasonably similar, but they differ
by $\sim$40$^\circ\!$ from that of the Cas--Tau members. We conclude
that the star-forming region in Tau--Aur and the old Cas--Tau
association are distinct stellar groups.

\subsection*{\centerline{\normalsize\sl 7.3. Camelopardalis~OB1}}

Morgan et al.\ (1953) investigated the luminous star content
in Camelopardalis, and found evidence for an `aggregate', Cam~OB1, of
8 early-type stars at a distance of $\sim$900~pc. Ruprecht (1966)
extended this list with 2 stars. The 10 members are spread over an
area of $15^\circ\! \times 3^\circ\!$ on the sky. The corresponding
physical dimensions, $240~{\rm pc} \times 50~{\rm pc}$, make a common
origin for these stars implausible (cf.\ Garmany \& Stencel 1992).
Humphreys (1978) compiled a list of 19 luminous stars for Cam~OB1
based on positions, photometric distances and, when available, radial
velocities. She derived a distance of $\sim$1~kpc. However, these
stars are distributed over an even larger area on the sky than the 10
stars of Ruprecht. Haug (1970) found 39 luminous Cam~OB1 members in an
area of $4^\circ\! \times 2^\circ\!$ at $\sim$1~kpc, associated with the
HII region \Sh202 located at $(\ell,b)$$\sim$$(140^\circ\!,2^\circ\!)$
at $\sim$800~pc (Brand \& Blitz~1993). Digel et al.\ (1996) also
associated Cam~OB1 with \Sh202 based on the velocity of the stellar
association and the CO emission from the cloud complex containing
\Sh202. The region around \Sh202 is an active star-forming complex
containing outflow objects and HII regions (Blitz, Fich \& Stark 1982;
Snell et al.\ 1984; Campbell, Persson \& McGregor 1986), as well as
the R association Cam~R1 ($\sim$870~pc; Racine 1968). Some of the
Cam~OB1 members according to Morgan et al., Ruprecht, and Humphreys
are identified as Cam~R1 members by Racine.

The Hipparcos parallaxes of the classical members are consistent with
a distance of $\sim$1~kpc (Figure~20). The proper motions of the
classical members are small, and do not allow identification of a
moving group. The field also contains three open clusters: \NGC1502
(at $\sim$900~pc), \IC1848 (at $\sim$2.5~kpc), and \IC 1805 (at
$\sim$2.2~kpc) (Mermilliod 1998).

Camelopardalis embraces a star-forming region of considerable
dimensions, which clearly requires a detailed\break study of its internal
structure and kinematics, comparable to what Hipparcos allows us to do
for the nearest associations. This will be complicated by the fact
that Camelopar\-dalis lies in the direction of the Perseus spiral arm,
so that the large number of early-type stars and giants might be
caused by accidental alignments along the line of sight.
 
\begin{figure}[t]
\centerline{
\psfig{file=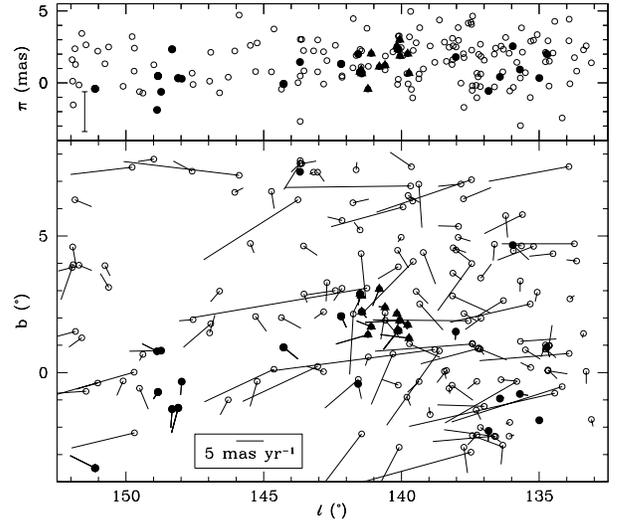,width=8truecm,silent=}}
\caption{\small Positions and proper motions (bottom) and
parallaxes (top) of the classical Camelopardalis~OB1 members in the
Hipparcos Catalogue, filled symbols, and all OB-type stars with
parallaxes $\pi < 5$~mas, open circles. Filled triangles denote the
Cam~OB1 members from Haug (1970). The concentrations at $(\ell, b)$$
\sim$$(143\fdg5, 7\fdg5)$, $(137\fdg5, 1\fdg0)$, and $(135\fdg0,
1\fdg0)$ are three distant open clusters: \FNGC1502, \FIC1848, and
\FIC1805, respectively.}
\end{figure}

\section*{\centerline{\normalsize 8. LACERTA, CEPHEUS, CYGNUS AND SCUTUM}} 

The Cepheus and Cygnus region is rich in luminous young stars,
molecular cloud complexes and active star-forming regions. The most
striking example is the Cygnus~X complex, which is $\sim$$10^\circ\!$
in diameter, at a distance of $\sim$1~kpc (e.g., Wendker, Higgs \&
Landecker 1991). Ruprecht (1966) lists 5 OB associations in Cepheus
and 9 in Cygnus. All associations in Cygnus, except Cyg~OB4 and 7, are
most likely physically connected to the Cygnus Superbubble and are
located beyond 1~kpc. We start by studying Lac~OB1, which has
remarkably little associated interstellar medium. Then we discuss
Cep~OB2, 3, and 4, a new group in Cepheus, which we refer to as
Cep~OB6, as well as Cyg~OB4 and Cyg~OB7. We conclude this section with
the one remaining nearby association of Table~1, Sct~OB2.

\subsection*{\centerline{\normalsize\sl 8.1. Lacerta~OB1}}

\begin{figure*}[t]
\centerline{
\psfig{file=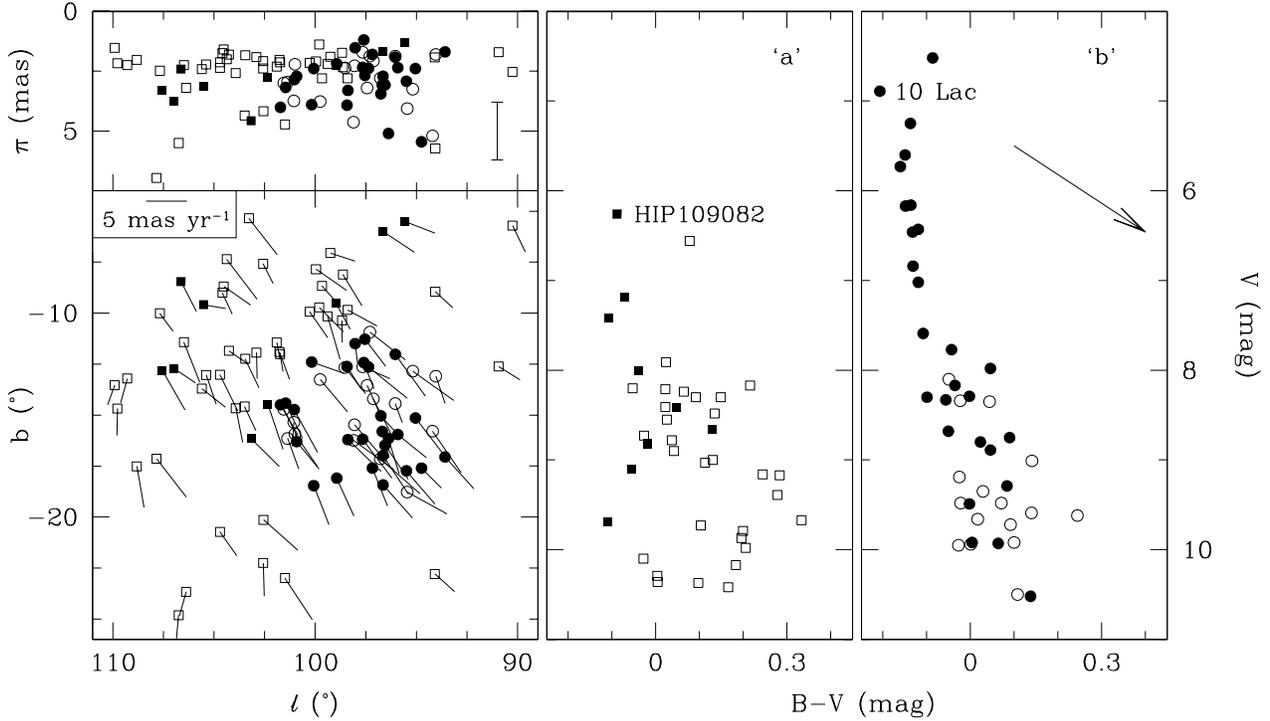,width=17truecm,silent=}}
\caption{\small Left: positions and proper motions (bottom)
and parallaxes (top) for the 96 Lacerta~OB1 members. Circles (27
filled: OB stars; 18 open: later-type) indicate stars inside the
circle with radius $5^\circ\!$ centered on $(\ell, b) = (97\fdg0,
-15\fdg5)$, subgroup `b'. Squares (9 filled: OB stars; 42 open:
later-type) denote stars outside the circle, subgroup `a'. Middle:
color-magnitude diagram, not corrected for reddening, for subgroup
`a'. Twelve stars (the star without a spectral type, the carbon star,
all 3 M-, and 7 of the 8 K stars) have $B - V > 0\fm4$. Right:
similar diagram for subgroup `b'. \FHIP113145 (A2) and
\FHIP113237 (K2) have $B - V > 0\fm4$, and are not shown.}
\end{figure*}

{\it Pre-Hipparcos}: Blaauw's (1952a) detection of expansion of Per~OB2
confirmed Ambartsumian's hypothesis that OB associations are young,
unbound groups. Stimulated by these results, Blaauw \& Morgan (1953;
cf.\ Blaauw 1952c) found similar evidence for a young aggregate
containing 29 candidate O--B5 members in the field $60^\circ\!\! < \!
\ell^I \! < \! 75^\circ\!$, $-20^\circ\!\! < \! b^I \! < \! -5^\circ\!$:
Lac~OB1. Among the members were the O9V star 10~Lac (\HIP111841), and
the $\beta$ CMa-type variables 12~Lac (\HIP112031) and 16~Lac
(\HIP113281). The B2V and B3V stars gave a distance of 460~pc. Based
on 25 members, a disputed expansion age of 4.2~Myr was derived:
because of systematic errors in the right ascension proper motion
components, the analysis used declination values alone, which possibly
contained systematic errors as well (van Herk 1959). A re-analysis
with new data by Woolley \& Eggen (1958) could not confirm the
expansion of Lac~OB1 (cf.\ Steffey 1973). Lesh (1969) used new proper
motions for 10 stars (Delhaye, unpublished) to obtain an expansion age
of $2.5 \pm 0.5~{\rm Myr}$.

Based on proper motions and radial velocities (Delhaye \& Blaauw,
unpublished), Blaauw (1958) divided Lac~OB1 in an old dispersed
subgroup `a' (15 stars; 16--25~Myr), and a younger concentrated
subgroup `b' (11 stars; 12--16~Myr, see Blaauw 1964a, 1991). The
Hertzsprung--Russell diagram was consistent with this division: the
members of subgroup `a' were somewhat brighter on average, and
consequently older, than those of `b', while also lacking the earliest
spectral types. These results were confirmed by Crawford (1961) and
Lesh (1969). Lesh also derived different distances for the subgroups:
368~pc for `a' and 603~pc for `b' (cf.\ Crawford \& Warren 1976: 417
and 479~pc, respectively). Nonetheless, the reality of `a' has always
been controversial (e.g., Lesh 1969). Consequently, most studies
focused on `b', often referred to as Lac~OB1. Ruprecht's (1966) field,
$96\fdg0 < \ell < 98\fdg0$, $-18\fdg7 < b < -15\fdg6$, indeed refers
to subgroup `b' exclusively. Besides polarization studies (e.g.,
Krzem\'{\i}nski \& Oskanjan 1961), radial velocity (e.g., Blaauw \&
van Albada 1963; Bijaoui, Lacoarret \& Granes 1981) and rotational
velocity (e.g., Abt \& Hunter 1962) studies, several photometric and
spectroscopic investigations were conducted (e.g., Harris 1955;
Seyfert \& Hardie 1957; Blaauw 1958; Hardie \& Seyfert 1959; Crawford
1960, 1961; Adelman 1968, 1973; Lesh 1968; Crawford \& Warren 1976;
Levato \& Abt 1976; Guetter 1976).

\smallskip
{\it New results}: We find 96 Hipparcos members: 1 O, 35 B, 46~A, 1 F,
8 K, 3 M-type, 1 star without spectral classification (\HIP111762),
and the carbon star \HIP116681 at the edge of our field. We confirm
16 of the 29 Blaauw \& Morgan (1953) members (among which are 10, 12,
and 16~Lac). The two B2V stars \HIP110790 and \HIP110849 labeled as
`Field star?' in Blaauw \& Morgan's table~1 are both selected as
members. We also confirm \HD215227 (\HIP112148, B5:ne; Harris 1955),
\HD213976 (\HIP111429; Odenwald 1988; Olano, Walmsley \& Wilson 1994),
but reject \HD211835 (\HIP110177, B3:Ve; e.g., Crampton 1968),
\HD215304 and \HD213421 (\HIP112215, A1V, and\break \HIP111106, Ap,
respectively; Crawford \& Warren 1976). Only 1 of the 4 (\HD209961,
\HIP109082, B2V) spectroscopic binaries discussed by van Albada \&
Klomp (1969) is selected. Several other studies have put forward
`special' candidate members (e.g., Crawford 1961; Slettebak, Bahner \&
Stock 1961; Adelman 1968; Kodaira, Greenstein \& Oke 1970; Levato \&
Abt 1976; Guetter 1976; Malaroda 1981; Garmany \& Stencel 1992): we
confirm membership of the emission-line star \HD216851 (\HIP113226,
B3V:n), the shell star \HD213801 (\HIP111337, B9V), and the red 
supergiants \HD213310 and \HD216946 (\HIP111022, M0II, and \HIP113288,
K5Ibvar, respectively). We do not confirm most candidate members with
peculiar and/or magnetic spectra. The very metal-poor horizontal
branch star candidate member \BD+39~4926 (\HIP112418, B8) has a
negative parallax.

Our field (Table~A1) comprises the classical subgroups `a' and
`b'. Fourteen of the 17 O--B5 members lie in subgroup `b', which for
convenience we represent as a circle centered on $(\ell, b) = (97\fdg0,
-15\fdg5)$ with radius $5^\circ\!$ (Figure~21). The corresponding
color-magnitude diagram also confirms the existence of `b'. The main
sequence is narrow, and indicates that the effects of differential
reddening are small. The deviating star at the bright end of the main
sequence is 10~Lac. Although it is a spectral-type standard, it is
well-known for its peculiar properties (e.g., Lamers et al.\ 1997).
The brightest member is an evolved star (\HIP111104, B2IV).

Based on Hertzsprung--Russell diagrams and photometric distances,
several authors have concluded that most A-type stars observed in the
direction of Lacerta are not main-sequence association members (e.g.,
Seyfert \& Hardie 1957; Howard 1958; but see Coyne, Burley--Mead
\& Kaufman 1969 and Weaver 1970), although most studies were
incomplete at the visual magnitude of $\sim$A0V stars (e.g., Guetter
1976). The Hipparcos A-type members show a nearly uniform distribution
on the sky. The $\sim$25 expected A-type interlopers, the lack of
structure in the color-mag\-nitude diagram, and the presence of only a
few B stars in subgroup `a', make its existence unlikely.

We find a mean distance of 368$\pm$17~pc for all 96 members, and
358$\pm$22~pc for the 45 stars in subgroup `b'. These values are
significantly smaller than most previous distance estimates, including
the classical value given by Ruprecht (1966; cf.\ Table~1). We
suspect that this is caused by the significant modification of the
list of members.

\subsection*{\centerline{\normalsize\sl 8.2. Cepheus~OB2}}

{\it Pre-Hipparcos}: The association Cep~OB2 was discovered by
Ambartsumian (1949). Later investigations (Morgan et al.\ 1953;
Schmidt 1958; Ruprecht 1966; Humphreys\footnote{\HD207306 in
Humphreys' list has wrong coordinates; we suspect the correct entry
should be \HD207308. \HD207538 is listed twice.} 1978; Keller 1970)
confirmed its existence. The most detailed study to date is by
Simonson (1968), who identified 75 bright members, including the
runaway O6Iab star $\lambda$~Cep (\HIP109556), which we adopt as the
classical members. The clusters \NGC7160 and Trumpler~37 (Tr~37),
with its associated HII region \IC1396 (Patel et al.\ 1995), have
similar distances as Cep~OB2, $\sim$800~pc. Simonson \& van Someren
Greve (1976) suggested the division of Cep~OB2 into two subgroups. One
of these is Tr~37, with O6 as the earliest main-sequence member, one
of the youngest known open clusters, with an age of 3--7~Myr (e.g.,
Marschall, Comins \& Karshner 1990). However, its diameter of
$\sim$40~pc suggests that Tr~37 is gravitationally unbound (e.g., Kun
1986). Garrison \& Kormendy (1976) suggested that the bright star
$\mu$~Cep (\HIP107259; M2Ia) is a member of Tr~37. The main source of
excitation of \IC1396 is the O6 star \HIP106886, which is a
Trapezium-like system (e.g., Stickland 1995; Schulz, Bergh\"ofer \&
Zinnecker 1997). \IC1396 and neighbouring areas contain, besides a
large number of H$\alpha$ emission objects, most likely T Tauri stars
(e.g., Kun 1986; Kun \& P\'asztor 1990; Bal\'azs et al.\ 1996),
several globules, possibly containing embedded young stellar objects
(e.g., Duvert et al.\ 1990; Schwartz, Gyulbudaghian \& Wilking 1991).
The other subgroup is older, $\sim$10~Myr, consistent with the large
number of evolved massive stars which are spread over a large area,
and contains \NGC7160. This subgroup is surrounded by a
$9^\circ\!$-diameter infrared emission ring (Figure~22), which
possibly resulted from a supernova explosion (Kun, Bal\'azs \& T\'oth
1987). This supernova might have triggered star formation in this
ring, as suggested by the presence of several HII regions, and a
number of infrared sources which have the characteristics of embedded
young stellar objects (Bal\'azs \& Kun 1989).

\begin{figure*}[t]
\centerline{
\psfig{file=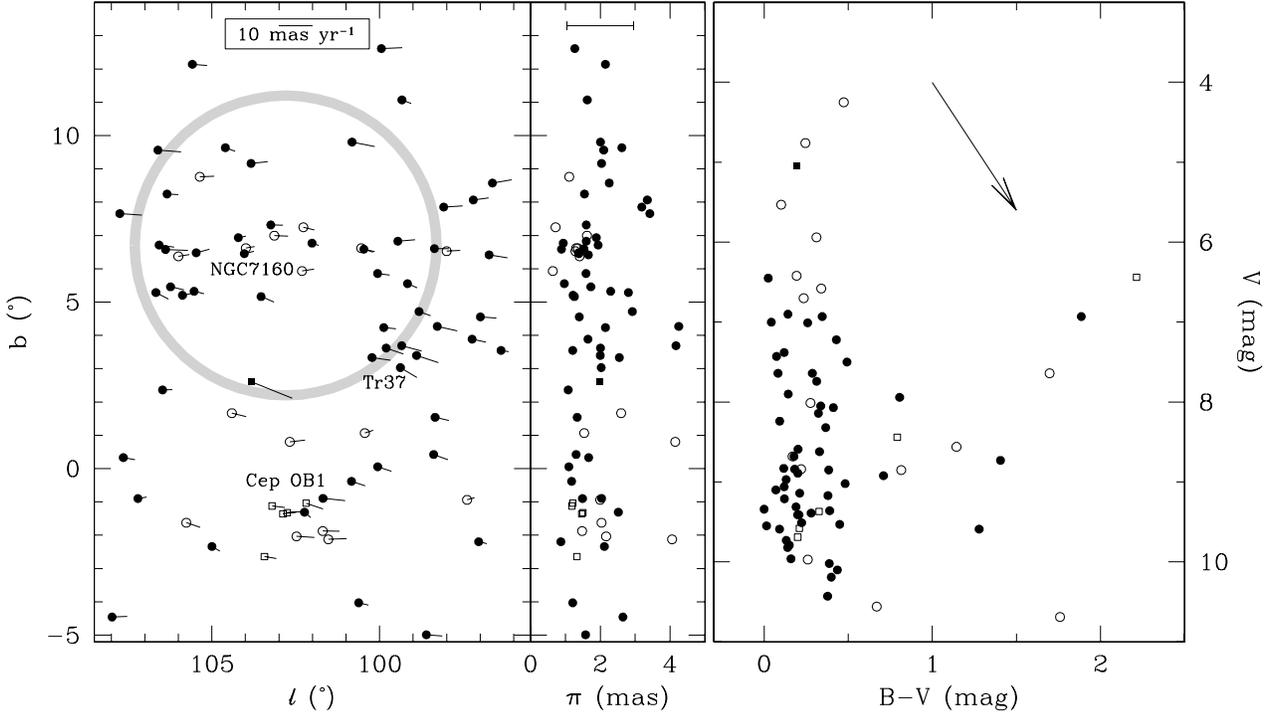,width=17truecm,silent=}}
\caption{\small Positions and proper motions (left),
parallaxes (middle), and color-magnitude diagram, not corrected for
reddening (right), of the Cepheus~OB2 members. Filled circles denote
members with luminosity classes IV, V and undefined. Open circles
denote stars with luminosity classes I, II and III. The filled square
indicates the runaway O star $\lambda$~Cep. The gray circle represents
the Cepheus ring. The approximate position of Cep~OB1 is indicated.
The 5 open squares denote Cep~OB1 members (see \S 8.2).}
\end{figure*}

\smallskip
{\it New results}: We select 76 members, 1 O, 56 B, 10 A, 5 F, 1 G, 2
K, and 1 M-type, in the Cep~OB2 field (Table~A1 and Figure~22). The
mean distance of Cep~OB2 is 615$\pm$35~pc, making it the most distant
association for which the Hipparcos measurements allow kinematic
member selection (Table~2). All members with spectral type F or later
have no luminosity class information or are (super)giants. Most of
these may well be interlopers (Table~A2). Of the 75 stars selected as
Cep~OB2 members by Simonson (1968), 61 are contained in the Hipparcos
Catalogue. We select 20 of these as secure member, and also confirm the
physical relation of the clusters Tr~37 and \NGC7160 with Cep~OB2.
Neither $\mu$~Cep nor \HIP106886 are selected as members. However,
\HIP106886 is physically connected to \IC1396 and definitely belongs
to Tr~37 and thus Cep~OB2; our result may be affected by the multiple
nature of this object. Most of the members around \NGC7160 are
evolved massive stars (Figure~22) supporting the claim that this part
of the association is older than Tr~37 where no evolved stars are
selected.

Besides the two classical subgroups we find one other clump at $(\ell,
b)$$ \sim$$(102\fdg0, -1\fdg5)$. The mean parallax for this clump,
$\langle \pi \rangle = 1.6 \pm 0.3$~mas, is similar to that of all
members, $\langle \pi \rangle = 1.9 \pm 0.1$~mas. It lies outside the
Cepheus ring as well as outside the classical Cep~OB2 field (Ruprecht
1966). Six of the 11 stars in the clump have radial velocities, taken
from the Hipparcos Input Catalogue, of $\sim$$-20~{\rm km~s}^{-1}$,
similar to the other Cep~OB2 members, whereas 5 stars,
\HIP109996 (B1II), 110362 (B0.5IV:n), 110431 (F2Ib), 110504 (G8Ia var),
and 111071 (B0IVn), have large radial velocities, $v_{\rm rad} <
-67~{\rm km~s}^{-1}$. These are, in fact, classical members of Cep~OB1
at $\sim$3.6~kpc (Ruprecht 1966; Humphreys 1978). We have omitted
them from the list in Table~C1. Fifteen other stars in the field
$96^\circ\!\! < \! \ell \! < \! 108^\circ\!$ and $-5^\circ\! < \! b
\! < \! 0^\circ\!$ have radial velocities $v_{\rm rad} < -50~{\rm
km~s}^{-1}$; 9 of these are also classical Cep~OB1 members. Selection
of some Cep~OB1 members as Cep~OB2 members can partly be explained
because the associations have similar projected kinematics: for
distant objects, the observed proper motion is determined by
differential Ga\-lactic rotation only and is, to first order,
independent of distance. However, the radial velocities vary linearly
with distance. Thus, Cep~OB1 and Cep~OB2 members have similar proper
motions, but different radial velocities. Furthermore, some of the
Cep~OB1 members have observed parallaxes which are both consistent
with the distance of Cep~OB1 ($\sim$0.3~mas) as well as Cep~OB2
($\sim$1.8~mas).

The color-magnitude diagram of Cep~OB2 (Figure~22) shows a broad
main-sequence, with a number of evolved stars located at its tip, as
expected. The width of the main sequence, as well as its overall shift
towards positive $B\!-\!V$, indicates that (foreground) reddening is
significant. The earliest spectral type suggests an age of
$\sim$5~Myr. A more accurate age determination will be possible once
intermediate band photometry is available for all members.

\subsection*{\centerline{\normalsize\sl 8.3. Cepheus~OB3}}

Blaauw, Hiltner \& Johnson (1959) made the first detailed photometric
investigation of the association Cep~OB3.\break They found 40 early-type
members at $\sim$725~pc. Blaauw (1964a) found evidence for two
subgroups, with ages of 4 and 8~Myr. Several photometric studies
(Crawford \& Barnes 1970; Garrison 1970; Jordi, Trullols \&
Galad\'{\i}--Enr\'{\i}quez 1996) refined the Blaauw et al.\ 
membership list and extended it to fainter stars. Garmany (1973)
suggested an expansion age of 0.72~Myr, based on the relative motion
of the two subgroups. A detailed summary of all previous membership
studies was given by Jordi et al.\ (1996), who obtained ages of
5.5 and 7.5~Myr for the two subgroups.
Simonson \& van Someren Greve (1976) found an expanding HI shell
centered on the young subgroup in Cep~OB3, but did not detect
significant HI associated with the older subgroup. Sargent (1977,
1979) found several clumps in the Cep~OB3 molecular cloud. Some of
these, Cep~A and Cep~F, show signs of recent star formation (Hughes
1988). Cep~B, the hottest CO component, is located closest to
Cep~OB3. The interaction of the early-type stars and the molecular
cloud is clearly visible as the HII region \Sh155. Elmegreen \& Lada
(1977) considered Cep~OB3 as one of the examples of sequential star
formation. Panagia \& Thum (1981) suggested that the younger subgroup
of Cep~OB3 originated from the Cep B/\Sh155 complex. Testi et al.\
(1995) presented evidence for young stars embedded in the Cep~B cloud.

\begin{figure}[t]
\centerline{
\psfig{file=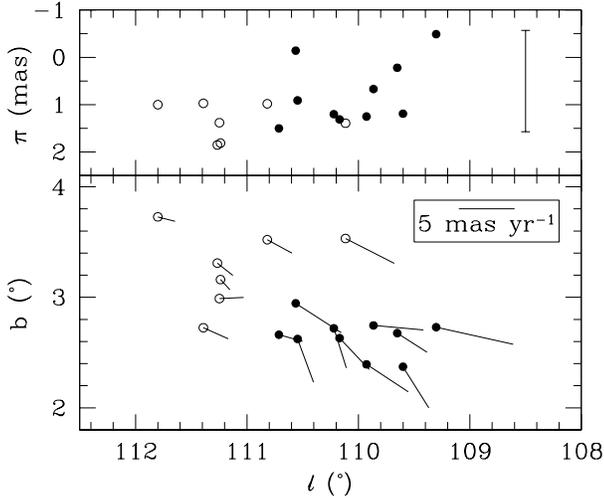,width=8truecm,silent=}}
\caption{\small Positions and proper motions (bottom) and
parallaxes (top) of the classical Cepheus~OB3 members contained in the
Hipparcos Catalogue (Blaauw et al.\ 1959). The open circles show the
classical members of the older subgroup of Cep~OB3, and the filled
circles indicate the younger subgroup.}
\end{figure}

Seventeen of the 40 Cep~OB3 members compiled by Blaauw et al.\ are
contained in the Hipparcos Catalogue. Figure~23 shows that there is
some evidence for two subgroups with a small difference in the
magnitude of their mean proper motions. However, our selection
procedure is unable to identify a moving group in this field.

\subsection*{\centerline{\normalsize\sl 8.4. Cepheus~OB4}}

Cep~OB4 was discovered by Blanco \& Williams (1959), who noticed the
presence of 16 early-type stars in a small region, including the
cluster Berkeley~59. Cep~OB4 is related to a dense, irregular
absorption cloud containing several emission regions, including the
dense HII region \Sh171 (e.g., Herschel 1833; Lozinskaya, Sitnik \&
Toropova 1987). A detailed description of the association and related
objects was given by MacConnell (1968). He identified 42 members
earlier than B8 at $\sim$845~pc. Furthermore, he found 11 H$\alpha$
emission-line objects within the dark cloud, some of which may be T
Tauri stars (e.g., Cohen \& Kuhi 1976). Based on the absence of
supergiants, an earliest spectral type of O7V, and the gravitational
contraction time of a B8 star, he estimated an age between 0.6 and
6~Myr. 

\begin{figure}[t]
\centerline{
\psfig{file=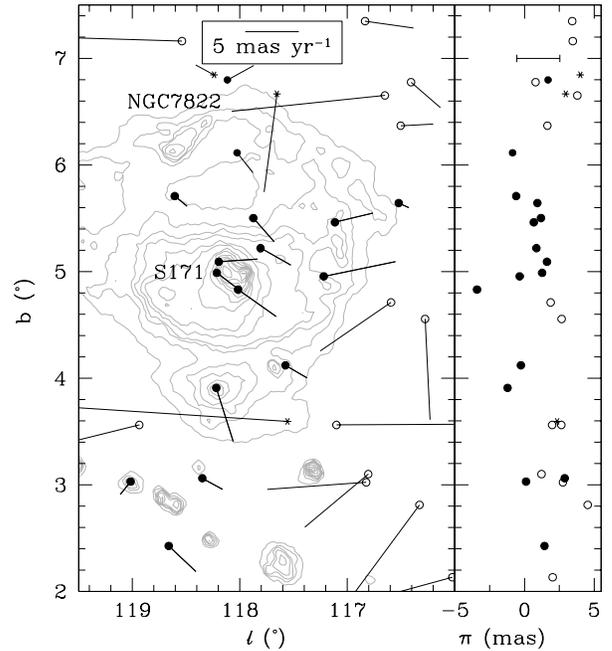,width=8truecm,silent=}}
\caption{\small Positions and proper motions (left) and
parallaxes (right) of the early-type stars in the Cepheus~OB4
region. The Cep~OB4 members as found by MacConnell (1968) are
indicated by filled circles. The three asterisks are MacConnell
members which, based on their proper motion or parallax, do not belong
to Cep~OB4. Open circles are the remaining stars earlier than A0, with
$\pi < 5$~mas. The gray contours show the IRAS 60$\mu$m flux.}
\end{figure}

\begin{figure*}[t]
\centerline{
\psfig{file=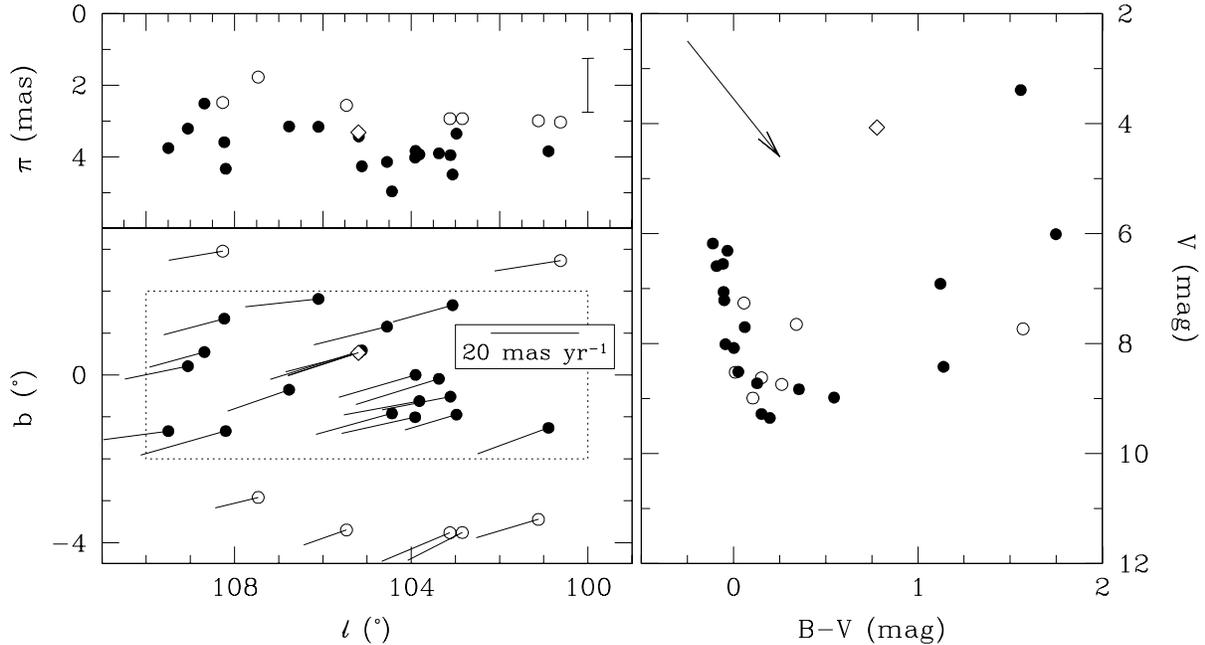,width=16truecm,silent=}}
\caption{\small Left: positions and proper motions
(bottom) and parallaxes (top) for the new moving group in Cepheus,
which we call Cepheus~OB6. Right: color-magnitude diagram, not
corrected for reddening. Open circles denote stars which are
kinematically selected to belong to Cep~OB6 but fall outside the area
$100^\circ\!\! < \! \ell \! < \! 110^\circ\!$ and $-2^\circ\!\! < \! b
\! < \! 2^\circ\!$. The filled circles represent 20 members of
Cep~OB6. The open diamond denotes $\delta$~Cep.}
\end{figure*}

Lozinskaya et al.\ (1987) found two expanding shells in Cep~OB4: one
shell, of radius $\sim$$0\fdg7$, connects \NGC7822 and
\Sh171. Most Cep~OB4 members are located inside this shell; their
energy input into the interstellar medium can account for its observed
size and expansion velocity of $\sim$$10~{\rm km~s}^{-1}$. The other
shell, of radius $\sim$1$\fdg5$, is centered on \Sh171 and has an
expansion velocity of $\sim$30--40~km~s$^{-1}$; it may be the result
of a supernova explosion or of the stellar wind of a massive star
which so far has escaped detection.

Only 19 of the 42 classical members of Cep~OB4 are listed in the
Hipparcos Catalogue (Figure~24). We suspect this is caused by a
combination of crowding effects and the large extinction towards
Cep~OB4, $A_V \ga 3^{\rm m}$ (MacConnell 1968). Based on their proper
motions and parallaxes, three MacConnell stars (\HIP117724, 118192,
118194) are not associated with Cep~OB4. The parallaxes of the other
classical members are consistent with a distance of 800--1000~pc. Our
member selection procedure fails to detect the association.

\subsection*{\centerline{\normalsize\sl 8.5. Cepheus~OB6}}

{\it New results:} Hoogerwerf et al.\ (1997) reported the discovery of
a new moving group in the Cepheus region, based on a subset of the
Hipparcos Catalogue. Here we use all stars in the Catalogue in the
field $100^\circ\!\! < \! \ell \! < \! 110^\circ\!$ and $-4^\circ\!\!
< \! b \! < \! 3^\circ\!$ to examine this moving group in more
detail. A poor sampling of this group can be expected as it was not
part of our (or any other) 1982 proposal to the Hipparcos Input
Catalogue (\S 2.2). We started by applying the Spaghetti method, and
found a significant peak in velocity space at $(U, V, W) =
(-14.00,-28.58,-5.67)~{\rm km~s}^{-1}$. We then searched for stars
consistent with this velocity and corresponding convergent point, and
found 27 co-movers. These showed a modest central concentration at
$(\ell,b)$$\sim$$ (104\fdg0,$ $-0\fdg5)$. We therefore changed the field
to $100^\circ\!\! < \! \ell \! < \! 110^\circ\!$ and $-2^\circ\!\! <
\! b \! < \! 2^\circ\!$ (Figure~25). This resulted in 20 members: 6 B,
7 A, 1 F, 2 G, 3 K-type, and 1 star without spectral classification.
Eleven of these have no luminosity class information, one is a
main-sequence star, while the remaining 8 are classified as
giants. The brightest member is the K1Ibv supergiant $\zeta$~Cep
(\HIP109492). We expect only a few interlopers (Table~A2). The
color-magnitude diagram is very narrow and strengthens the evidence
that these stars form a moving group. We suspect it is an old OB
association: the earliest spectral type is B5III, suggesting an age of
$\sim$50~Myr. We provisionally refer to this group as Cep~OB6.

One of the red supergiants in Cep~OB6 is \HIP110991, the prototype
classical Cepheid $\delta$~Cep. Its parallax, $\pi = 3.32\pm0.58$~mas
($301^{+64}_{-45}$~pc), is consistent with the mean distance of
Cep~OB6 of 270$\pm$12~pc. Previous distance estimates of $\delta$~Cep
agree well with the mean distance of Cep~OB6, e.g., 240$\pm$24~pc
(Mourard et al.\ 1997), $2.80\pm0.96$~mas (Gatewood, de Jonge \&
Stephenson 1993),\break 246$\pm$14~pc (Fernley, Skillen \& Jameson 1989),
and\break 290$\pm$17~pc (Gieren, Barnes \& Moffett 1993).

\subsection*{\centerline{\normalsize\sl 8.6. Cygnus~OB4}}

Membership of the association Cyg~OB4 was studied by Morgan et al.\
(1953), Ruprecht{\footnote{Ruprecht's list contains a typographical
error: \HD201349 should be \HD202349.}} (1966) and Humphreys
(1978). All three quote the same distance of 1~kpc, originally derived
by Morgan et al., based on photometric distances of 4 luminous
members: 3 main sequence stars earlier than B1, and the B9Iab
supergiant $\sigma$~Cyg. All 4 stars are included in the Hipparcos
Catalogue and have parallaxes of $\sim$1~mas, consistent with the
distance estimate of 1~kpc. However, Figure~26 shows no evidence of
common motion: the proper motions differ considerably in magnitude. It
also shows that it is impossible to make any kinematic member
selection based on the Hipparcos measurements.

\begin{figure}[t]
\centerline{
\psfig{file=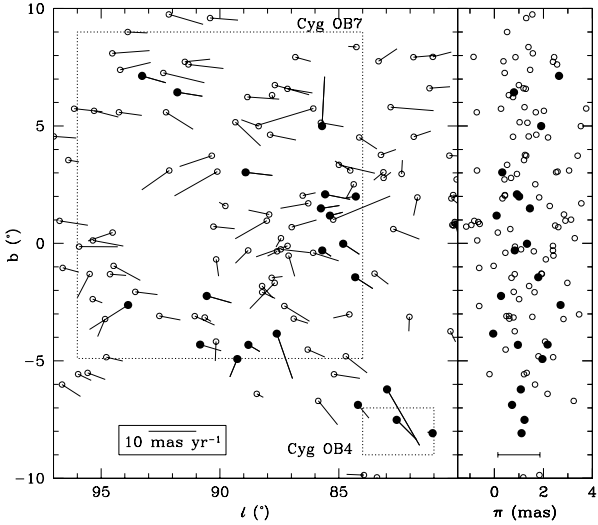,width=8truecm,silent=}}
\caption{\small Positions and proper motions (left) and
parallaxes (right) of the classical Cygnus~OB4 (Ruprecht 1966) and
Cygnus~OB7 (Schmidt--Kaler 1961; Ruprecht 1966; Humphreys 1978)
members (filled circles). The open circles represent all stars
earlier than B5, with parallaxes $\pi < 4$~mas. The dotted lines
indicate the association boundaries given by Ruprecht. Two of the
classical Cyg~OB4 members fall outside the field. The lack of stars
around $(\ell,b)$$\sim$$(92\fdg5,2\fdg0)$ is due to the foreground dark
cloud \FKh141 (see \S 8.7).}
\end{figure}

\begin{figure}[t]
\centerline{
\psfig{file=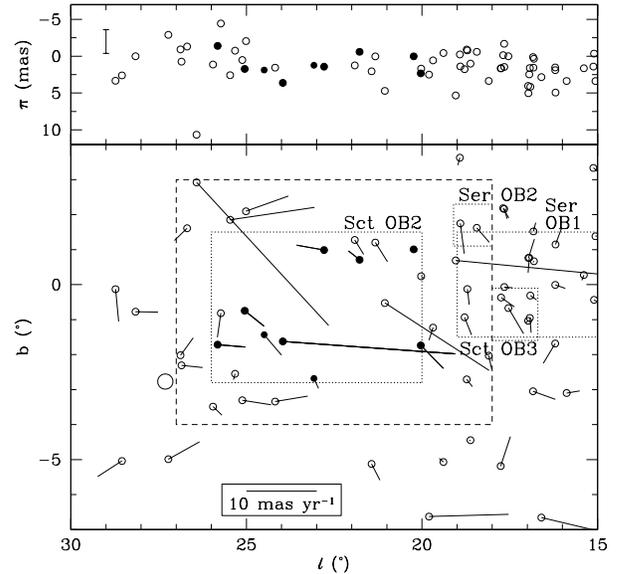,width=8truecm,silent=}}
\caption{\small Positions and proper motions (bottom) and
parallaxes (top) of all Hipparcos O--B5 stars (17 O; 45 B0--B5) in and
around the Scutum~OB2 field. Filled circles denote 9 classical
photometric Sct~OB2 members (see \S 8.8). The large circle denotes
the open cluster \FNGC6705. The dashed rectangle denotes our Sct~OB2
field. The dotted rectangles refer to the IAU boundaries of all OB
associations in the field (Ruprecht 1966): Ser~OB1: $D = 1.7~{\rm
kpc}$, Ser~OB2: $D = 2.0~{\rm kpc}$, Sct~OB3: $D = 1.6~{\rm
kpc}$.}
\end{figure}

\subsection*{\centerline{\normalsize\sl 8.7. Cygnus~OB7}}

Based on a survey of bright stars by Hiltner (1956), Schmidt (1958)
put Cyg~OB7 on the list of associations, and derived a distance of
740~pc. Subsequent studies by Schmidt--Kaler (1961), Ruprecht (1966)
and Humphreys (1978) resulted in a list of 18 stars associated with
Cyg~OB7, located in a field of $12^\circ\! \times 14^\circ\!$. The
dark cloud \Kh141 (Khavtassi 1960), `the Northern Coalsack', lies in
the middle of the Cyg~OB7 field, at a distance of $\sim$400~pc. It
contains several T Tauri stars. Continuum sources seen projected
onto the cloud are most likely HII regions behind the cloud complex
(Simonson \& van Someren Greve 1976).

The Hipparcos parallaxes and proper motions for 17 of the 18 classical
Cyg~OB7 members do not show evidence of common motion (Figure~26). As
for Cyg~OB4, we do not detect a moving group. However, the line of
sight towards Cygnus runs parallel to the local spiral arm. Some of
the previous claims for OB associations based on large numbers of
supergiants and/or early-type stars in this direction, e.g., Cyg~OB4
and Cyg~OB7, may therefore be the result of chance projections.

\subsection*{\centerline{\normalsize\sl 8.8. Scutum~OB2}}

The first reference to the association Sct~OB2 is by Ambartsumian
(1949), who estimated a distance of 1250~pc, and suggested a possible
physical connection to the open cluster \NGC6705 (Meurers \&
Mikulitsch 1968). In his catalog of OB associations, Schmidt (1958)
gave a distance of 730~pc. Ruprecht (1966) listed 6 photometric
members (cf.\ Keller 1970). Humphreys (1978) found another 10
photometric members, and derived a distance of $\sim$1000~pc based on
4 stars. Rei\-chen et al.\ (1990) found 17 photometric members, and
suggested the presence of two groups: one at 510~pc (3 stars), the
other at 1170~pc (14 stars). Of the combined list of 29 classical
members, 9 are included in the Hipparcos Catalogue.

Figure~27 shows all O--B5 stars observed by Hipparcos in the field
$15^\circ\!\! \leq \! \ell \! \leq \! 30^\circ\!$, $-7^\circ\!\! \leq
\! b \! \leq \! 4^\circ\!$. Most stars have parallaxes $\pi \la
2~{\rm mas}$; only 3 of the 62 stars have $\pi \geq 5~{\rm mas}$. No
kinematic signature of Sct~OB2 is evident.

\section*{\centerline{\normalsize 9. MEAN DISTANCES AND MOTIONS}}

We postpone a full investigation of the physical properties of the
nearby associations and their members, and the implications for
scenarios for the formation of the Solar neighbourhood, to a future
investigation. Here we limit ourselves to a derivation of the mean
distances and mean space motions, present an improved map of the
nearby associations (Figure~29), and discuss the kinematics of the
Gould Belt.

\subsection*{\centerline{\normalsize\sl 9.1. Mean distances}}

We have derived the mean distances to all associations and subgroups
for which our kinematic selection method has identified members in the
Hipparcos Catalogue. The mean distances are derived from the mean
parallax of the secure members, and are corrected for the bias
introduced by discarding stars with negative parallaxes in the member
selection (\S 3.6 and Appendix B). The results are summarized in
Table~2, where we give the mean distance determined for the early- and
late-type members separately (specified in Table~A1), and for the
entire set of members. The errors are derived from the formal errors
in the mean parallaxes. We adopt as the best estimate of the mean
distance, underlined in Table~2, the value derived for all members,
except for Vel~OB2, Tr~10, Col~121 and Per~OB2 for which we only
consider the early-type members. In these groups the number of
selected late-type stars is similar to the expected number of
interlopers.

\begin{figure}[h]
\centerline{
\psfig{file=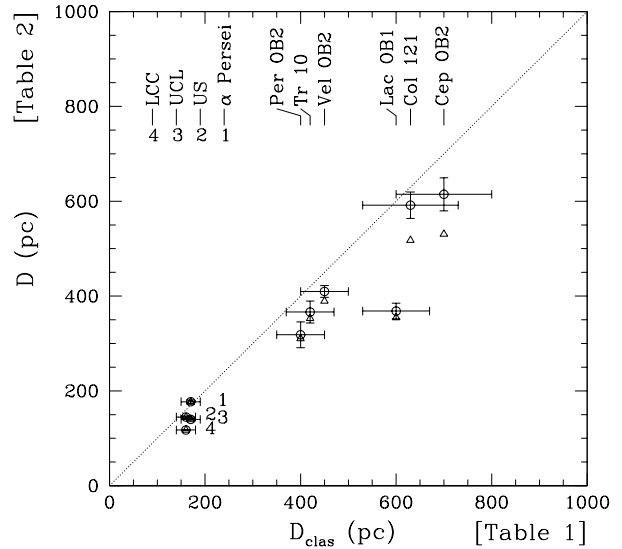,width=8truecm,silent=}}
\caption{\small Distances $D$ derived from the mean parallax 
of the Hipparcos members of the OB associations (Table~2) versus the
classical distances $D_{\rm clas}$ (Ruprecht 1966; Table~1), open
circles. For three associations $D_{\rm clas}$ was taken from another
source: Col~121 (Feinstein 1967), Vel~OB2 (Brandt et al.\ 1971) and
Tr~10 (Lyng\aa\ 1959, 1962). The distances $D$ have been corrected for
systematic effects due to our selection procedure and the Hipparcos
completeness limit (\S 3.6 and Appendix~B). The open triangles
indicate the distances {\it uncorrected} for these systematic
effects.}
\end{figure}

Figure~28 compares the resulting distances to the distances in
Table~1, which are taken from a variety of sources. These generally
are not the most recent (or best) values, but are the most widely
used, and are mostly based on early calibrations of $M_V$ versus
spectral type for upper main-sequence stars. The open triangles are
the Hipparcos mean distances, uncorrected for the bias described
above. The open circles include this correction, $D_{\rm sys}$, which
is based on our Monte Carlo simulations (\S 3.6). $D_{\rm sys}$ is
negligible for the nearest associations, but changes the distance
estimate by $\sim$15~per cent for Col~121 and Cep~OB2. With the
exception of Lac~OB1 (cf.\ \S 8.1), the pre- and corrected
post-Hipparcos distances are correlated remarkably tightly, with the
latter showing a modest systematic offset towards smaller
distances. This tight relation is perhaps surprising at first, as in
many cases our kinematic member selection method has halved the number
of classical members, while at the same time at least tripling the
total number of members. However, the early distance estimates were
often based on a few of the brightest stars, and these have in most
cases been confirmed as members.

\begin{figure*}[t]
\centerline{
\psfig{file=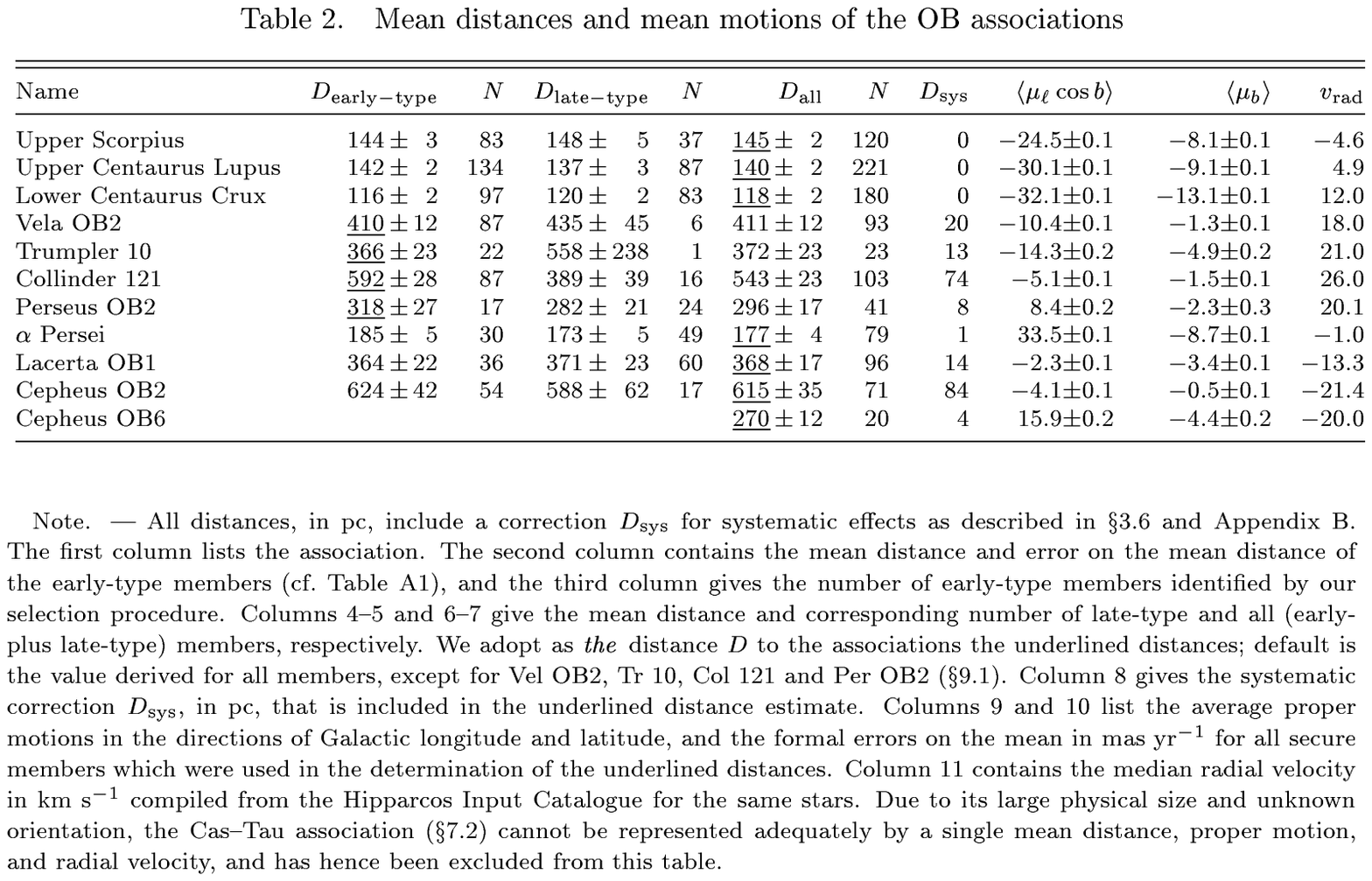,width=15truecm,silent=}}
\end{figure*}

A small systematic offset in distance modulus may have a number of
reasons, the improved member lists being one (especially for the more
distant associations). Another possibility is that the correction for
interstellar extinction is inaccurate, perhaps caused by local
variations in the value of the ratio $R$ of total to selective
extinction. We suspect however that the principle reason for the
discrepancies is likely to be found in the calibration of the upper
main sequence, which is based to a large extent on the bright members
of the nearest associations (e.g., Blaauw 1956). Hipparcos parallaxes
for individual stars, and for open clusters, have given similar
indications that stars on the upper main sequence may be a few tenths
of a magnitude fainter than assumed previously (cf.\ Lamers et al.\
1997; Mermilliod et al.\ 1997). This is not surprising, as the
classical calibration is based on a patching together of the upper
main sequences of nearby young groups --- including many of those
studied here. The short lifetimes of the most massive stars make it
likely that the calibration stars have evolved away from the zero-age
main sequence, so that they are a little brighter than a main-sequence
star of similar spectral type. A more quantitative investigation of
the calibration of the upper main sequence requires homogeneous
spectral types and intermediate band photometry, and is beyond the
scope of this paper.

\subsection*{\centerline{\normalsize\sl 9.2. Mean motions}}

Table~2 also lists the mean proper motions of the kinematically
selected nearby OB associations. The means are based on the individual
measurements for all secure members, except for Vel~OB2, Tr~10,
Col~121 and Per~OB2, for which they are based on the early-type
members only (as in \S 9.1). The errors are the formal errors on the
mean. The Hipparcos Input Catalogue lists radial velocities from a
variety of sources. They are not available for all secure association
members, and we therefore list the median radial velocity in Table~2,
and do not attach an error estimate to it. The Spaghetti method
predicts a mean radial velocity for each moving group based on the
positions, proper motions and parallaxes of the members (\S 3.2). The
uncertainties on these predicted values are considerable, and all we
can conclude is that they are consistent with the measured median
values.

\begin{figure}[h]
\centerline{
\psfig{file=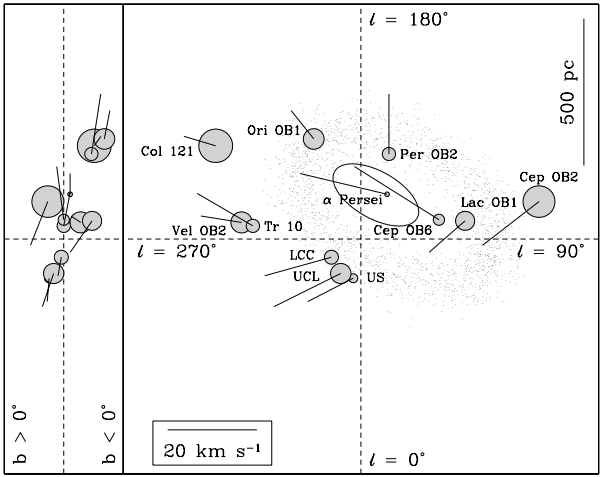,width=8truecm,silent=}}
\caption{\small Locations of the kinematically detected OB
associations projected onto the Galactic plane (right) and a
corresponding cross section (left) (cf.\ Figure~1). The gray circles
indicate the physical dimensions as obtained from the angular
dimensions and mean distances, on the same scale. The lines represent
the streaming motions, derived from the average proper motions, mean
distances and median radial velocities of the secure members,
corrected for `standard' Solar motion and Galactic rotation (see
\S 9.2). The ellipse around the $\alpha$~Persei cluster indicates the
Cas--Tau association. The small dots schematically represent the Olano
(1982) model of the Gould Belt.}
\end{figure}

We have derived the mean space motions of the nearby OB associations
in km~s$^{-1}$ from the mean proper motions, mean distances, and the
median radial velocities. We also determined values for the OB stars
in Ori~OB1 although we used an {\it ad hoc$\,$} selection procedure
(\S 6.4). The Ori~OB1 radial velocity of $23.0$~km~s$^{-1}$ is taken
from Morrell \& Levato (1991), while the mean proper motion $\langle
\mu_\ell \cos b \rangle = 0.8$~mas~yr$^{-1}$, $\langle \mu_b \rangle =
0.1$~mas~yr$^{-1}$ corresponds to the mean proper motion
in equatorial coordinates as given in eq.~(\ref{orioneq}).
Figure~29 shows the result, after subtraction of Solar motion (Dehnen
\& Binney 1998) and differential Galactic rotation (Feast \& Whitelock
1997). Some of the associations in Figure~29 seem to fit a coherent
pattern of expansion and rotation, which is very similar to that
derived by Lindblad et al.\ (1997; their figure~4) and Torra et al.\
(1997) from Hipparcos measurements of OB stars with ages less than
$\sim$30~Myr. This large-scale kinematic feature is known as Gould's
Belt, which is the flat system of early-type stars within $\sim$500~pc
(Gould 1874), associated with a large structure of interstellar
matter, including reflection nebulae, dark clouds, and HI. Its most
striking feature is a tilt of $\sim$18$^\circ\!$ with respect to the
Galactic plane. We refer the reader to P\"oppel (1997) for a
comprehensive review of this structure, with emphasis on the role and
characteristics of the interstellar medium.

Following a suggestion by Blaauw (1965), who studied the mean space
motions of the nearby OB associations, Lindblad (1967) interpreted the
observations of the local HI gas (`feature A') associated with Gould's
Belt in terms of a ballistically expanding ring. Subsequent kinematic
models combined with new observations confirmed this picture (e.g.,
Lindblad et al.\ 1973; Elmegreen 1982; Olano 1982).
However, these models seem inadequate for a description of the space
motions of the clumpy distribution of early-type stars, as no unique
expansion center and/or expansion age can be defined for them (e.g.,
Blaauw 1965; Lesh 1968, 1972; Stothers \& Frogel 1974; Frogel \&
Stothers 1977; Westin 1985; Comer\'on \& Torra 1994). The lack of a
homogeneous set of accurate radial velocities for all local early-type
stars prevents an optimal exploitation of the Hipparcos data, and a
full kinematic model is not yet available. Even so, we can use the
mean motions of the OB associations to shed light on the systemic
motions in the Gould Belt.

The associations Sco~OB2, Ori~OB1, Per~OB2, and Lac~OB1 are thought to
be components of the Gould Belt (e.g., Olano 1982).
The pattern displayed in Figure~29 seems to be shared by Tr~10, and
perhaps also by Vel~OB2 and Col~121, suggesting that these may all
belong to the same coherent structure. The new distance of Lac~OB1
reduces the extent of the Belt in the direction
$\ell$$\sim$$90^\circ\!$.\break Cep~OB2 is not located in the plane of the
Belt, and its motion is parallel to the Galactic plane. We conclude
that it does not seem to belong to the Belt. The Cas--Tau complex
which surrounds the $\alpha$~Persei cluster and shares its motion, as
well as Cep~OB6, are located inside the main ring of associations,
have a different space motion, and are both significantly older at
$\sim$50~Myr.

\section*{\centerline{\normalsize 10. SUMMARY AND FUTURE WORK}}

\subsection*{\centerline{\normalsize\sl 10.1. Main conclusions}}

We have carried out a comprehensive census of the stellar content of
the known OB associations within 1~kpc from the Sun (Table~1), based
on the globally accurate Hipparcos positions, proper motions and
parallaxes. We have used a combination of de Bruijne's (1998)
refurbished convergent point method and the `Spaghetti method' of
Hoogerwerf \& Aguilar (1998) to search for moving groups in fields
along the Galactic plane centered on nearby OB associations. The
selection procedure is objective and reliable. Monte Carlo simulations
provide an estimate of the expected number of interlopers. The
resulting census significantly improves the description of the
ensemble of OB associations in the Solar neigbourhood.

We have found a clear kinematic signature of a moving group in twelve
cases, all at distances less than 650~pc. These are the three
subgroups Upper Scorpius, Upper Centaurus Lupus, and Lower Centaurus
Crux of Sco~OB2, Vel~OB2, Tr~10, Col~121, Per~OB2, $\alpha$~Persei
(Per~OB3), Cas--Tau, Lac~OB1, Cep~OB2, and a new group in Cepheus,
which we have designated as Cep~OB6. Astrometric evidence for moving
groups in the fields of R~CrA, CMa~OB1, Mon~OB1, Ori~OB1, Cam~OB1,
Cep~OB3, Cep~OB4, Cyg~OB4, Cyg~OB7, and Sct~OB2 is inconclusive.
Previous photometric and spectroscopic studies indicate that many of
these are physical groups, but they are either at distances where the
Hipparcos parallaxes are of limited use and the number of members
bright enough to be included in the Hipparcos Catalogue is small, or
they have unfavorable kinematics, so that the group proper motion does
not stand out from that of the field stars in the Galactic disk (e.g.,
Ori~OB1).

Our member selection procedure has extended the firm\-ly established
kinematic members of the nearby OB associations to fainter magnitudes
than before. The previously established member
lists for the earliest spectral types have been corrected and
improved, and many members with later spectral type have been added.
These include evolved giants and supergiants, amongst which the famous
Wolf--Rayet stars $\gamma^2$~Vel (WR11) in Vel~OB2 and EZ~CMa (WR6) in
Col~121, and $\delta$~Cep in Cepheus~OB6. In some cases the new
members extend well into the regime of the F dwarfs, which is the
range where the stars in these young groups are expected to be in the
final pre-main sequence phase. Our list of astrometric members of
Sco~OB2 indeed contains such objects, providing a direct link between
the classical high-mass stellar content, and the lower-mass objects
which are still approaching the main sequence. This suggests strongly
that the mass functions of all OB associations extend to low masses,
so that their individual total masses are likely to be of order a few
thousand ${\rm M}_\odot$ or more.

Six of the moving groups we have detected do not appear in the
classical list of nearby OB associations (Table~1; Ruprecht 1966).
For Upper Centaurus Lupus and Lower Centaurus Crux this is due to the
absence of O stars. Tr~10 and Col~121 were known as open clusters, but
we have shown that they are in fact extended moving groups with all
the characteristics of unbound OB associations. The sparse group of
stars previously known as Vel~OB2 has similarly been transformed into
a full-fledged association, which is identical to Col~173, and to the
young stellar group postulated by Sahu (1992) to explain the observed
properties of an IRAS shell in this region. Cep~OB6 is a serendipitous
discovery in our Cep~OB2 field, and is an (old) OB association that
was completely overlooked. We have also confirmed the existence of
Cas--Tau, which was previously considered as a doubtful association,
and shown that it is connected kinematically to the $\alpha$~Persei
cluster. Cas--Tau can be considered as an extended halo of
$\alpha$~Persei, or the cluster as a bound component in the otherwise
unbound complex. It follows that the number of unbound young stellar
groups in the Solar neighbourhood may be significantly larger than
thought previously, especially those without bright O and early B-type
stars. Such groups are older than the classical OB associations, but
may well form the missing link between the youngest groups such as
Upper Scorpius and Ori~OB1, and the old dispersed associations like
Cas--Tau and Cep~OB6.

The measured distances of the associations, based on the mean
Hipparcos parallaxes of the astrometric members, and corrected for the
known biases (\S 9.1), are systematically smaller than the previous
photometric determinations. Whereas part of this effect must be caused
by the much improved membership lists, especially for the more distant
groups, we suspect that the upper main sequence may be a few tenths of
a magnitude fainter than assumed previously.

We have used the mean distances of the 12 moving groups to obtain an
improved map of the Solar neighbourhood. Together with the mean proper
motions, and the median radial motion derived from the radial
velocities in the Hipparcos Input Catalogue, this also provides a more
reliable description of the motions in the Gould Belt.

Elmegreen (1993) suggested that fragmentation and (the first
generation of recent) star formation in the Solar neighbourhood was
triggered by the passage of the Carina spiral arm $\sim$60~Myr ago,
and that Cas--Tau is one of the associations that formed in this
process. The high-mass stars in Cas--Tau then blew a large bubble
which resulted in Lindblad's ring, i.e., the Gould Belt, out of which
$\sim$20~Myr ago the molecular clouds and associations in
Scorpio--Centaurus--Lupus--Crux, Orion, Perseus, and Lacerta were
formed as the second generation. The present star formation seen in
Taurus--Auriga, and in Ophiuchus, is regarded as the third
generation. The results of our census are in qualitative agreement
with this picture. The traditional boundary between bound open cluster
and unbound expanding association is maybe not as clear-cut as is
commonly assumed (cf.\ Elmegreen \& Efremov 1998), and the total
amount of mass in the young stellar groups may have been
underestimated. A reassesment of the star formation history in the
Solar neighbourhood is clearly warranted.

\subsection*{\centerline{\normalsize\sl 10.2. Future work}}

The discovery of Cep~OB6 in our Cep~OB2 field suggests that there
might be other previously unidentified nearby associations. A
systematic search in regions of the strip $-30^\circ\!\! \le \! b \!
\le \! 30^\circ\!$ not covered in the current study, might reveal such
groups. We suspect this will not yield a rich harvest because the
sampling of possible members is likely to be sparse. Whereas many
candidate members of the known associations were included deliberately
in the Hipparcos Catalogue, in particular in the magnitude range where
the Catalogue is incomplete, this will generally not be the case in
other areas. We have generally ignored the known open clusters. Their
membership is well-defined in many cases, and is based on relative
proper motions from photographic studies in small fields. The limit on
stellar number density of about 3 per square degree results in the
Catalogue being significantly incomplete in the areas of open
clusters, even for the brightest members. Even so, it is interesting
to investigate whether extended halos can be found around other
clusters.

We have presented color-magnitude diagrams based on the $V$ and $B\!-\!V$
data reported in the Hipparcos catalog, but have not used these data
in the member selection. We have also not discussed the physical
parameters of the member stars in detail, mostly because the required
homogeneous multi-color photometry (and spectral classification) is
incomplete (\S 2.3). Now that the astrometric member selection has
reduced the number of candidates in a given field by an order of
magnitude, it is relatively straightforward to obtain the missing
photometry, and we are carrying out a program to do so. This is
particularly interesting in Sco~OB2, where the faintest astrometric
members can be connected directly to the populations of pre-main
sequence objects discovered through X-ray searches. This will provide
accurate ages, initial mass functions (Brown 1998), and the energy and
momentum input into the interstellar medium, and will allow a detailed
investigation of the influence of the young stellar group on the
surrounding distribution of gas and dust (e.g., de Geus 1992). The
Leiden--Dwingeloo HI survey (Hartmann \& Burton 1997) delineates the
HI distribution and kinematics with unprecedented accuracy over the
entire sky north of $\delta = -30^\circ\!$. It has already been used to
study the shell around Ori~OB1 (Brown et al.\ 1995). It will be
interesting to look for similar structures around other nearby
associations.

It is highly desirable to complement the uniform astrometric
measurements provided by Hipparcos with homogeneous radial velocities
with accuracies of $\sim$$3~{\rm km~s}^{-1}$ or better. While this is
relatively easy for spectral types F and later, it is a considerable
challenge for earlier spectral types, because of the modest number of
suitable spectral lines, the fast stellar rotation, and the large
number of spectroscopic binaries (e.g., Verschueren et al.\ 1997; Brown 
\& Verschueren 1997). An unbiased member selection based
only on radial velocities is still impractical, but measurement of the
radial velocity of the proper motion members identified here is
feasible. This will allow removal of a significant number of remaining
interlopers: for a typical measurement error in radial velocity of
3~km~s$^{-1}$, our Monte Carlo simulations (\S 3.4) suggest that the
expected reduction of the number of interlopers is typically about a
factor of 4. We are in the process of obtaining radial velocities for
this purpose. These will also provide further information on the
distribution of spectroscopic binaries in these groups. The many new
astrometric binaries discovered by Hipparcos are also a source for a
follow-up study.

The Hipparcos proper motions have an accuracy of
$\sim$1~mas~yr$^{-1}$, which corresponds to $\sim$$1~{\rm km~s}^{-1}$
at the distance of Sco~OB2. It is natural to ask whether the data can
resolve an overall expansion. This has been attempted from the ground
for a number of associations, and has led to the derivation of
kinematic ages, defined as the time since the group occupied a minimum
volume (e.g., Blaauw 1978). However, the relative motion of the Sun
with respect to the association also introduces a virtual expansion or
contraction in the measured proper motions, and hence unambiguous
measurement of internal motions and expansion ages requires radial
velocities for all group members that are significantly more accurate
than the internal dispersion of the group. Even then, it will not be
easy to define an expansion age (Brown et al.\ 1997). It follows that
all previously determined kinematic ages should be considered as
provisional.

The Hipparcos parallaxes are not sufficiently accurate to resolve the
internal structure of even the nearest associations. This would be
interesting, as it might help to delineate substructure, and hence
shed light on the details of the star formation process throughout an
interstellar cloud. However, it is possible to improve the individual
distance estimates further by using the proper motions (and radial
velocities) of the established members to compute secular parallaxes
(e.g., Dravins et al.\ 1997).

Hipparcos has provided a much improved description of the ensemble of
young stellar groups in the Solar neighbourhood, out to a distance of
$\sim$650~pc. Like the Hipparcos Catalogue, the lists of members given
in Table~C1 are incomplete beyond $V$$\sim$$7\fm3$, and may have
excluded a few genuine members (including some long-period binaries)
that are brighter (\S 3.5). Our lists extend to $V$$\sim$$10\fm5$, and
include a few pre-main sequence objects in the nearest groups. It is
natural to ask whether the member lists can be completed, and extended
to even fainter stars, by use of available large-scale ground-based
studies which can now be put on the Hipparcos reference
system. Unfortunately, the space motions of the young groups are not
large, and as a result the proper motions of the group members do not
differ very much from those of the field stars except in a few cases,
such as the $\alpha$~Persei cluster. For example, if we degrade the
vector point diagram for the Upper Scorpius field (Figure~5) to, e.g.,
the PPM accuracy of 3--4~mas~yr$^{-1}$ (R\"oser \& Bastian 1991;
Bastian et al.\ 1993; R\"oser, Bastian \& Kuzmin 1994), the group
hardly stands out. Reliable extension of our lists requires proper
motions with accuracies of order 2~mas~yr$^{-1}$ or better. These are
provided by the ACT Catalog (Urban, Corbin \& Wycoff 1998) and the TRC
Catalogue (H{\o}g et al.\ 1998), through a combination of the
positions in the Astrographic Catalog with the Tycho positions to
$V$$\sim$$11^{\rm m}$. The individual stellar positions in these two
catalogs have modest accuracy, but the $\sim$80~yr epoch difference
results in quite accurate proper motions. The first installment of the
ACT and TRC catalogs both contain about 1 million measurements. The
completion of the second edition of the Tycho Catalogue (see H{\o}g
1997) will provide proper motions of similar quality for about 2.5
million objects to $V$$\sim$$12^{\rm m}$. The ACT/TRC will also allow
us to investigate the astrometric membership for those long-period
binaries for which the Hipparcos measurements are suspect (\S
3.5). However, the 1~mas accuracy parallaxes obtained by Hipparcos
play a crucial role in culling interlopers from the membership lists
(Figure~4). With few exceptions, such parallaxes are not available for
fainter association members. A complete study of all OB associations
in the Solar neighbourhood (extent, distance, structure, kinematics)
has to await the future GAIA space astrometry mission (e.g., Perryman,
Lindegren \& Turon 1997; Gilmore et al.\ 1998).

\medskip
It is a pleasure to thank Luis Aguilar, Eug\`{e}ne de Geus, Rudolf Le
Poole, Jan Lub, Michael Perryman, and No\"el Robichon for stimulating
discussions and detailed comments on the manuscript. Ernest Stavro
Blofeld, Jan Brand, Ma\-rijn Franx, Frank Israel, Jan van Paradijs, and
Cor de Vries made significant contributions to the 1982 {\tt SPECTER}
proposal. This work has made extensive use of SIMBAD, ADS, and
SKYVIEW, and was supported in part by ESA, NWO and the Leids Kerkhoven
Bosscha Fonds.

%
%
\section*{\centerline{\normalsize REFERENCES}}
\begin{list}{}{
\setlength{\labelwidth}{0truecm}
\setlength{\labelsep}{0truecm}
\setlength{\leftmargin}{0.3truecm}
\setlength{\rightmargin}{0truecm}
\setlength{\itemindent}{-0.3truecm}
\setlength{\parsep}{-0.02truecm}
\setlength{\itemsep}{0truecm}
\small}

\item Abt H.A., Hunter J.H., 1962, ApJ, 136, 381

\item Abt H.A., Landolt A.U., Levy S.G., Mochnacki S., 1976, AJ, 81, 541

\item Adelman S.J., 1968, PASP, 80, 329

\item Adelman S.J., 1973, PASP, 85, 676

\item Abt H.A., Hunter J.H., 1962, ApJ, 136, 381

\item Abt H.A., Landolt A.U., Levy S.G., Mochnacki S., 1976, AJ, 81, 541

\item Adelman S.J., 1968, PASP, 80, 329

\item Adelman S.J., 1973, PASP, 85, 676

\item van Albada T.S., Klomp M., 1969, BAN, 20, 208

\item van Albada T.S., Sher D., 1969, BAN, 20, 204

\item Alcal\'a J.M., et al., 1996, A\&AS, 119, 7

\item Alcal\'a J.M., Chavarr\'{\i}a--K.\ C., Terranegra L., 1998, A\&A, 330, 1017

\item van Altena W.F., Cudworth K.M., Johnston K., Lasker B., Platais I.,
Russell J.L., 1993, in {\sl Workshop on databases for Galactic
structure}, eds A.G.D.\ Philip, B.\ Hauck, A.R.\ Upgren (Schenectady:
L.\ Davis Press), p.\ 250

\item Alter G., Ruprecht J., Van\'ysek V., 1970, {\sl Catalogue of Star
Clusters and Associations, $2^{\rm nd}$ Edition}, eds G.\ Alter, B.\
Bal\'azs, J.\ Ruprecht (Budapest: Akad\'emiai Kiad\'o)

\item Ambartsumian V.A., 1947, in {\sl Stellar Evolution and Astrophysics},
Armenian Acad.\ of Sci.\ (German translation, 1951, Abhandl.\
Sowjetischen Astron., 1, 33)

\item Ambartsumian V.A., 1949, Dokl.\ Akad.\ Nauk SSR, 68, 22

\item Ambartsumian V.A., 1954, IAU Trans., 8, 665

\item Andreazza C.M., Vilas--Boas J.W.S., 1996, A\&AS, 116, 21

\item Artyukhina N.M, 1972, Sov.\ Astron., 16, 317

\item Bal\'azs L.G., Kun M., 1989, AN, 310, 385

\item Bal\'azs L.G., Garibjanyan A.T., Mirzoyan L.V., Hambaryan V.V.,
Kun M., Front\'o A., Kelemen J., 1996, A\&A, 311, 145

\item Barnard E.E., 1927, {\sl A photographic atlas of selected regions of 
the Milky Way}, Washington: Carnegie Inst.

\item Bastian U., R\"oser S., Yagudin L.I., Nesterov V.W., 1993, PPM
Star Catalogue III, IV., Heidelberg, Spektrum Akad.\ Verlag

\item Becker W., 1963, ZfA, 57, 117

\item Bertiau F.C., 1958, ApJ, 128, 533

\item Bijaoui A., Lacoarret M., Granes P., 1981, A\&AS, 45, 483

\item Blaauw A., 1944, BAN, 10, 29

\item Blaauw A., 1946, {\sl PhD Thesis}, Groningen Univ.\ (Publ.\ Kapteyn
Astron.\ Lab., 52, 1)

\item Blaauw A., 1952a, BAN, 11, 405

\item Blaauw A., 1952b, BAN, 11, 414

\item Blaauw A., 1952c, AJ, 57, 199

\item Blaauw A., 1956, ApJ, 123, 408

\item Blaauw A., 1958, AJ, 63, 186

\item Blaauw A., 1960, in {\sl Present Problems Concerning the Structure and
Evolution of the Galactic System} (Nuffic Intern.\ Summer Course, The
Hague), 3, p.\ 1

\item Blaauw A., 1961, BAN, 15, 265

\item Blaauw A., 1964a, ARA\&A, 2, 213

\item Blaauw A., 1964b, in {\sl The Galaxy and the Magellanic Clouds}, IAU
Symp.\ 20, eds F.J.\ Kerr \& A.W.\ Rodgers, p.\ 50

\item Blaauw A., 1965, Proc.\ Royal Neth.\ Acad.\ of Sciences, 74, 54

\item Blaauw A., 1978, in {\sl Problems of Physics and Evolution of the
Universe}, ed.\ L.V.\ Mirzoyan (Yerevan), p.\ 101

\item Blaauw A., 1991, in {\sl The Physics of Star Formation and Early Stellar
Evolution}, eds C.J.\ Lada \& N.D.\ Kylafis, NATO ASI Ser.\ C,
Vol.\ 342, p.\ 125

\item Blaauw A., 1993, in {\sl Massive Stars: Their Lives in the
Interstellar Medium}, eds J.P.\ Cassinelli \& E.B.\ Churchwell, ASP
Conf.\ Ser., 35, p.\ 207

\item Blaauw A., van Albada T.S., 1963, ApJ, 137, 791

\item Blaauw A., van Albada T.S., 1964, preprint 

\item Blaauw A., Morgan W.W., 1953, ApJ, 117, 256

\item Blaauw A., Morgan W.W., Bertiau F.C., 1955, ApJ, 121, 557

\item Blaauw A., Hiltner W.A., Johnson H.L., 1959, ApJ, 130, 69

\item Blanco V.M., Williams A.D., 1959, ApJ, 130, 482

\item Blitz L., 1978, {\sl PhD Thesis}, Columbia Univ.

\item Blitz L., Fich M., Stark A.A., 1982, ApJS, 49, 183

\item Bohnenstengel H.--D., Wendker H.J., 1976, A\&A, 52, 23

\item Bok B.J., 1934, Harvard College Obs.\ Circ., 384, 1

\item Borgman J., Blaauw A., 1964, BAN, 17, 358 

\item Boss B., 1910, AJ, 26, 163

\item Botley C.M., 1980, Observatory, 100, 211

\item Braes L.L.E., 1962, BAN, 16, 297

\item Brand J., Blitz L., 1993, A\&A, 275, 67

\item Brandner W., K\"ohler R., 1998, ApJ, 499, L79

\item Brandner W., Alcal\'a J.M., Kunkel M., Moneti A., Zinnecker H., 1996,
A\&A, 307, 121

\item Brandt J.C., Stecher T.P., Crawford D.L., Maran S.P., 1971, ApJ, 163,
L99

\item Brown A., 1950, ApJ, 112, 225

\item Brown A., 1987, ApJ, 322, L31

\item Brown A.G.A., 1998, in {\sl The Stellar Initial Mass Function}, 
eds G.F.\ Gilmore \& D.\ Howell, ASP Conf.\ Ser., 142, p.\ 45

\item Brown A.G.A., Verschueren W., 1997, A\&A, 319, 811

\item Brown A.G.A., de Geus E.J., de Zeeuw P.T., 1994, A\&A, 289, 101 

\item Brown A.G.A., Hartmann D., Burton W.B., 1995, A\&A, 300, 903

\item Brown A.G.A., Dekker G., de Zeeuw P.T., 1997, MNRAS, 285, 479

\item Brown A.G.A., Arenou F., van Leeuwen F., Lindegren L., Luri X.,
1997b, ESA SP--402, p.\ 63

\item Brown A.G.A., Walter F.M., Blaauw A., 1998, in {\sl The Orion Complex
Revisited}, eds M.J.\ McCaughrean \& A.\ Burkert, ASP Conf.\ Ser., in
press 

\item de Bruijne J.H.J., 1998, MNRAS, submitted

\item de Bruijne J.H.J., Hoogerwerf R., Brown A.G.A., Aguilar L.A., de
Zeeuw P.T., 1997, ESA SP--402, p.\ 575
 
\item van Bueren H.G., 1952, BAN, 11, 385

\item Burrows D.N., Singh K.P., Nousek J.A., Garmire G.P., Good J., 1993,
ApJ, 406, 97

\item Buscombe W., 1963, MNRAS, 126, 29

\item Campbell B., Persson S.E., McGregor P.J., 1986, ApJ, 305, 336

\item Canavaggia R., Fribourg M.--L., 1934, Bull.\ Astron., 2e Serie, Tome
9, 259

\item Cappa de Nicolau C.E., P\"oppel W.G.L., 1991, A\&AS, 88, 615

\item Catalog of Open Clusters, 1997, \hfill\break
{\sl http://heasarc.gsfc.nasa.gov/W3Browse/all/openclust.html}

\item Cernicharo J., Bachiller R., Duvert G., 1985, A\&A, 149, 273

\item {\v C}ernis K., 1990, Ap\&SS, 166, 315

\item {\v C}ernis K., 1993, Baltic Astronomy, 2, 214

\item Chen H., Grenfell T.G., Myers P.C., Hughes J.D., 1997, ApJ, 478, 295

\item Chereul E., Cr\'ez\'e M., Bienaym\'e O., 1997, ESA SP--402, p.\ 545

\item Clari\'a J.J., 1974a, AJ, 79, 1022

\item Clari\'a J.J., 1974b, A\&A, 37, 229

\item Cohen M., Kuhi L.V., 1976, ApJ, 210, 365

\item Cohen M., Kuhi L.V., 1979, ApJS, 41, 743

\item Collinder P., 1931, Ann.\ Obs.\ Lund, 2, No.\ 1 

\item Comer\'on F., Torra J., 1994, A\&A, 281, 35

\item Comer\'on F., Torra J., G\'omez A.E., 1998, A\&A, 330, 975

\item Coyne G., Burley--Mead J., Kaufman M., 1969, AJ, 74, 103

\item Crampton D., 1968, AJ, 73, 338

\item Crawford D.L., 1960, AJ, 65, 487

\item Crawford D.L., 1961, ApJ, 133, 860

\item Crawford D.L., 1963, ApJ, 137, 523

\item Crawford D.L., Barnes J.V., 1970, AJ, 75, 952

\item Crawford D.L., Warren W.H., 1976, PASP, 88, 930

\item Cudworth K.M., 1998, in {\sl Proper Motions and Galactic Astronomy},
ed.\ R.M.\ Humphreys, ASP Conf.\ Ser., 127, p.\ 91

\item Dame T.M., Ungerechts H., Cohen R.S., de Geus E.J., Grenier I.A., 
May J., Murphy D.C., Nyman L.--{\AA}., Thaddeus P., 1987, ApJ, 322, 706

\item Dehnen W., Binney J.J., 1998, MNRAS, 298, 387

\item Delhaye J., Blaauw A., 1953, BAN, 12, 72

\item Digel S.W., Lyder D.A., Philbrick A.J., Puch D., Thaddeus P., 1996,
ApJ, 458, 561

\item Dravins D., Lindegren L., Madsen S., Holmberg J., 1997, ESA SP--402, p.\ 733 

\item Duvert G., Cernicharo J., Bachiller R., G\'omez--Gonz\'alez J., 1990,
A\&A, 233, 190

\item Eddington A.S., 1910, MNRAS, 71, 43

\item Eddington A.S., 1914, {\sl Stellar movements and the structure of the
universe}, London: MacMillan \& Co.

\item Eggen O.J., 1961, Royal Obs.\ Bull., No.\ 41

\item Eggen O.J., 1978, PASP, 90, 436

\item Eggen O.J., 1980, ApJ, 238, 627

\item Eggen O.J., 1981, ApJ, 247, 507

\item Eggen O.J., 1982, ApJS, 50, 199

\item Eggen O.J., 1983, AJ, 88, 197

\item Eggen O.J., 1986, AJ, 92, 1074

\item Elmegreen B.G., 1982, in {\sl Submillimeter wave astronomy}, eds
J.E.\ Beckman \& J.P.\ Phillips, p.\ 3

\item Elmegreen B.G., 1993, in {\sl Protostars and Planets III}, eds E.H.\
Levy \& J.I.\ Lunine (Tucson: Univ.\ of Arizona Press), p.\ 97

\item Elmegreen B.G., Efremov Y.N., 1998, in {\sl The Orion Complex
Revisited}, eds M.J.\ McCaughrean \& A.\ Burkert, ASP Conf.\ Ser., in
press

\item Elmegreen B.G., Lada C.J., 1977, ApJ, 214, 725

\item ESA, 1989, The Hipparcos Mission, ESA SP--1111

\item ESA, 1997, The Hipparcos and Tycho Catalogues, ESA SP--1200

\item Feast M., Whitelock P., 1997, MNRAS, 291, 683

\item Feigelson E.D., Lawson W.A., 1997, AJ, 113, 2130

\item Feinstein A., 1967, ApJ, 149, 107
 
\item Fernley J.A., Skillen I., Jameson R.F., 1989, MNRAS, 237, 947

\item Fitzgerald M.P., Harris G.L.H., Reed B.C., 1990, PASP, 102, 865

\item Fredrick L.W., 1954, AJ, 59, 321

\item Fredrick L.W., 1956, AJ, 61, 437

\item Fresneau A., 1980, AJ, 85, 66

\item Frink S., R\"oser S., Neuh\"auser R., Sterzik M.F., 1997, A\&A, 325, 613

\item Frogel J.A., Stothers R., 1977, AJ, 82, 890

\item Gaposchkin S., Greenstein J.L., 1936, Harvard College Obs.\ Bull., 904, 8

\item Garmany C.D., 1973, AJ, 78, 185

\item Garmany C.D., Stencel R.E., 1992, A\&AS, 94, 211
 
\item Garrison R.F., 1967, ApJ, 147, 1003

\item Garrison R.F., 1970, AJ, 75, 1001

\item Garrison R.F., Kormendy J., 1976, PASP, 88, 865

\item Gatewood G., de Jonge J.K., Stephenson B., 1993, PASP, 105, 1101

\item de Geus E.J., 1992, A\&A, 262, 258

\item de Geus E.J., de Zeeuw P.T., Lub J., 1989, A\&A, 216, 44

\item de Geus E.J., Lub J., van der Grift E., 1990, A\&AS, 85, 915

\item Gieren W.P., Barnes T.G., Moffett T.J., 1993, ApJ, 418, 135

\item Gilmore G.F., et al., 1998, in {\sl Astronomical Interferometry}, SPIE
Proc.\ 3350, eds R.D.\ Reasenberg \& M.\ Shao, in press
[astro--ph/9805180]

\item Gim\'enez A., Clausen J.V., 1994, A\&A, 291, 795

\item Gingrich C.H., 1922, ApJ, 56, 139 

\item Glaspey J.W., 1971, AJ, 76, 1041

\item Glaspey J.W., 1972, AJ, 77, 474

\item Glass I.S., Penston M.V., 1975, MNRAS, 172, 227

\item Goss W.M., Manchester R.N., Brooks J.W., Sinclair M.W., Manefield
G.A., Danziger I.J., 1980, MNRAS, 191, 533

\item Goudis C., 1982, {\sl The Orion Complex: A case study of interstellar matter}, 
Ap\&SSL, p.\ 90

\item Gould B.A., 1874, Proc.\ AAAS, p.\ 115

\item Grasdalen G.L., Strom K.M., Strom S.E., 1973, ApJL, 184, L53

\item Greene T.P., Young E.T., 1992, ApJ, 395, 516

\item Greenstein J.L., 1948, ApJ, 107, 375 

\item Gregorio--Hetem J., L\'epine J.R.D., Quast G.R., Torres C.A.O., de la
Reza R., 1992, AJ, 103, 549

\item Grillmair C.J., Freeman K.C., Irwin M., Quinn P.J., 1995, AJ, 109, 2553

\item Guetter H.H., 1976, AJ, 81, 1120

\item Guetter H.H., 1977, AJ, 82, 598

\item Guillout P., Sterzik M.F., Schmitt J.H.M.M., Motch C., Egret D., Voges
W., Neuh\"auser R., 1998, A\&A, 334, 540

\item Guti\'errez--Moreno A., Moreno H., 1968, ApJS, 15, 459
 
\item Hardie R.H., Crawford D.L., 1961, ApJ, 133, 843

\item Hardie R.H., Seyfert C.K., 1959, ApJ, 129, 601

\item Hardie R.H., Seyfert C.K., Grenchik R.T., 1957, AJ, 62, 143

\item Harju J., Haikala L.K., Mattila K., Mauersberger R., Booth R.S., Nordh
H.L., 1993, A\&A, 278, 569

\item Haro G., 1953, ApJ, 117, 73

\item Harris D.L., 1955, ApJ, 121, 554

\item Harris D.L., 1956, ApJ, 123, 371

\item Harris D.L., Morgan W.W., Roman N.G., 1954, ApJ, 119, 622 

\item Harris D.L., Morgan W.W., Roman N.G., 1955, AJ, 60, 53

\item Hartmann D., Burton W.B., 1997, {\sl Atlas of Galactic Neutral
Hydrogen}, Cambridge Univ.\ Press

\item Hartquist T.W., Morfill G.E., 1983, ApJ, 266, 271

\item Hauck B., Mermilliod M., 1990, A\&AS, 86, 107

\item Haug U., 1970, A\&AS, 1, 35
 
\item Heckmann O., L\"ubeck K., 1958, ZfA, 45, 243

\item Heckmann O., Dieckvoss W., Kox H., 1956, AN, 283, 109

\item Heeschen D.S., 1951, ApJ, 114, 132

\item Herbig G.H., 1954, PASP, 66, 19

\item Herbig G.H., Rao N.K., 1972, ApJ, 174, 401

\item Herbst W., 1980, in {\sl Star Clusters}, IAU Symp.\ 85, ed.\
J.E.\ Hesser, p.\ 33

\item Herbst W., Assousa G.E., 1977, ApJ, 217, 473

\item Herbst W., Racine R., Richer H.B., 1977, PASP, 89, 663

\item Herbst W., Racine R., Warner J.W., 1978, ApJ, 223, 471

\item Herbst W., Miller D.P., Warner J.W., Herzog A., 1982, AJ, 87, 98

\item van Herk G., 1959, AJ, 64, 348

\item Herschel J.F.W., 1833, Phil.\ Trans., p.\ 481

\item Herschel J.F.W., 1847, {\sl Results of Astron.\ Observations made
during the years 1834--1838 at the Cape of Good Hope}, London, p.\ 385

\item Hiltner W.A., 1956, ApJS, 2, 389

\item Hoag A.A., Johnson H.L., Iriarte B., Mitchell R.I., Hallam K.L., 
Sharpless S., 1961, Publ.\ Naval Obs., 17, 345

\item H{\o}g E., 1997, ESA SP--402, p.\ 25

\item H{\o}g E., Kuzmin A., Bastian U., Fabricius C., Kuimov K., Lindegren
L., Makarov V.V., R\"oser S., 1998, A\&A, 335, L65

\item Hoogerwerf R., Aguilar L.A., 1998, MNRAS, submitted (HA)

\item Hoogerwerf R., de Bruijne J.H.J., Brown A.G.A., Lub J., Blaauw
A., de Zeeuw P.T., 1997, ESA SP--402, p.\ 571

\item Howard W.E., 1958, AJ, 63, 50

\item Howarth I.D., Schmutz W., 1995, A\&A, 294, 529

\item Hubble E., 1922a, ApJ, 56, 162 

\item Hubble E., 1922b, ApJ, 56, 400 
 
\item van der Hucht K.A., et al., 1997, New Astronomy, 2, 245

\item Hughes J.D., Hartigan P., Clampitt L., 1993, AJ, 105, 571

\item Hughes J.D., Hartigan P., Krautter J., Kelemen J., 1994, AJ, 108, 1071

\item Hughes V.A., 1988, ApJ, 333, 788

\item Humphreys R.M., 1978, ApJS, 38, 309

\item Jones D.H.P., 1971, MNRAS, 152, 231

\item Jordi C., Trullols E., Galad\'{\i}--Enr\'{\i}quez D., 1996, A\&A,
312, 499

\item Jung J., Bischoff M., 1971, Bull.\ Inf.\ Centre Donn.\ Stell., 2, 8

\item Kalas P., Jewitt D., 1997, Nature, 386, 52

\item Kaper L., van Loon J.T., Augusteijn T., Goudfrooij P., Patat F.,
Waters L.B.F.M., Zijlstra A.A., 1997, ApJ, 475, L37 (err.\ ApJ, 479,
L153)

\item Kapteyn J.C., 1911, Trans.\ Int.\ Solar Union, 3, 215

\item Kapteyn J.C., 1914, ApJ, 40, 43

\item Kapteyn J.C., 1918, ApJ, 47, pp 104, 146, 255

\item Keller H.--U., 1970, Wien.\ Ann., 29, No.\ 3

\item Kenyon S.J., Dobrzycka D., Hartmann L., 1994, AJ, 108, 1872

\item Khavtassi D.S., 1960, {\sl Atlas of Galactic Dark Nebulae},
Abastumani Astrophys.\ Obs.

\item Klochkova V.G., Kopylov I.M., 1985, Bull.\ Spec.\ Astrophys.\ Obs., 20, 3

\item Knacke R.F., Strom K.M., Strom S.E., Young E.T., Kunkel W., 1973, ApJ,
179, 847

\item Kodaira K., Greenstein J.L., Oke J.B., 1970, ApJ, 159, 485

\item Koyama K., Hamaguchi K., Ueno S., Kobayashi N., Feigelson E.D., 1996,
PASJ, 48, L87

\item Krelowski J., Megier A., Strobel A., 1996, A\&A, 308, 908

\item Krzem\'{\i}nski W., Oskanjan V., 1961, Act.\ Astr., 11, 1

\item Kulikovsky P., 1940, Bull.\ Sternberg State Astr.\ Inst., No.\ 2

\item Kun M., 1986, Ap\&SS, 125, 13

\item Kun M., P\'asztor L., 1990, Ap\&SS, 174, 13 

\item Kun M., Bal\'azs L.G., T\'oth I., 1987, Ap\&SS, 134, 211 

\item Kutner M.L., Dickman R.L., Tucker K.D., Machnik D.E., 1979, ApJ, 232, 724

\item Lada C.J., Gottlieb C.A., Litvak M.M., Lilley A.E., 1974, ApJ, 194, 609

\item Lada C.J., Alves J., Lada E.A., 1996, AJ, 111, 1964

\item Lada E.A., Lada C.J., 1995, AJ, 109, 1682

\item Ladd E.F., Lada E.A., Myers P.C., 1993, ApJ, 410, 168

\item Lamers H.J.G.L.M., Harzevoort J.M.A.G., Schrijver H., Hoogerwerf R.,
Kudritzki R.P., 1997, A\&A, 325, L25
 
\item van Leeuwen F., 1985, in {\sl Dynamics of Star Clusters}, IAU
Symp.\ 113, eds J.\ Goodman \& P.\ Hut, p.\ 579

\item van Leeuwen F., 1994, in {\sl Galactic and Solar System Optical
Astrometry: Observation and Application}, eds L.V.\ Morrison \&
G.F.\ Gilmore (Cambridge Univ.\ Press), p.\ 223

\item Lesh J.R., 1968, ApJS, 17, 371

\item Lesh J.R., 1969, AJ, 74, 891

\item Lesh J.R., 1972, A\&AS, 5, 129
 
\item Levato H., Abt H.A., 1976, PASP, 88, 141

\item Levato H., Malaroda S., 1975, PASP, 87, 173

\item Levato H., Malaroda S., Morrell N., Solivella G., 1987, ApJS, 64, 487

\item Lindblad P.O., 1967, BAN, 19, 34

\item Lindblad P.O., Grape K., Sandqvist Aa., Schober J., 1973, A\&A, 24, 309

\item Lindblad P.O., Palou\v{s} J., Lod\'en K., Lindegren L., 1997, ESA
SP--402, p.\ 507

\item Lindegren L., 1989, ESA SP--1111, Vol.\ 3, p.\ 311 

\item Lindegren L., 1997, ESA SP--402, p.\ 13

\item Lissauer J.L., 1997, Nature, 386, 18

\item Loren R.B., 1976, ApJ, 209, 466

\item Loren R.B., 1979, ApJ, 227, 832

\item Lozinskaya T.A., Sitnik T.G., Toropova M.S., 1987, Sov.\ Astron., 31, 493 

\item Lub J., Pel J.W., 1977, A\&A, 54, 137 

\item Lynds B.T., 1962, ApJS, 7, 1

\item Lynds B.T., 1969, PASP, 81, 496

\item Lyng\aa\ G., 1959, AfA, 2, 379

\item Lyng\aa\ G., 1962, AfA, 3, 65

\item Lyng\aa\ G., Wramdemark S., 1984, A\&A, 132, 58
 
\item Machnik D.E., Hettrick M.C., Kutner M.L., Dickman R.L., Tucker K.D.,
1980, ApJ, 242, 121

\item MacConnell D.J., 1968, ApJS, 16, 275

\item Maeder A., Meynet G., 1994, A\&A, 287, 803

\item Malaroda S., 1981, PASP, 93, 614

\item Margulis M., et al., 1990, ApJ, 352, 615

\item Markarian B.E., 1952, Proc.\ Acad.\ Sci.\ Armenian SSR, 15, 13

\item Marraco H.G., 1978, A\&A, 70, L61

\item Marraco H.G., Rydgren A.E., 1981, AJ, 86, 62

\item Marschall L.A., Comins N.F., Karshner G.B., 1990, AJ, 99, 1536

\item Mathieu R.D., 1986, in {\sl Highlights of Astronomy}, Vol.\ 7, p.\ 481

\item May J., Murphy D.C., Thaddeus P., 1988, A\&AS, 73, 51

\item McCaughrean M.J., Burkert A., 1998, {\sl The Orion Complex Revisited},
ASP Conf.\ Ser., in press

\item Mermilliod J.--C., 1998, {\sl http://obswww.unige.ch/webda/}

\item Mermilliod J.--C., Mermilliod M., 1994, {\sl Catalog of Mean UBV Data on
Stars}, Springer Verlag

\item Mermilliod J.--C., Turon C., Robichon N., Arenou F., Lebreton Y.,
1997, ESA SP--402, p.\ 643

\item Meurers J., Mikulitsch W., 1968, Wien Ann., 27, No.\ 5

\item Meynet G., Mermilliod J.--C., Maeder A., 1993, A\&AS, 98, 477

\item Morgan W.W., Sharpless S., Osterbrock D., 1952a, Sky \& Telescope,
11, 138

\item Morgan W.W., Sharpless S., Osterbrock D., 1952b, AJ, 57, 3

\item Morgan W.W., Whitford A.E., Code A.D., 1953, ApJ, 118, 318

\item Morrell N., Levato H., 1991, ApJS, 75, 965

\item Mourard D., Bonneau D., Koechlin L., Labeyrie A., Morand F.,
Stee Ph., Tallon--Bosc I., Vakili F., 1997, A\&A, 317, 789
 
\item Murphy D.C., May J., 1991, A\&A, 247, 202

\item Murphy D.C., Cohen R., May J., 1986, A\&A, 167, 234

\item Nakano M., Wiramihardja S.D., Kogure T., 1995, PASJ, 47, 889

\item Neuh\"auser R., 1997, Science, 276, 1363

\item Neuh\"auser R., Brandner W., 1998, A\&A, 330, L29

\item Odenwald S.F., 1988, ApJ, 325, 320

\item Ogura K., 1984, PASJ, 36, 139

\item Olano C.A., 1982, A\&A, 112, 195

\item Olano C.A., Walmsley C.M., Wilson T.L., 1994, A\&A, 290, 235

\item \"Opik E.J., 1953, IrAJ, 2, 219

\item Panagia N., Thum C., 1981, A\&A, 98, 295

\item Pannekoek A., 1929, Publ.\ Astron.\ Inst.\ Amsterdam, No.\ 2, p.\ 63

\item Patel N.A., Goldsmith P.F., Snell R.L., Hezel T., Xie T., 1995,
ApJ, 447, 721

\item P\'{e}rez M.R., 1991, RMxAA, 22, 99

\item Perryman M.A.C., Lindegren L., Turon C., ESA SP--402, p.\ 743

\item Perryman M.A.C., et al., 1998, A\&A, 331, 81

\item Petrie R.M., 1958, MNRAS, 118, 80

\item Petrie R.M., 1962, MNRAS, 123, 501

\item Pinsonneault M.H., Stauffer J., Soderblom D.R., King J.R., Hanson
R.B., 1998, ApJ, 504, 170

\item Plaskett J.S., 1928, MNRAS, 88, 395

\item Plaskett J.S., Pearce J.A., 1934, MNRAS, 94, 679

\item Plummer H.C., 1913, MNRAS, 73, 492

\item P\"oppel W.G.L., 1997, Fund.\ of Cosmic Phys., 18, 1

\item Poveda A., Ruiz J., Allen C., 1967, Bol.\ de los Obs.\ Tonantzintla y
Tacubaya, Vol.\ 4, No.\ 28, 86

\item Preibisch T., 1997, A\&A, 324, 690

\item Preibisch T., Zinnecker H., Herbig G.H., 1996, A\&A, 310, 456

\item Preibisch T., Guenther E., Zinnecker H., Sterzik M., Frink S., R\"oser
S., 1998, A\&A, 333, 619

\item Prosser C.F., 1992, AJ, 103, 488

\item van Rensbergen W., Vanbeveren D., de Loore C., 1996, A\&A, 305, 825

\item Racine R., 1968, AJ, 73, 233
 
\item Rajamohan R., 1976, Pram\~ana, 7, 160

\item Rasmuson N.H., 1921, Lund Medd., Ser.\ II, 26, 1

\item Rasmuson N.H., 1927, Lund Medd., Ser.\ II, 47b, 1

\item Reichen M., Lanz T., Golay M., Huguenin D., 1990, Ap\&SS, 163, 275

\item Reynolds R.J., Ogden P.M., 1978, ApJ, 224, 94

\item Robichon N., Arenou F., Turon C., Mermilliod J.--C., Lebreton Y., 1997, 
ESA SP--402, p.\ 567 
 
\item Roman N.G., Morgan W.W., 1950, ApJ, 111, 426

\item R\"oser S., Bastian U., 1991, PPM Star Catalogue I, II., Heidelberg, 
Spektrum Akad.\ Verlag

\item R\"oser S., Bastian U., Kuzmin A., 1994, A\&AS 105, 301

\item Rossano G.S., 1978, AJ, 83, 234

\item Ruprecht J., 1966, IAU Trans., 12B, 348

\item Ruprecht J., Bal\'azs B., White R.E., 1981, {\sl Catalogue of Star
Clusters and Associations, Supplement to $2^{\rm nd}$ Edition}, ed.\
B.\ Bal\'azs (Budapest: Akad\'emiai Kiad\'o) \hfill\break
{\sl http://obswww.unige.ch/webda/}

\item Rydgren A.E., 1971, PASP, 83, 656

\item Sahu M.S., 1992, {\sl PhD Thesis}, Groningen Univ.

\item Sahu M.S., Blaauw A., 1994, The Messenger, 76, 48

\item Sancisi R., 1970, A\&A, 4, 387

\item Sancisi R., 1974, in {\sl Galactic radio astronomy}, IAU Symp.\ 60, 
eds F.J.\ Kerr \& S.C.\ Simonson, p.\ 115

\item Sancisi R., Goss W.M., Anderson C., Johansson L.E.B., Winnberg A., 1974, 
A\&A, 35, 445

\item Sandqvist Aa., Tomboulides H., Lindblad P.O., 1988, A\&A, 205, 225

\item Sargent A.I., 1977, ApJ, 218, 736

\item Sargent A.I., 1979, ApJ, 233, 163

\item Schaerer D., Schmutz W., Grenon M., 1997, ApJ, 484, L153

\item Schmidt K.H., 1958, AN, 284, 76

\item Schmidt--Kaler T., 1961, ZfA, 53, 28

\item Schreur J.J., 1970, AJ, 75, 38

\item Schulz N.S., Bergh\"ofer T.W., Zinnecker H., 1997, A\&A, 325, 1001

\item Schwartz R.D., Gyulbudaghian A.L., Wilking B.A., 1991, ApJ, 370, 263

\item Sciortino S., Damiani F., Favata F., Micela G., 1998, A\&A, 332, 825

\item Seyfert C.K., Hardie R.H., 1957, AJ, 62, 146

\item Seyfert C.K., Hardie R.H., Grenchik R.T., 1960, ApJ, 132, 58

\item Simonson S.C., 1968, ApJ, 154, 923

\item Simonson S.C., van Someren Greve H.W., 1976, A\&A, 49, 343

\item Skinner S.L., Itoh M., Nagase F., 1998, New Astronomy, 3, 37

\item Slawson R.W., Hill R.J., Landstreet J.D., 1992, ApJS, 82, 117

\item Slettebak A., Bahner K., Stock J., 1961, ApJ, 134, 195

\item Smart W.M., 1936, MNRAS, 96, 568

\item Smart W.M., 1939, MNRAS, 100, 60

\item Smith H., Eichhorn H., 1996, MNRAS, 281, 211

\item Snell R.L., Scoville N.Z., Sanders D.B., Erickson N.R., 1984, ApJ,
284, 176

\item Snow T.P., Hanson M.M., Seab C.G., Saken J.M., 1994, ApJ, 420, 632

\item Stauffer J.R., Hartmann L.W., Burnham J.N., Jones B.F., 1985, ApJ, 289, 247

\item Stauffer J.R., Hartmann L.W., Jones B.F., 1989, ApJ, 346, 160

\item Stauffer J.R., Hartmann L.W., Prosser C.F., Randich S., Balachandran
S., Patten B.M., Simon T., Giampapa M., 1997, ApJ, 479, 776
 
\item Steffey P.C., 1973, PASP, 85, 520

\item Stickland D.J., 1995, Observatory, 115, 180

\item Stock J., 1984, RMxAA, 9, 127

\item Stothers R., Frogel J.A., 1974, AJ, 79, 456

\item Straka W.C., 1971, in {\sl The Gum Nebula and Related Problems}, eds
S.P.\ Maran, J.C.\ Brandt, T.P.\ Stecher, NASA SP--332, p.\ 126

\item Straka W.C., 1973, ApJ, 180, 907

\item Strom K.M., Strom S.E., Grasdalen G.L., 1974, ApJ, 187, 83

\item Strom S.E., Grasdalen G.L., Strom K.M., 1974, ApJ, 191, 111

\item Strom S.E., Strom K.M., Carrasco L., 1974, PASP, 86, 798

\item Taylor K.N.R., Storey J.W.V., 1984, MNRAS, 209, 5P

\item Testi L., Olmi L., Hunt L., Tofani G., Felli M., Goldsmith P.,
1995, A\&A, 303, 881

\item Tian K.P., van Leeuwen F., Zhao J.L., Su C.G., 1996, A\&AS, 118, 503

\item Torra J., G\'omez A.E., Figueras F., Comer\'on F., Grenier S.,
Mennessier M.O., Mestres M., Fern\'andez D., 1997, ESA SP--402, p.\
513

\item Trullols E., Jordi C., 1997, A\&A, 324, 549
 
\item Turner D.G., 1976, ApJ, 210, 65

\item Turon C., et al., 1992, Hipparcos Input Catalogue, ESA SP--1136

\item Upton E.K.L., 1971, in {\sl The Gum Nebula and Related Problems},
eds S.P.\ Maran, J.C.\ Brandt, T.P.\ Stecher, NASA SP--332, p.\ 119

\item Urban S.E., Corbin T.E., Wycoff G.L., 1998, AJ, 115, 2161 

\item Vasilevskis S., Sanders W.L., Balz A.G.A., 1965, AJ, 70, 797

\item Verschueren W., David M., Brown A.G.A., 1996, in {\sl The Origins,
Evolution, and Destinies of Binary Stars in Clusters}, eds E.F.\
Milone \& J.--C.\ Mermilliod, ASP Conf.\ Ser., 90, p.\ 131

\item Verschueren W., Brown A.G.A., Hensberge H., David M., Le Poole R.S.,
de Geus E.J., de Zeeuw P.T., 1997, PASP, 109, 868

\item Vrba F.J., Strom S.E., Strom K.M., 1976, AJ, 81, 317

\item Wackerling L.R., 1972, PASP, 84, 827

\item Walker M.F., 1956, ApJS, 2, 365

\item Walker M.F., 1969, ApJ, 155, 447

\item Walter F.M., Boyd W.T., 1991, ApJ, 370, 318
 
\item Walter F.M., Vrba F.J., Mathieu R.D., Brown A., Myers P.C., 1994, AJ, 107, 692 

\item Walter F.M., Vrba F.J., Wolk S.J., Mathieu R.D., Neuh\"auser R., 1997,
AJ, 114, 1544

\item Walter F.M., Wolk S.J., Sherry W., 1998, in {\sl Cool Stars, Stellar
Systems, and the Sun}, eds R.\ Donahue \& J.\ Bookbinder, in press

\item Warren W.H., Hesser J.E., 1977a, ApJS, 34, 115

\item Warren W.H., Hesser J.E., 1977b, ApJS, 34, 207

\item Warren W.H., Hesser J.E., 1978, ApJS, 36, 497

\item Weaver W.B., 1970, AJ, 75, 938

\item Welin G., 1979, A\&A, 79, 334

\item Wendker H.J., Higgs L.A., Landecker T.L., 1991, A\&A, 241, 551

\item Westin T.N.G., 1985, A\&AS, 60, 99

\item Whiteoak J.B., 1961, MNRAS, 123, 245

\item Wichmann R., Bastian U., Krautter J., Jankovics I., Rucinski S.M.,
1997a, ESA SP--402, p.\ 359

\item Wichmann R., Krautter J., Covino E., Alcal\'a J.M., Neuh\"auser R.,
Schmitt J.H.M.M., 1997b, A\&A, 320, 185

\item Wielen R., Schwan H., Dettbarn C., Jahrei{\ss} H., Lenhardt H., 1997,
ESA SP--402, p.\ 727

\item Wilking B.A., Harvey P.M., Joy M., Hyland A.R., Jones T.J., 1985, ApJ,
293, 165

\item Wilking B.A., Greene T.P., Lada C.J., Meyer M.R., Young E.T., 1992,
ApJ, 397, 520

\item Wilking B.A., McCaughrean M.J., Burton M.G., Giblin T., Rayner J.T.,
Zinnecker H., 1997a, AJ, 114, 2029

\item Wilking B.A., Schwartz R.D., Fanetti T.M., Friel E.D., 1997b, PASP, 109, 549

\item Woolley R.v.d.R., Eggen O.J., 1958, Observatory, 78, 149

\item de Zeeuw P.T., Brand J., 1985, in {\sl Birth and Evolution of Massive
Stars and Stellar Groups}, eds W.\ Boland \& H.\ van Woerden
(Dordrecht: Reidel), p.\ 95

\item de Zeeuw P.T., Brown A.G.A., Verschueren W., 1994, in {\sl Galactic
and Solar System Optical Astrometry: Observation and Application},
eds L.V.\ Morrison \& G.F.\ Gilmore (Cambridge Univ.\ Press), p.\
215

\item de Zeeuw P.T., Brown A.G.A., de Bruijne J.H.J., Hoogerwerf R., Le
Poole R.S., Lub J., Blaauw A., 1997, ESA SP--402, p.\ 495

\end{list}

%
%
\appendix

\section*{\centerline{\normalsize A. RECIPE FOR MEMBERSHIP DETERMINATION}}

Here we specify the parameters of our astrometric member selection
procedure for a moving group, following the principles set out in \S\S
3.1--3.3. Table~A1 lists the values for these parameters for the 12
associations and subgroups which we have identified in the Hipparcos
Catalogue (\S\S 4--8).

First, we fix the parameters $D_{\rm guess}$, $D_{\rm range}$,
$\sigma_{\rm int}$, $\epsilon_{\rm min}$, $t_{\rm min}$ and $S_{\rm
min}$. Here $D_{\rm guess}$ is the initial guess for the mean
distance (in pc) of the association, and $D_{\rm range}$ denotes the
range in distance (in pc) over which we expect to find association
members (\S 3.2). The default value for $D_{\rm range}$ is 100~pc. If
the association was detected by de Zeeuw et al.\ (1997), this distance
is taken as the initial guess; otherwise, $D_{\rm guess}$ is taken
from de Zeeuw et al.\ (1994). The quantity $\sigma_{\rm int}$ is the
internal velocity dispersion in the association (one-dimensional, in
${\rm km}~{\rm s}^{-1}$). We choose $\sigma_{\rm int} = 3.0~{\rm
km}~{\rm s}^{-1}$ (cf.\ Mathieu 1986; Tian et al.\ 1996; \S 3.3).
The quantity $\epsilon_{\rm min}$ is the stop criterion for the
convergent point method, as defined by de Bruijne (1998); it sets an
acceptable value of $\chi^2$ (\S 3.1) by requiring that $\epsilon\geq
\epsilon_{\rm min}$, where
\begin{equation}
\epsilon\!=\! {1 \over \Gamma({N\over 2}\!-\!1)} \! 
           \int_{\chi^2}^\infty \!
           x^{{N\over 2}-2} {\rm e}^{-x} \, {\rm d} x.
\end{equation}
We set $\epsilon_{\rm min}$ equal to 0.954, based on a Monte Carlo
analysis (de Bruijne 1998). The parameter $t_{\rm min}$ is the
minimum ratio of total proper motion and related $1\sigma$ error
required for a star to enter the convergent point method
(eq.~\ref{deftpm}). Finally, the quantity $S_{\rm min}$ is the Spaghetti
method membership criterion limit for a star to be selected as member
(\S 3.2); it was fixed at $S_{\rm min} = 0.1$ based on a Monte Carlo
analysis (Hoogerwerf \& Aguilar 1998).

\begin{figure*}[t]
\centerline{
\psfig{file=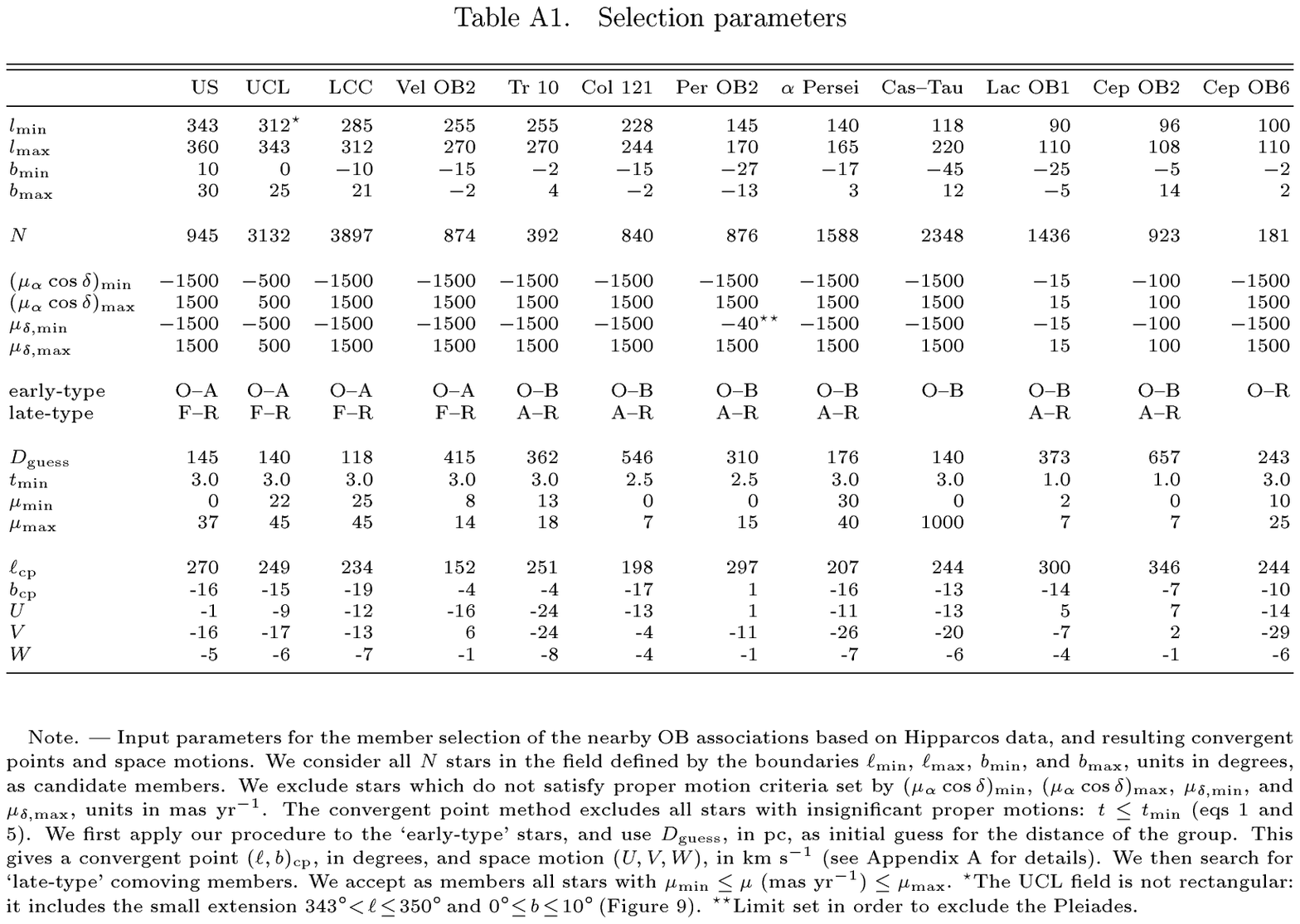,width=15truecm,silent=}}
\end{figure*}

Before starting the member selection, we discard all stars which do
not satisfy $(\mu_\alpha \cos \delta)_{\rm min} \leq \mu_\alpha \cos 
\delta \leq$\break $(\mu_\alpha \cos \delta)_{\rm max}$ and $\mu_{\delta 
{\rm ,min}} \leq \mu_\delta \leq \mu_{\delta {\rm ,max}}$. We also
check the quality of the astrometric parameters: stars with 15~per
cent or more rejected data, or a goodness-of-fit indicator larger than
3.0 (fields H$29$ and H$30$ in the Hipparcos Catalogue, respectively),
are discarded. Typically, this excludes less than 0.5~per cent of the
total sample. We then apply the convergent point and Spaghetti method
independently to a sample of early-type stars, e.g., all OB stars (see
Table~A1 for the precise definition of `early-type'). The results of
the two methods are combined by intersecting the individual membership
lists. Next, we consider the vector point diagram for all stars in
the intersection. This vector point diagram contains association
members as well as a small number of field stars. These field stars
have astrometric properties which make them indistinguishable from the
association members. However, the association members generally
define a clear concentration in this diagram whereas the field stars
have a much broader distribution. In order to remove as many field
stars as possible while, at the same time, rejecting as few as
possible genuine members, we define a lower and upper cutoff in $\mu
\equiv \sqrt{\mu_{\ell}^2 \cos^2 b + \mu_b^2}$, $\mu_{\rm min}$ and
$\mu_{\rm max}$, respectively, such that the range $\mu_{\rm min} \leq
\mu \leq \mu_{\rm max}$ mainly corresponds to the association members.
We construct the resulting parallax distributions to make sure that
our proper motion cutoffs efficiently separate the foreground and
background stars from the association members. We then define as
secure members all stars in the intersection of the individual
membership lists which lie within the range $\mu_{\rm min}
\leq \mu \leq \mu_{\rm max}$.

The convergent point and Spaghetti method are then applied to the
secure early-type members to determine their convergent point and
space motion, respectively. This convergent point is used to define
the membership probability, $P = p_{\rm cp}$, for each star (cf.\ \S
3.1). We then fix the convergent point and space motion to those of
the secure early-type members and apply both methods to the remaining
(late-type) stars in the field to search for `comoving members'. The
individual membership lists are again intersected. Using the same
proper motion cutoffs as for the early-type members, we define the
secure late-type members and their membership probabilities.

Given the final list of all secure members we determine the
association distance $D$ and the related value for the systematic
underestimate $D_{\rm sys} \geq 0$ (\S 3.6). If the corrected
distance $D_{\rm cor} \equiv D + D_{\rm sys}$ deviates more than
25~per cent of the value of $D_{\rm guess}$, we repeat the whole
member selection process starting with $D_{\rm guess} = D_{\rm
cor}$. This turned out to be required only in the case of Cep~OB2 (\S
8.2), where we replaced the initial $D_{\rm guess} = 559~$pc by
$D_{\rm guess} = 657~$pc.

\begin{figure*}[t]
\centerline{
\psfig{file=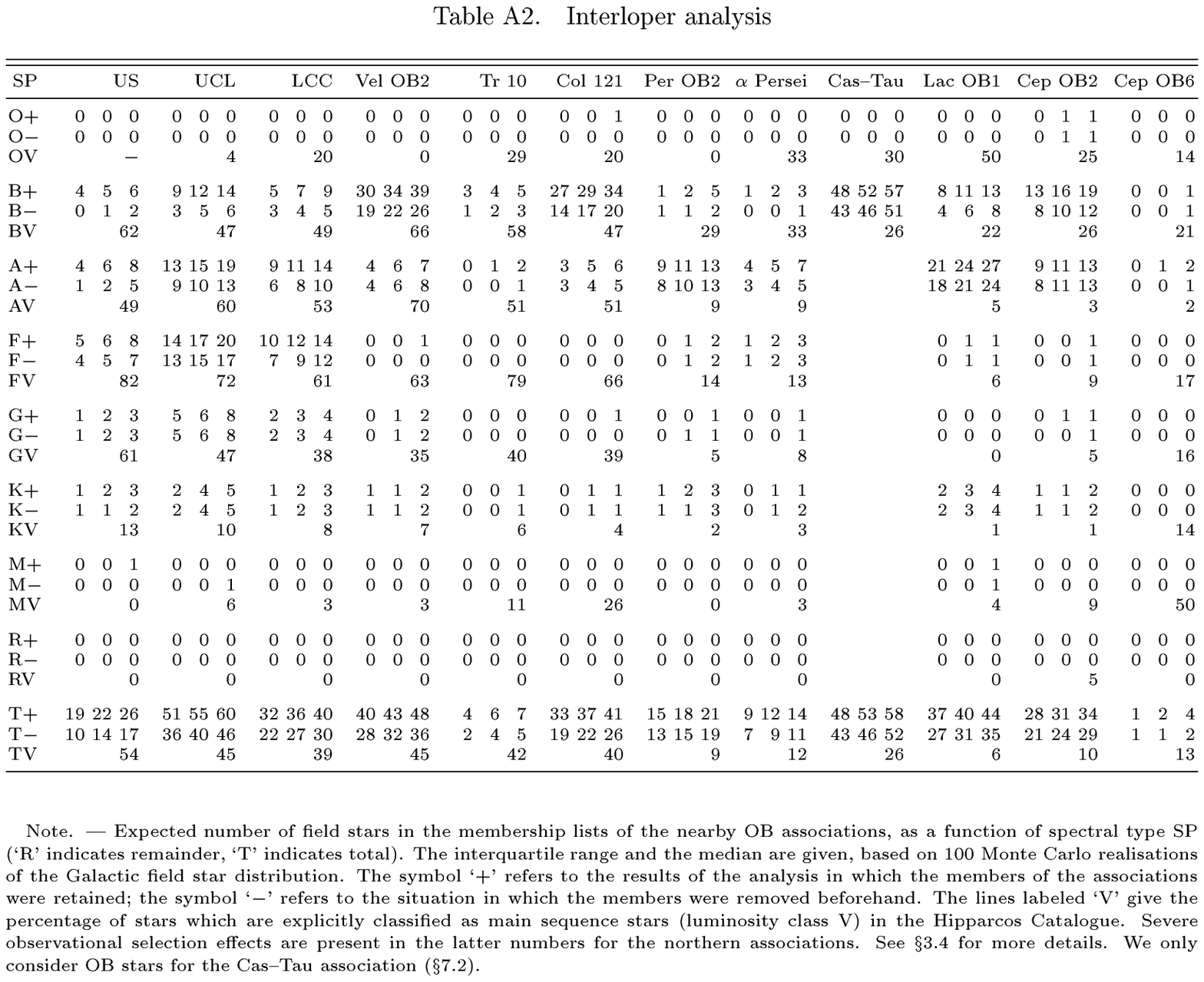,width=15truecm,silent=}}
\end{figure*}

Table~A1 also gives the convergent points and space velocities found
by the convergent point and Spaghetti method, respectively. These
values must be interpreted with great care. Most importantly, the
convergent point and space motion may contain an unknown component due
to expansion or contraction of the association (cf.\ \S 3.7).
Furthermore, the uncertainty of the space motion in the line-of-sight
direction is large, and can be up to several tens of km~s$^{-1}$;
compare the large difference in the $V$-component of the space motion
of $\alpha$~Persei and Cep~OB6 (Table~A1) with their similar mean
motion displayed in Figure~29 (\S 9.2). Moreover, the space motions
are biased due to the use of the parallax in the calculation of the
tangential velocity (Hoogerwerf \& Aguilar 1998). The typical
uncertainty in the convergent points is several degrees along the
great circle connecting the convergent point with the association, and
much smaller, less than a degree, in the perpendicular direction (de
Bruijne 1998).

Our procedure to estimate the expected number of interlopers in the
list of association members produced by the above selection method is
described in detail in \S 3.4. Table~A2 summarizes the results as a
function of spectral type, for the 12 fields where the Hipparcos
measurements allow identification of a moving group.

\section*{\centerline{\normalsize B. MEAN DISTANCES} }

The conversion from parallax to distance introduces a bias in the
distance (Smith \& Eichhorn 1996; Brown et al.\ 1997b). Consider a
sphere of radius $R$ with its center at a distance $D_0$ ($R < D_0$),
with a homogeneous distribution of stars. An observer who measures
distances $D$ to the individual stars in the group will derive a
number density as a function of distance, $F(D)\, {\rm d} D$, with
\begin{equation}
  F(D) = {3 D^2\over 2 R^3} 
     \Bigl( 1 - {D_0^2 + D^2 - R^2 \over 2 D_0 D} \Bigr).
\end{equation}
The mean distance $\langle D \rangle$ of the group, computed by
integrating $D F(D)\, {\rm d} D$, then is
\begin{equation}
\langle D \rangle = D_0 \Bigl(1 + {1 \over 5}{R^2 \over D_0^2} \Bigr).
\end{equation}
Thus, the mean of the individual distances is an overestimate of the
true mean distance $D_0$, and it depends on the angular size of the
group.

The number density distribution as a function of parallax, 
$F^\star(\pi)\, {\rm d} \pi$, is given by
\begin{equation}
F^\star(\pi) = {1 \over \pi^2} F({1\over\pi}) = {3 \over 2 R^3} \Bigl
( {1\over \pi^4} \!-\!  {1\over 2 D_0 \pi^5} \!+\!  {R^2 - D_0^2\over2
D_0 \pi^3} \Bigr),
\end{equation}
and the mean parallax $\langle \pi \rangle$ is
\begin{equation}
\langle \pi \rangle = {1 \over D_0}.
\end{equation}
The errors in the Hipparcos parallaxes are distributed as a Gaussian.
As this is a symmetric function, convolution of the parallax
distribution, and then computing the mean parallax of the group, again
produces $1/D_0$. It follows that the mean parallax of a spherical
homogeneous group is an unbiased estimator of the mean distance of the
group. Any spherically symmetric density distribution can be written
as the sum of homogeneous spheres with different radii. The inverse
of the mean parallax therefore is an unbiased estimator of the
distance for all spherical star clusters.

The OB associations studied in this paper are gravitationally unbound,
and not likely to be exactly spherical. We have computed the observed
mean parallax for ellipsoidally stratified groups, as a function of
viewing angle, and find that in this case there is a small bias in the
mean parallax. However, to an accuracy of about 1 per cent, the
corresponding mean distance is equal to the true distance for
realistic shapes and for all viewing angles. We conclude that the
inverse of the mean parallax is an essentially unbiased estimator of
the distance to the associations.

The calculation of mean distances for open clusters based on Hipparcos
data must be done with care, as it needs to take into account the
correlated measurements for stars located in a small area. This can be
achieved by using Hipparcos intermediate data. Mermilliod et al.\
(1997) have done so, and find a distance of 184.2$\pm$7.5~pc for 32
members of the $\alpha$~Persei cluster, which is the highest-density
group investigated in our census. Our method yields the same distance
for the early-type stars, which gives confidence that the distances of
the low-density associations obtained in this paper are unbiased.
This agrees with the conclusions of, e.g., Lindegren (1989) and
Pinsonneault et al.\ (1998).

\begin{figure}[t]
\centerline{
\psfig{file=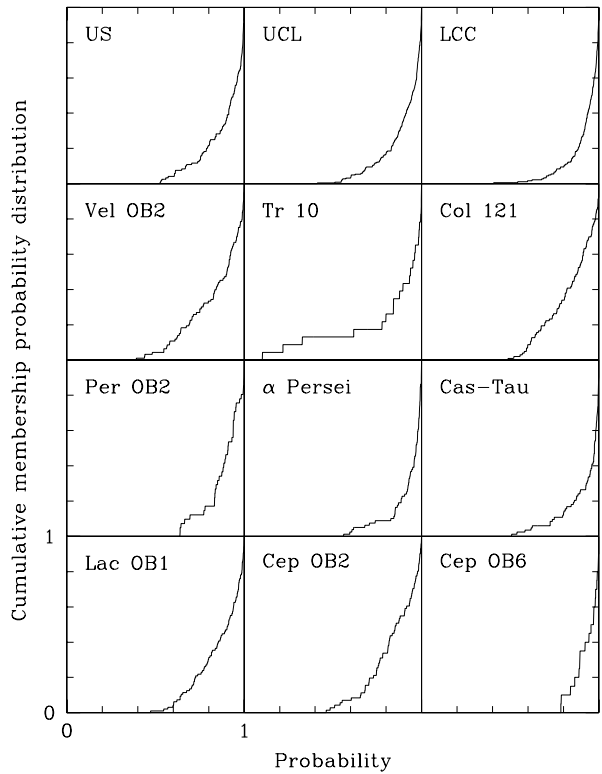,width=8truecm,silent=}}
\caption{\small Cumulative membership probability
distributions of the secure members of the associations listed in
Table~C1.}
\end{figure}

\section*{\centerline{\normalsize C. MEMBERSHIP LISTS}}

Rather than give lengthy lists with properties of the selected member
stars, and cross references to the different numbering systems used in
previous studies, we simply tabulate in Table~C1 the entry numbers in
the Hipparcos Catalogue for all secure members found by our selection
procedure, together with the membership probability defined in
eq.~(6). All other properties, including HR and HD numbers, as well as
spectral types, $V$ and $B\!-\!V$ photometry, and indication of
multiplicity, can be found in the Hipparcos Catalogue. Radial
velocities can be found in the Hipparcos Input Catalogue, or in
SIMBAD.

Many classical proper motion members of the nearby associations
are not confirmed by our selection procedure, and hence do not appear
in Table C1. However, as described in \S 3.5, some of these `rejects'
are in fact long-period binaries, for which the Hipparcos measurements
do not necessarily reflect the mean space motion but rather an
instantaneous orbital motion. The ground-based proper motions used in
the classical studies generally are based on a much longer
timespan. It is likely that a significant number of these astrometric
binaries are in fact members. They are identified in the main text in
\S\S 4--8.

Figure~C1 illustrates the cumulative membership probabilities for all
secure members. In general, $\sim$45--65~per cent of all members have
membership probabilities $P \geq 90$~per cent, while $\sim$10--15~per
cent have $P \geq 99.7$~per cent. The median membership probabilities
for $\alpha$~Persei and Cep~OB6 are higher than for the other
associations, but the small number of Cep~OB6 members (20 stars)
combined with the special member selection for this case (\S 8.5) call
for caution.

\begin{figure*}[t]
\centerline{
\psfig{file=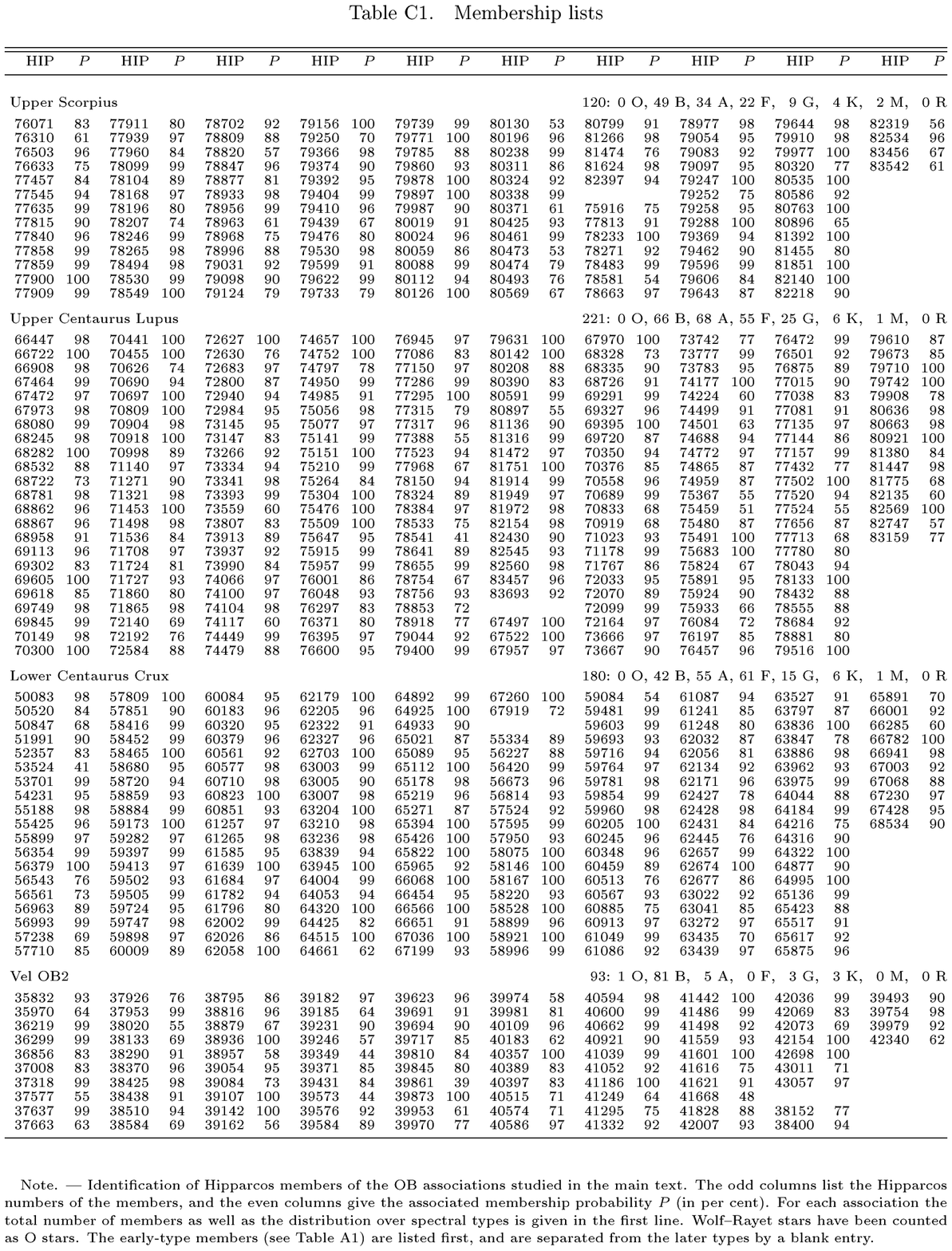,width=15truecm,silent=}}
\end{figure*}
\begin{figure*}[t]
\centerline{
\psfig{file=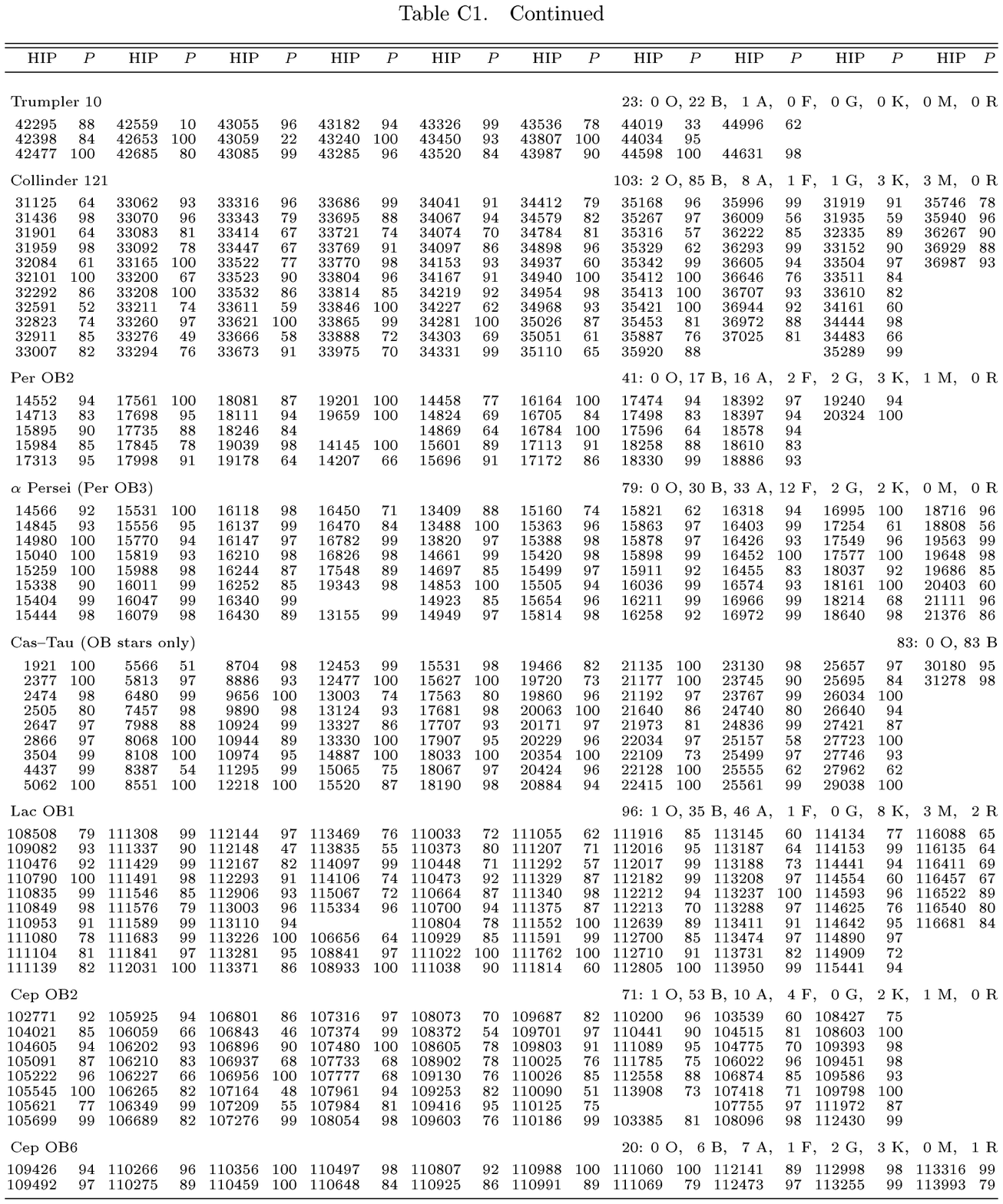,width=15truecm,silent=}}
\end{figure*}

\end{document}